\documentclass[11pt,letterpaper]{article}

\usepackage{article-template}
\usepackage{tikz}
\usetikzlibrary{arrows.meta,decorations.markings}

\renewcommand{\thesubfigure}{(\alph{subfigure})}
\newcommand{\panellabel}[1]{%
  \begingroup\phantomsubcaption\phantomsection\label{#1}\thesubfigure\endgroup}

\newcommand{\BR}{\operatorname{BR}}

\newcommand{\E}{\mathbb E}
\newcommand{\V}{\mathcal V}
\newcommand{\cE}{\mathcal E}
\newcommand{\R}{\mathbb R}
\newcommand{\C}{\mathcal C}
\newcommand{\cP}{\mathcal P}
\newcommand{\cA}{\mathcal A}
\newcommand{\cQ}{\mathcal Q}
\newcommand{\cX}{\mathcal X}

\newcommand{\rhoFP}{$\rho$-FP}
\newcommand{\vc}{\mathbf{c}}
\newcommand{\e}{\mathbf{e}}
\newcommand{\f}{\mathbf{f}}
\newcommand{\q}{\mathbf{q}}

\newcommand{\vv}{\mathbf{v}}
\newcommand{\vu}{\mathbf{u}}
\newcommand{\z}{\mathbf{z}}
\newcommand{\1}{\mathbf{1}}
\newcommand{\0}{\mathbf{0}}

\usepackage{booktabs,tabularx,pifont}
\newcommand{\cmark}{\ding{51}}
\newcommand{\xmark}{\ding{55}}

\newtheorem{thm}[theorem]{Theorem}
\newtheorem{lem}[theorem]{Lemma}
\newtheorem{defi}[theorem]{Definition}
\newtheorem{prop}[theorem]{Proposition}
\newtheorem{coro}[theorem]{Corollary}
\newtheorem{rmk}[theorem]{Remark}

\newtheorem{asp}[theorem]{Assumption}
\newtheorem{exm}[theorem]{Example}

\title{\LARGE\bfseries
Fictitious Play When Agents Are Not Expected-Utility Maximizers:
A Geometric Analysis in $2$-Action Games}

\author{
Hao Liang$^\dagger$
\and Solaleh Mohammadi$^\dagger$
\and Kaiqing Zhang$^\dagger$
}

\date{November 12, 2026}

\begin{document}
\maketitle

\begingroup
\renewcommand{\thefootnote}{}
\begin{NoHyper}
\footnotetext{$^\dagger$University of Maryland, College Park.
Emails: \texttt{\{haoliang,solalehm,kaiqing\}@umd.edu}.}
\end{NoHyper}
\endgroup
\setcounter{footnote}{0}

\begin{abstract}
Fictitious play (FP) is one of the earliest and most fundamental learning procedures in games,  where each agent %
independently chooses a best response to the empirical distribution of opponents' past actions by maximizing the \emph{expected} payoff with respect to it. However, learning agents may not always be expected-utility maximizers: they may use \emph{nonlinear} evaluation functionals to account for the uncertainty induced by the opponents, e.g., via risk measures. In this work, we study nonlinear fictitious play (NL-FP), in which each pure action is evaluated by a general, possibly nonlinear functional of the payoff distribution induced by the opponent's empirical belief. To analyze NL-FP, we first identify a single-crossing property (SCP) of evaluation functionals that preserves the classical geometry of FP in $2 \times 2$ games. Furthermore, we conduct a novel geometric analysis of the convergence behavior of NL-FP in  $2 \times 2$ games, even when the SCP fails. Beyond $2 \times 2$ games, we show that the convergence of NL-FP in $2 \times n$ ($n\ge3$) games may fail even for continuous-time  NL-FP, when the two-action player uses conditional value-at-risk (CVaR) %
in a $2 \times 3$ game. In addition, we identify a finite-crossing condition that ensures convergence of the selected continuous-time dynamics and the discrete-time iterates %
in general \(2\times n\) games, which also provides new insights into the global landscape of continuous-time classical FP in such games.
\end{abstract}

\section{Introduction}
\label{sec:intro}
Fictitious play (FP) is one of the earliest and most fundamental learning procedures in games \citep{brown1951iterative}. It is a canonical model of adaptive and independent learning in repeated game playing, requiring neither coordination among agents\footnote{We use ``agent'' and ``player'' interchangeably throughout.}  nor knowledge of opponents'  payoffs.  The idea behind  FP is simple: each agent repeatedly \emph{best responds} to the empirical distribution of opponents' past actions, i.e., the empirical {belief} about the opponent's mixed strategy. Specifically, agent $i$ selects its action $a^i_k$ at iteration $k\ge0$ according to

\begin{equation}
\label{eqt:fp}
a_k^i\in\arg\max_{a^i}\mathbb{E}\!\left[r^i\!\left(a^i,A^{-i}\right)\right],\quad A^{-i} \sim \hat{\pi}_k^{-i},
\end{equation}
where $r^i$ is the payoff function for agent $i$, and
$\hat{\pi}_k^{-i}$ is the empirical belief based
on the opponent's actions  before iteration $k$
(see also Equation~\eqref{eqt:blf}). The convergence of empirical beliefs to the Nash equilibrium set was established in two-player zero-sum games \citep{robinson1951iterative},  $2\times2$ games \citep{miyasawa2010convergence,metrick1994fictitious},
nondegenerate $2\times n$ games \citep{berger2005fictitious},
and potential games \citep{monderer1996fictitious}, under
the assumptions and tie-breaking conventions of the
respective results.
Despite these positive results, nonconvergence %
of FP is known to occur even in $3 \times 3$ games \citep{shapley1963some}.

Classical FP evaluates each action by its \textit{expected} payoff against the empirical distribution of opponents’ past actions.  In particular, each action’s evaluation is \textit{linear} in the empirical belief. Such {linear} evaluation is considered natural, and is consistent with  the seminal results by \citet{vonNeumannMorgenstern1944}, \citet{nash1951noncooperative}, and \citet{harsanyi1967games} that laid the foundations of the
\emph{expected-utility maximization}  framework in game theory.

More generally, however, agents may evaluate payoff distributions using \textit{nonlinear} alternatives that account for features beyond their means, such as downside risk, tail outcomes, and probability weighting. A substantial literature developed such alternatives, including coherent and convex risk measures \citep{artzner1999coherent,follmer2002convex}, optimized certainty equivalents \citep{ben2007old}, conditional value-at-risk (CVaR) \citep{rockafellar2000optimization}, mean-semivariance and other downside-risk criteria \citep{estrada2007mean}, spectral risk measures \citep{acerbi2002spectral}, rank-dependent expected utility \citep{quiggin1982theory}, and cumulative prospect theory (CPT) \citep{tversky1992advances}.
Within the literature of game theory and multi-agent learning, prior work studied equilibrium existence, computation, and learning in games with {non-expected-utility}   preferences \citep{fiat2010players},  mean-variance equilibria and stochastic FP \citep{slumbers2023game}, CPT-based learning \citep{phade2023learning}, and
risk-sensitive equilibrium concepts \citep{mazumdar2025tractable}.

Nonetheless, when it comes to fictitious play as a learning procedure, these developments did not address the following basic question:
\emph{How does the choice of evaluation functional shape the convergence and geometry %
of FP?} In this paper, we study \textit{nonlinear} fictitious play (NL-FP), which retains the empirical-belief updates of classical FP but evaluates each action through a general, possibly nonlinear functional of its induced payoff distribution (Equation~\eqref{eqt:rho_fp}). This formulation includes classical FP as a special case, and allows us to examine which properties of the evaluation functional preserve convergence and which can undermine it.
 The following fundamental questions regarding NL-FP naturally arise:
\begin{enumerate}[leftmargin=*]
    \item \textit{Does nonlinear FP converge in $2\times n$ games under general evaluations?}

    \item \textit{When does nonlinear FP converge, and when can convergence fail?}

    \item \textit{When convergence occurs, where does nonlinear FP converge to?}
\end{enumerate}

Towards answering these questions, we focus on the canonical model of learning in $2$-action games, i.e., games in which at least one player
has two actions.  The convergence of classical FP in nondegenerate $2\times n$ games \citep{berger2005fictitious}, together with the nonconvergence result in $3\times 3$ games \citep{shapley1963some}, makes the former a natural setting and starting point. We begin by introducing \emph{$\rho$-Nash equilibrium} ($\rho$-NE), where $\rho=(\rho^1,\rho^2)$ denotes the players' evaluation functionals. This notion generalizes the standard Nash equilibrium and reduces to it when both players evaluate payoffs by expectation.
In $2\times2$ games, the response of each player is determined by a scalar \textit{advantage function}. We first show that the \textit{single-crossing property} (SCP) of the advantage function preserves the geometry of classical FP, establishing the convergence of NL-FP. We further identify several sufficient conditions on evaluation functionals,  under which the advantage function satisfies SCP. However, we also show that several functionals commonly used in finance and decision theory indeed violate SCP. Hence, one of our major contributions is to further establish the convergence of NL-FP  even when SCP fails.  Furthermore,
we also show that such positive convergence results can be fragile beyond $2\times2$ games, even when only the two-action player is risk-averse.
We summarize our key  contributions as follows, and
list our main results   in Table \ref{tab:convergence-summary}.

\begin{itemize}[leftmargin=*]
    \item We identify the SCP of advantage functions as a key condition for preserving the convergence of classical FP \citep{metrick1994fictitious}, and provide sufficient conditions on the evaluation functionals to identify SCP.
    In particular, we verify the SCP of several broad classes of nonlinear functionals. However, we  show that SCP may not hold for several functionals common in finance and decision theory.

    \item We prove the convergence of the continuous-time NL-FP dynamics
    and the discrete-time NL-FP iterates under a \textit{finite-crossing condition} in $2 \times 2$ games, via a novel geometric analysis of a finite planar grid. %

    \item We show that the convergence of continuous-time NL-FP in  $2\times n$ games does not hold in general, by constructing a nonconvergent  counterexample where the two-action player uses CVaR in a $2 \times 3$ game, in sharp contrast to the $2 \times 2$ case.

    \item We further identify a finite-crossing condition that
ensures convergence of the continuous-time NL-FP dynamics
and the discrete-time NL-FP iterates
in $2\times n$ games when the row best response depends on
a linear statistic of the opponent's strategy. Our geometric analysis
reveals the basin--separatrix landscape
and, under stronger simple interior column-crossing
assumptions, provides an alternative proof of the classical
FP convergence result \citep{berger2005fictitious}.

\end{itemize}

\begin{table}[t]
\centering
\small
\renewcommand{\arraystretch}{1.2}
\caption{Convergence of nonlinear fictitious play. %
\cmark: convergence to a $\rho$-NE under the stated core
assumptions; \xmark: nonconvergence can occur;
\textemdash: not established in this work.}
\label{tab:convergence-summary}
\begin{tabularx}{\textwidth}{@{}lXccl@{}}
\toprule
Game & Conditions
& \shortstack{Continuous}
& \shortstack{Discrete}
& Reference \\
\midrule
$2\times2$
& Strict single-crossing property (Definition \ref{def:scp})
& \cmark & \cmark
& Prop.~\ref{prop:sscp} \\

$2\times2$
& Single-crossing property (Definition \ref{def:scp})
& \cmark & \cmark
& Prop.~\ref{prop:scp} \\

$2\times2$
& Finite-crossing (Assumption \ref{asp:2by2})
& \cmark & \cmark
& Thms.~\ref{thm:ct_conv_2by2}, \ref{thm:dt_conv_2by2} \\

\midrule

$2\times3$
& CVaR row player and risk-neutral column player
& \xmark & \textemdash
& Prop. \ref{prop:cvar_nonconvergence} \\

$2\times n$
& Finite-crossing and nondegeneracy (Assumptions~\ref{asp:mean_rho} and~\ref{asp:fin_sw})
& \cmark & \cmark
& Thms.~\ref{thm:conv_2byn}, \ref{thm:conv_dt_2byn} \\

\bottomrule
\end{tabularx}

\smallskip
\begin{minipage}{\textwidth}
\footnotesize
Continuous-time positive results use the selection rules stated
in the respective results; the finite-grid analysis uses equilibrium
stopping. Discrete-time positive results use
pure best responses, with the tie-breaking restrictions stated
in the respective results.
\end{minipage}
\end{table}

\paragraph{Organization.} The rest of the paper is organized as follows. Section \ref{sec:pre} introduces the game model, nonlinear
evaluation functionals, $\rho$-NE%
, and the continuous-time dynamics and discrete-time iterates of NL-FP.
Section~\ref{sec:scp} studies the convergence of NL-FP under the SCP in $2\times2$ games. Section~\ref{sec:ct} establishes the convergence of continuous-time NL-FP under the finite-crossing assumption and characterizes the global landscape and local geometry in $2\times 2$ games. Section~\ref{sec:dt} proves the convergence of discrete-time NL-FP under the finite-crossing assumption in  $2\times 2$ games. Section~\ref{sec:2byn} extends the analysis to $2\times n$ games, presenting a counterexample in a $2\times 3$ game, a new geometric analysis of classical FP in $2\times n$ games, and sufficient conditions for continuous- and discrete-time NL-FP convergence in $2\times n$ games. Section~\ref{sec:con} concludes the paper.  We defer a detailed discussion of related work to Appendix~\ref{app:lit}.

\section{Preliminaries}
\label{sec:pre}

\paragraph{Notation.} Let $a<b$ be two real numbers. We denote by $\mathscr{D}([a,b])$ and $\mathscr{D}$ the space of all cumulative distribution functions (CDFs) supported on $[a,b]$ and the space of all CDFs on the reals, respectively. %
For a positive integer $n$, $[n]:=\{1,2,\cdots,n\}$; we also set $[0]:=\varnothing$. Let $\Delta_n:=\{\vv\in\R^n|\1_n^\top \vv=1, \vv \ge \0_n\}$ be the probability simplex in $\R^n$. $\1_n$ and $\0_n$ denote the $n$-dimensional vector whose components are all ones and all zeros, respectively. $\e_i$ denotes the one-hot vector whose $i$-th component is 1.  We denote by $(u,v;p)$ a two-point  distribution where $u$ occurs with probability $p$ and $v$ otherwise. For a vector $\vv\in\R^n$ and $\q\in\Delta_n$, we use $(\vv,\q)$ to denote the discrete probability distribution $\sum_{i\in[n]} q_i\delta_{v_i}$.  For two random variables $X\sim F$ and $Y\sim G$, we say that $Y$ dominates $X$  if $\forall x\in\mathbb{R},\ F(x)\ge G(x)$, and we write  $Y\succeq X$ or $X\preceq Y$. We write $X \overset{d}{=} Y$ if $X$ and $Y$ have the same probability distributions. We denote the support of a probability distribution $F$ by $\text{supp}(F)$. In the following, we start with introducing the key setting: $2\times 2$ and $2\times n$ bimatrix games.

\subsection{Matrix games}

\paragraph{$2\times 2$ bimatrix games.}
We consider a $2\times 2$ bimatrix game $G$:
\[
\begin{array}{c|cc}
 & \textbf{L} \quad (\text{w.p.} \ q) & \textbf{R} \quad (\text{w.p.} \ 1-q)\\ \hline
\textbf{U} \quad (\text{w.p.} \ p) &(a,e)&(b,f)\\
\textbf{D} \quad (\text{w.p.} \ 1-p) &(c,g)&(d,h)
\end{array},
\]
where $p\in[0,1]$ and $q\in[0,1]$ denote the probability of choosing $\textbf{U}$ for player 1 and $\textbf{L}$ for player 2, respectively. For fixed $q\in[0,1]$, player 1's payoff distributions are $(a,b;q)$ under $\textbf{U}$ and $(c,d;q)$ under $\textbf{D}$, respectively. For fixed $p\in[0,1]$, player 2's payoff distributions are $(e,g;p)$ under $\textbf{L}$ and $(f,h;p)$ under $\textbf{R}$, respectively.

\paragraph{$2\times n$  bimatrix games.}
Let $(A,B)$ be a two-player bimatrix game where $A\in \R^{2\times n}, B\in\R^{n\times 2}$ are the deterministic payoff matrix for player 1 and player 2, respectively. For convenience, we write $A=(A_{1\cdot},A_{2\cdot})^\top, B=(B_{j\cdot})_{j\in[n]}$, where $A_{i\cdot}\in\R^n, B_{j\cdot}\in\R^2$ are the transposed row vectors of $A$ and $B$, respectively.
Let \(p\in[0,1]\) denote the probability with which player 1 uses row action \(1\), and let \(\q\in\Delta_n\) be player 2's mixed strategy. The payoff distribution for player 1 choosing action $i\in\{1,2\}$ if player 2 chooses mixed strategy $\q$ is $(A_{i\cdot},\q)$. Analogously, $(B_{j1},B_{j2};p)$ is the payoff distribution for player 2 choosing action $j\in[n]$ if player 1 chooses action 1 with probability $p$.

We introduce the setup for two-player strategic-form games, which will be useful in defining the best-response correspondence and Nash equilibrium.

\paragraph{Two-player strategic-form games.} A two-player strategic-form game is specified by the tuple $G=\langle \mathcal{A}^1,\mathcal{A}^2,r^1,r^2\rangle$. Here \(\mathcal{A}^i\) is the finite action set of player \(i\), and
$r^i:\mathcal{A}\to\mathbb{R}$ is the payoff function of player \(i\), where $\mathcal{A}:=\mathcal{A}^1\times\mathcal{A}^2$. The two players simultaneously choose actions \(a^i\in\mathcal{A}^i\), forming the action profile \(a:=(a^1,a^2)\), after which player \(i\) receives payoff \(r^i(a^1,a^2)\). We allow each player to independently randomize over her actions. A mixed strategy of player \(i\) is a probability distribution $\pi^i\in\Delta(\mathcal{A}^i)$,
where \(\pi^i(a^i)\) denotes the probability of choosing action \(a^i\). The mixed-strategy profile of the two players is denoted by $\pi:=(\pi^1,\pi^2)$. The random payoff for player $i$ choosing pure action $a^i$ if the opponent player playing mixed strategy $\pi^{-i}$ is $r^i(a^i,A^{-i}),\ A^{-i} \sim \pi^{-i}$, where the randomness comes from $\pi^{-i}$. The payoff distribution of $r^i(a^i,A^{-i})$ is given by $r^i(a^i,A^{-i}) \sim (r^i(a^i,\cdot), \pi^{-i})$.

\subsection{Two-player games with  nonlinear evaluation functionals}

\paragraph{Nonlinear evaluation functionals.} An evaluation functional is a map $\rho:\mathscr X\to\mathbb R$ from a set of random variables to the reals. It is said to be \emph{distribution-invariant} or law-invariant if $\rho(X)=\rho(Y)$ when $X \overset{d}{=} Y$. Throughout this paper, we restrict attention to law-invariant evaluation functionals~\cite{artzner1999coherent,frittelli2002putting}.
Thus, we will overload notation and write $\rho(F)=\rho(X)$ for simplicity. In addition, we use $\mu$ to denote the mean functional.

Throughout, we follow the convention of reward maximization (instead of loss minimization) for the evaluation functionals.
Risk measures (RMs) are among the most prominent examples of nonlinear evaluation functionals, which are widely used in mathematical finance and operations research. We use subsets of the following axiomatic properties of an evaluation functional $\rho$ to define different classes of risk measures.
\begin{asp}[Axiomatic properties of risk measures]
    \label{asp:rm}
   (a) normalization: \(\rho(0)=0\); (b) monotonicity: \(X\preceq Y\Rightarrow\rho(X)\leq\rho(Y)\); (c) translation invariance: \(\rho(X+c)=\rho(X)+c\) for all \(c\in\mathbb{R}\); (d) superadditivity: \(\rho(X+Y)\geq\rho(X)+\rho(Y)\); (e) positive homogeneity: \(\rho(\lambda X)=\lambda\rho(X)\) for all \(\lambda\geq0\); and (f) concavity: \(\rho(\lambda X+(1-\lambda)Y)\geq\lambda\rho(X)+(1-\lambda)\rho(Y)\) for all \(\lambda\in[0,1]\).
\end{asp}
\begin{defi}[Risk measures]
\label{def:rm}
    A functional satisfying Assumption \ref{asp:rm} (a)--(c) is called a (monetary) risk measure under the maximization convention, and it is coherent if it additionally satisfies Assumption \ref{asp:rm} (d)--(e). Concave monetary risk measures replace Assumption \ref{asp:rm} (d)--(e) with the weaker condition Assumption \ref{asp:rm} (f).
\end{defi}

The convention is equivalent, up to a sign change, to the standard risk-measure convention for financial positions: defining \(R(X):=-\rho(X)\) yields \(X\preceq Y\Rightarrow R(X)\geq R(Y)\) and \(R(X+c)=R(X)-c\), and transforms superadditivity and concavity of \(\rho\) into subadditivity and convexity of \(R\), respectively \cite{artzner1999coherent,follmer2002convex}.
In addition to RMs, rank-dependent expected utility (RDEU) \cite{quiggin1982theory} and CPT \cite{tversky1992advances} are also prominent evaluation functionals that model nonlinear preferences in behavioral economics.

In a strategic-form game $\langle \mathcal{A}^1,\mathcal{A}^2,r^1,r^2\rangle$, each player is an \emph{expected-utility-maximizing} agent that evaluates each action via the expectation of the payoff distribution. Specifically, player $i$ evaluates  a pure action $a^i$ against an opponent using mixed strategy $\pi^{-i}$ by the expectation of the random payoff $r^i(a^i,A^{-i})$
\[
V^i(a^i;\pi^{-i})
:=
\E_{A^{-i}\sim\pi^{-i}}\!\left[r^i(a^i,A^{-i})\right].
\]
In contrast, when the players use nonlinear evaluation functionals $\rho=(\rho^1,\rho^2)$,  player $i$ evaluates  a pure action $a^i$ against an opponent playing mixed strategy $\pi^{-i}$ using $\rho^i$ rather than the mean
\begin{equation}
\label{eqt:v_rho}
    V^i_{\rho}(a^i;\pi^{-i}):=
\rho^i\!\left(r^i(a^i,A^{-i})\right)=\rho^i\!\left(r^i(a^i,\cdot),\pi^{-i}\right),
\end{equation}
where we pass between random variables and payoff distributions in the second equality using the law-invariance of $\rho^i$.  We define the \textit{pure-action} best-response set for each player $i$ under $\rho$ by
\begin{equation}
\label{eqt:br_rho}
    \operatorname{BR}^i_\rho(\pi^{-i})
:=\arg\max_{a^i\in\mathcal A^i}V^i_\rho(a^i;\pi^{-i})
\subseteq\mathcal A^i.
\end{equation}
Furthermore, we write $\operatorname{BR}^i$ for the
\textit{mixed} best-response correspondence:
\begin{equation}
\label{eqt:br}
    \operatorname{BR}^i(\pi^{-i})
:=\left\{\nu^i\in\Delta(\mathcal A^i):
\operatorname{supp}(\nu^i)\subseteq\operatorname{BR}^i_\rho(\pi^{-i})\right\}=\operatorname{co}\left\{\e_{a^i}:a^i\in\operatorname{BR}^i_\rho(\pi^{-i})\right\},
\end{equation}
where $\operatorname{supp}(\nu^i):=\{a^i\in\mathcal A^i:\nu^i(a^i)>0\}$ is the support, $\operatorname{co}$ denotes the convex hull, and $\e_{a^i}$ assigns probability one to action $a^i$. Thus, $\operatorname{BR}^i_\rho(\pi^{-i})$ is a set of pure actions, and $\operatorname{BR}^i(\pi^{-i})\subseteq\Delta(\mathcal A^i)$ is a set of probability distributions supported on those actions. The dependence of the mixed correspondence on $\rho$ is left implicit throughout.
By Equation~\eqref{eqt:br}, for each player $i$,
\[
\operatorname{supp}(\pi^i)\subseteq
\operatorname{BR}^i_\rho(\pi^{-i})
\quad\Longleftrightarrow\quad
\pi^i\in\operatorname{BR}^i(\pi^{-i}).
\]
\begin{defi}[$\rho$-Nash equilibrium]
\label{def:rho_ne}
A mixed-strategy profile $\pi^\star_\rho$  is a $\rho$-Nash equilibrium ($\rho$-NE) if $\pi^{\star,i}_\rho\in \operatorname{BR}^i(\pi^{\star,-i}_\rho)$ for every player $i\in[2]$.
\end{defi}

The notion of $\rho$-NE reduces to the standard notion of NE when specializing $\rho$ to be a mean functional.  We defer the existence and characterization of $\rho$-NE in $2 \times 2$  games, and their generalizations to  $2 \times n$ games to Section \ref{sec:rho_ne}. We denote by $(G,\rho)$ the game $G$ in which both players use functional $\rho=(\rho^1,\rho^2)$. Thus, $(G,\mu)$ denotes the game $G$ under the mean functional  $\mu$, i.e., the risk-neutral game. When the meaning is clear from context, we refer to a $\rho$-NE simply as an equilibrium.

\subsection{Nonlinear fictitious play}
In classical FP, player $i$ maintains a belief on the opponent’s strategy, denoted by $\hat{\pi}_k^{-i} \in \Delta(\cA^{-i})$. With the initial belief counted as one observation, the empirical belief after $k$ observed actions is
\begin{equation}
\hat{\pi}_k^{-i} := \frac{1}{k+1}\left(\hat{\pi}_0^{-i}
+\sum_{t=0}^{k-1}\e_{a_t^{-i}}\right), \qquad k\ge0,
\end{equation}
with arbitrary initial belief $\hat{\pi}_0^{-i}\in\Delta(\cA^{-i})$.
Since iterations are indexed from $k=0$, the belief is updated recursively according to
\begin{equation}
\label{eqt:blf}
\hat{\pi}_{k+1}^{-i}
=
\hat{\pi}_k^{-i}
+\frac{1}{k+2}
\bigl(\e_{a_k^{-i}}-\hat{\pi}_k^{-i}\bigr),
\qquad k\ge0.
\end{equation}
Importantly, player $i$ selects the action as
\begin{equation}
\label{eqt:action}
    a_k^i\in\arg\max_{a^i} \ \E_{A^{-i}\sim\hat{\pi}_k^{-i}}\!\left[r^i(a^i,A^{-i})\right],
\end{equation}
with an arbitrary tie-breaking rule. Each player selects the optimal action with respect to the mean value of each payoff distribution. NL-FP only modifies the action-selection step (Equation~\eqref{eqt:action}) by replacing the mean with a nonlinear evaluation functional $\rho$. We refer to the NL-FP under a specific evaluation $\rho$ as \emph{$\rho$-FP}. Specifically, player $i$ selects the action $a_k^i$ at iteration $k=0,1,\cdots$ according to
\begin{equation}
\label{eqt:rho_fp}
a_k^i\in\arg\max_{a^i} \ V^i_{\rho}(a^i;\hat{\pi}_k^{-i})= \arg\max_{a^i}\ \rho^i(r^i\!\left(a^i,A^{-i}\right)),\quad A^{-i} \sim \hat{\pi}_k^{-i}.
\end{equation}
Here, each player $i$ applies $\rho^i$ separately to each pure action $a^i$ and only to the payoff uncertainty induced by player $i$'s empirical belief about the opponent's action $A^{-i} \sim \hat{\pi}_k^{-i}$. Specifically, $\mu$-FP reduces to the classical FP.

\begin{rmk}
     We do not instead apply $\rho^i$ to the randomness caused by the \emph{joint} distribution generated by \emph{both} players' mixed strategies.
     Because $\rho^i$ may be nonlinear, these two constructions generally differ. Our formulation reflects the fictitious-play interpretation: a player's empirical mixed strategy records the player's action frequencies, while the empirical distribution of the opponent's actions determines the payoff distribution used to evaluate each pure action. This treatment parallels the action-dependent risk-aversion formulation of \citet{mazumdar2025tractable}, but differs from that of \citet{fiat2010players}, which applied the nonlinear criterion to the randomness generated by \emph{both} players' mixed strategies. In fact, under the  formulation in \citet{fiat2010players}, a (mixed-strategy) Nash equilibrium may fail to even  exist.
\end{rmk}
\paragraph{NL-FP in $2 \times 2$ games.}
The mixed-strategy profile is specified by $(p,q)\in[0,1]^2$ in $2 \times 2$ games.
 The belief $(p_k,q_k)$ is computed recursively according to
\begin{equation}
\label{eqt:blf_2by2}
    p_{k+1} =p_k+\frac{1}{k+2} (\mathbb{I}\{a_k^1=\textbf{U}\}-p_k),\quad q_{k+1} =q_k+\frac{1}{k+2} (\mathbb{I}\{a_k^2=\textbf{L}\}-q_k),
\end{equation}
with arbitrary initial belief $(p_0, q_0)\in[0,1]^2$. For player 1, we define the \textit{advantage function} $d^1:[0,1]\to\mathbb{R}$ of action \textbf{U} over action \textbf{D} given its opponent's strategy $q\in[0,1]$ and a functional $\rho^1$ as
\begin{equation}
    d^1_\rho(q):=\rho^1(a,b;q)-\rho^1(c,d;q),
\end{equation}
where $\rho^1(a,b;q)$ and $\rho^1(c,d;q)$ are the values of action $\textbf{U}$'s payoff distribution and action $\textbf{D}$'s payoff distribution evaluated under $\rho^1$, respectively. Analogously, the advantage function for player 2 is defined as $d^2_\rho(p):=\rho^2(e,g;p)-\rho^2(f,h;p)$. We may refer to advantage functions as \textit{advantages} for short. For simplicity, we may drop $\rho$ when it is clear from the context. %
The pure-action best-responses depend only on the \textit{sign} of the advantage functions:
\begin{equation}
\label{eqt:br_2by2}
\begin{aligned}
\BR^1_\rho(q)
&=
\begin{cases}
\{\textbf{U}\}, & d^1(q)>0,\\
\{\textbf{U},\textbf{D}\}, & d^1(q)=0,\\
\{\textbf{D}\}, & d^1(q)<0,
\end{cases}
& \quad
\BR^2_\rho(p)
&=
\begin{cases}
\{\textbf{L}\}, & d^2(p)>0,\\
\{\textbf{L},\textbf{R}\}, & d^2(p)=0,\\
\{\textbf{R}\}, & d^2(p)<0.
\end{cases}
\end{aligned}
\end{equation}
To express NL-FP in terms of action probabilities, we identify player 1's actions $\textbf{U},\textbf{D}$ with $1,0$, respectively, and player 2's actions $\textbf{L},\textbf{R}$ with $1,0$, respectively.
We use the same notation for the best-response sets in these scalar coordinates. For either player $i$ and opponent coordinate $s$,
\[
\BR^i_\rho(s)\subseteq\{0,1\},\qquad
\BR^i(s)=\operatorname{co}\bigl(\BR^i_\rho(s)\bigr)\subseteq[0,1].
\]
At indifference, the pure-action set is $\{0,1\}$ and the mixed-response set is $[0,1]$.
Unless otherwise stated, let $x(q)\in\BR^1_\rho(q)$ and $y(p)\in\BR^2_\rho(p)$ denote the selected pure best responses. Each selection is the unique element when its best-response set is a singleton; otherwise, a fixed \textit{deterministic} tie-breaking rule selects one element. Thus, $x(q),y(p)\in\{0,1\}$.
The selected actions  are
\begin{equation}
\label{eqt:rho_fp_2by2}
a_k^1 = x(q_k),\quad a_k^2 = y(p_k).
\end{equation}

Using the mixed best-response correspondences in
Equation~\eqref{eqt:br}, we write the continuous-time NL-FP dynamics
for $t\ge t_0>0$, where $t_0$ is an arbitrary initial time, as
\[
\dot p_t\in\frac1t\bigl(\operatorname{BR}^1(q_t)-p_t\bigr),
\qquad
\dot q_t\in\frac1t\bigl(\operatorname{BR}^2(p_t)-q_t\bigr).
\]
We obtain the best-response dynamics for $t\ge0$:
\begin{equation}
\label{eqt:cfp_2by2}
\dot p_t\in\operatorname{BR}^1(q_t)-p_t,
\qquad
\dot q_t\in\operatorname{BR}^2(p_t)-q_t.
\end{equation}
The continuous-time selection used below is specified in
Section~\ref{sec:ct}.
We defer the introduction of NL-FP in $2 \times n$ games to Section \ref{sec:2byn}.

\subsection{Existence and characterization of \texorpdfstring{$\rho$}{rho}-NE%
}
\label{sec:rho_ne}
Nash's theorem guarantees the existence of a mixed-strategy NE in every finite game evaluated by expected payoffs \citep{nash1951noncooperative}. This theorem does not directly establish the existence of a $\rho$-NE when $\rho$ is nonlinear. We first provide a decomposition result for the $\rho$-NE set in Proposition \ref{prop:ne_2by2}, and prove  the existence of $\rho$-NE in $2 \times 2$ games for generic nonlinear functionals under mild conditions in Proposition \ref{pro:rho-ne-existence}. In $2 \times 2$ games, we call the unit square $[0,1]^2$ in the $(p,q)$-plane the \textit{strategy square}.
\begin{asp}[Continuity]
\label{asp:cont}
    Fix an arbitrary positive integer $n$. $\rho^i(\vv,\q)$ is continuous with respect to the probability argument $\q\in\Delta_n$ for any $\vv\in\R^n$ and $i=1,2$.
\end{asp}
Assumption \ref{asp:cont} implies the continuity of the advantage functions with respect to the mixed strategies.

\begin{prop}[Characterization of $\rho$-NE in $2\times2$ games]
\label{prop:ne_2by2}
We define the \textit{interior} zero sets $\cP:=\{p\in(0,1):d^2(p)=0\},
\
\cQ:=\{q\in(0,1):d^1(q)=0\}.$%
 The set of $\rho$-Nash equilibria is
\[
\cE_\rho=(\cP\times \cQ)\cup \C_\rho\cup \cE_\rho^{\mathrm{edge}},
\]
where  $\C_\rho$ is called the pure-equilibrium set
\begin{equation}
\label{eqt:c_rho}
    \C_\rho:=
\left\{
(p,q)\in\{0,1\}^2:
(2p-1)d^1(q)\ge0,\;
(2q-1)d^2(p)\ge0
\right\},
\end{equation}
and $\cE_\rho^{\mathrm{edge}}$ is the equilibrium set on the open edges of the strategy square
\begin{equation}
\label{eqt:e_rho}
    \begin{aligned}
\cE_\rho^{\mathrm{edge}}
&:={}
\{(0,q):0<q<1,\ d^2(0)=0,\ d^1(q)\le0\}{}\cup
\{(1,q):0<q<1,\ d^2(1)=0,\ d^1(q)\ge0\}\\
&\qquad{}\cup
\{(p,0):0<p<1,\ d^1(0)=0,\ d^2(p)\le0\}{}\cup
\{(p,1):0<p<1,\ d^1(1)=0,\ d^2(p)\ge0\}.
\end{aligned}
\end{equation}
We refer to $\cE_\rho^{\mathrm{edge}}$ as the open-edge equilibrium set. In addition, $\cE_\rho\cap(0,1)^2=\cP\times \cQ$.\textbf{}
\end{prop}
The proof is provided in Appendix \ref{app:form_pf}. Proposition~\ref{prop:ne_2by2} shows that every $\rho$-NE in a $2\times2$ game falls into one of three classes. A \emph{pure $\rho$-NE} lies at a corner of $[0,1]^2$, where both players play pure actions. An \emph{open-edge $\rho$-NE} lies on an edge excluding its endpoints, where exactly one player mixes. An \emph{interior $\rho$-NE} lies in $(0,1)^2$, where both players mix.

\begin{coro}
    If additionally $d^i(0)d^i(1)\neq0$ for $i=1,2$, then
$\cE_\rho^{\mathrm{edge}}=\varnothing$, and hence $\cE_\rho=(\cP\times \cQ)\cup \C_\rho.$
\end{coro}

The decomposition in Proposition \ref{prop:ne_2by2} remains valid even when the equilibrium set is empty. This characterization will also be used to establish the existence of $\rho$-NE. %

\begin{prop}[Existence of $\rho$-NE in $2\times2$ games]
\label{pro:rho-ne-existence}
Suppose that for each $i\in\{1,2\}$: $d^i(0)d^i(1)<0$ implies that the advantage function $d^i$ has a zero $s_i \in (0,1)$, i.e., $d^i(s_i)=0$. %
Then, the game admits a $\rho$-NE. In particular, this condition %
holds when $d^1$ and $d^2$ are continuous. Therefore, a $\rho$-NE exists under Assumption~\ref{asp:cont}.
\end{prop}
The proof is provided in Appendix \ref{app:form_pf}. We now prove that the $\rho$-NE in $2 \times n$ games also exists for generic nonlinear functionals under the continuity of $\rho$. For $S\subseteq[n]$, let $\Delta(S):=
\{\mathbf{q}\in\Delta_n:\operatorname{supp}(\mathbf{q})\subseteq S\}.$ For a two-point payoff distribution $(u,v;p)$, we overload the notation and use $\rho_i(u,v;p)$ to denote its value under $\rho$. For player 2, we define
\[
G_j(p):=\rho^2(B_{j1},B_{j2};p),
\qquad
J(p):=\arg\max_{j\in[n]}G_j(p)=\operatorname{BR}^2_\rho(p).
\]
Player 2's mixed best-response correspondence is $\operatorname{BR}^2(p)
=
\Delta(J(p))\subseteq\Delta_n$.
Here $J(p)\subseteq[n]$ contains pure-action indices, and $\operatorname{BR}^2(p)$ contains all probability vectors supported on those indices. For the two-action row player, we again represent mixed responses by the probability of the first row action and encode the pure responses by $1$ and $0$.
\begin{lem}
\label{lem:con}
    Under Assumption \ref{asp:cont}, $d^1(\q):=\rho^1(A_{1\cdot},\q)-\rho^1(A_{2\cdot},\q)$ is continuous, and each column-action value $G_j:[0,1]\to\mathbb{R}$ is continuous.
\end{lem}

\begin{prop}[Characterization of  $\rho$-NE in $2 \times n$  games]
\label{prop:ne_2byn}
Under Assumption \ref{asp:cont},  $\mathcal E_\rho$ is nonempty.  It is given by the union of three sets
\begin{align*}
\mathcal E_\rho
&={}
\left\{(0,\mathbf{q}):\mathbf{q}\in\Delta(J(0)),\ d^1(\mathbf{q})\le0\right\}
\cup
\left\{(1,\mathbf{q}):\mathbf{q}\in\Delta(J(1)),\ d^1(\mathbf{q})\ge0\right\}\\
&\qquad\quad\cup
\left\{(p,\mathbf{q}):0<p<1,\ \mathbf{q}\in\Delta(J(p)),\
d^1(\mathbf{q})=0\right\}\neq \varnothing.
\end{align*}
\end{prop}
The proof of the proposition is provided in Appendix \ref{app:form_pf}.
Since the union of three sets $\mathcal E_\rho$ is nonempty, at least one of the three sets is nonempty.

\section{Single-Crossing Property Retains Convergence in \texorpdfstring{$2 \times 2$}{2 x 2} Games}
\label{sec:scp}
In this section, we focus on $2 \times 2$ games where both players do not have a dominant pure action. Equivalently, both advantage functions $d^i$ exhibit at least one sign change on \([0,1]\). The alternative case in which at least one player has a dominant action admits a direct convergence argument. We defer the corresponding convergence result (Proposition \ref{prop:dominant}) and its proof to Appendix \ref{app:dom}. We first consider advantage functions with a single sign change. We show that the geometric structure of \(\rho\)-FP retains that of classical FP \cite{metrick1994fictitious}.

\subsection{SCP preserves convergence}
\label{sec:scp_conv}
\begin{figure}[t]
        \centering
        \begingroup
\input{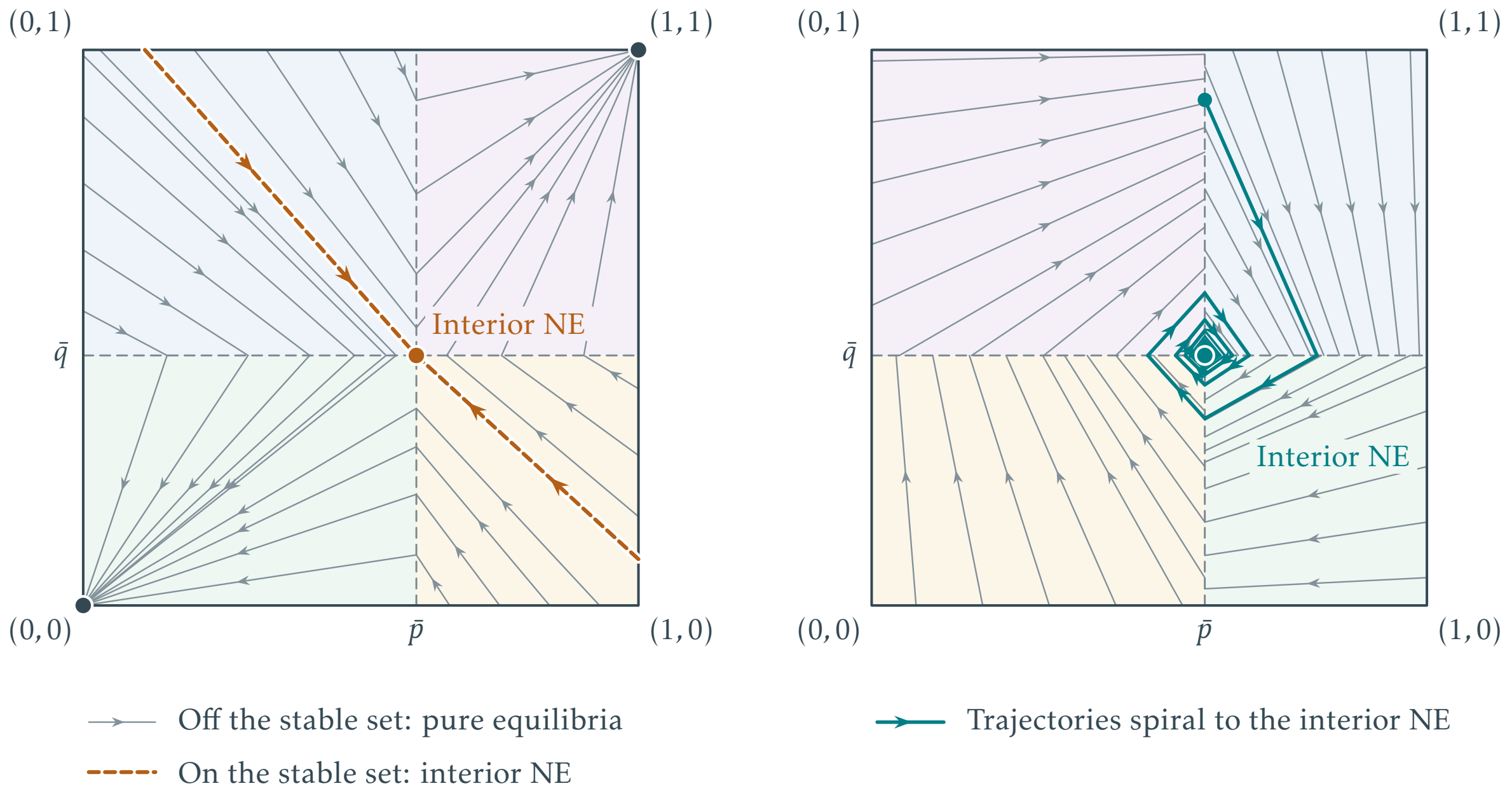}%
\def\fpP{0.6}%
\def\fpQ{0.45}%
\begin{tikzpicture}[
  x=\fpUnit,y=\fpUnit,font=\footnotesize,text=fpInk,
  line cap=round,line join=round
]
  \begin{scope}
    \node[font=\small\bfseries] at (0.5,-0.43)
      {\panellabel{fig:fp_geo_coordination} Coordination-type};
    \fill[fpBlueWash] (0,\fpQ) rectangle (\fpP,1);
    \fill[fpPurpleWash] (\fpP,\fpQ) rectangle (1,1);
    \fill[fpGreenWash] (0,0) rectangle (\fpP,\fpQ);
    \fill[fpGoldWash] (\fpP,0) rectangle (1,\fpQ);

    \foreach \s in {0.03,0.20,0.33,0.46,0.56} {
      \fpRay{\s}{1}{1}{0}{0}{\fpP}{\fpQ}{1}\fpToPure
    }
    \foreach \s in {0.53,0.64,0.76,0.88,0.99} {
      \fpRay{0}{\s}{1}{0}{0}{\fpP}{\fpQ}{1}\fpToPure
    }
    \foreach \s in {0.66,0.75,0.84,0.93} {
      \fpRay{\s}{0}{0}{1}{\fpP}{1}{0}{\fpQ}\fpToPure
    }
    \foreach \s in {0.16,0.27,0.39} {
      \fpRay{1}{\s}{0}{1}{\fpP}{1}{0}{\fpQ}\fpToPure
    }
    \fpFrame

    \draw[white,line width=2.7pt] ({1/9},1) -- (\fpP,\fpQ) -- (1,{1/12});
    \draw[fp stable] ({1/9},1) -- (\fpP,\fpQ);
    \draw[fp stable] (1,{1/12}) -- (\fpP,\fpQ);
    \node[fp equilibrium] at (0,0) {};
    \node[fp equilibrium] at (1,1) {};
    \node[fp equilibrium,fill=fpStable] at (\fpP,\fpQ) {};
    \node[anchor=south west,inner sep=2pt,fill=fpPurpleWash,
      text=fpStable] at (0.615,0.475) {Interior NE};

    \draw[fp flow] (0.01,-0.21) -- (0.13,-0.21);
    \node[fp key] at (0.17,-0.21) {Off the stable set: pure equilibria};
    \draw[draw=fpStable,line width=1.15pt,dash pattern=on 3pt off 1.8pt]
      (0.01,-0.30) -- (0.13,-0.30);
    \node[fp key] at (0.17,-0.30) {On the stable set: interior NE};
  \end{scope}

  \begin{scope}[shift={(1.42,0)}]
    \node[font=\small\bfseries] at (0.5,-0.43)
      {\panellabel{fig:fp_geo_spiral} Spiral-type};
    \fill[fpPurpleWash] (0,\fpQ) rectangle (\fpP,1);
    \fill[fpBlueWash] (\fpP,\fpQ) rectangle (1,1);
    \fill[fpGoldWash] (0,0) rectangle (\fpP,\fpQ);
    \fill[fpGreenWash] (\fpP,0) rectangle (1,\fpQ);

    \foreach \s in {0.54,0.65,0.76,0.87,0.98} {
      \fpRay{0}{\s}{1}{1}{0}{\fpP}{\fpQ}{1}
    }
    \foreach \s in {0.05,0.18,0.31,0.44,0.55} {
      \fpRay{\s}{\fpQ}{1}{1}{0}{\fpP}{\fpQ}{1}
    }
    \foreach \s in {0.65,0.73,0.81,0.89,0.97} {
      \fpRay{\s}{1}{1}{0}{\fpP}{1}{\fpQ}{1}
      \fpRay{\s}{\fpQ}{0}{0}{\fpP}{1}{0}{\fpQ}
    }
    \foreach \s in {0.53,0.64,0.75,0.86,0.97} {
      \fpRay{\fpP}{\s}{1}{0}{\fpP}{1}{\fpQ}{1}
    }
    \foreach \s in {0.05,0.15,0.25,0.35,0.43} {
      \fpRay{1}{\s}{0}{0}{\fpP}{1}{0}{\fpQ}
      \fpRay{\fpP}{\s}{0}{1}{0}{\fpP}{0}{\fpQ}
    }
    \foreach \s in {0.08,0.20,0.32,0.44,0.56} {
      \fpRay{\s}{0}{0}{1}{0}{\fpP}{0}{\fpQ}
    }
    \fpFrame

    \foreach \turn [remember=\fpNext as \fpTop (initially 0.91)] in {1,...,9} {
      \pgfmathsetmacro{\fpRight}{1-(1-\fpP)*\fpQ/\fpTop}
      \pgfmathsetmacro{\fpBottom}{\fpP*\fpQ/\fpRight}
      \pgfmathsetmacro{\fpLeft}{\fpP*(1-\fpQ)/(1-\fpBottom)}
      \pgfmathsetmacro{\fpNext}{1-(1-\fpQ)*(1-\fpP)/(1-\fpLeft)}
      \ifnum\turn<4\relax
        \tikzset{fp turn/.style={fp spiral arrow}}
      \else
        \tikzset{fp turn/.style={fp spiral,line width=0.7pt}}
      \fi
      \draw[fp turn] (\fpP,\fpTop) -- (\fpRight,\fpQ);
      \draw[fp turn] (\fpRight,\fpQ) -- (\fpP,\fpBottom);
      \draw[fp turn] (\fpP,\fpBottom) -- (\fpLeft,\fpQ);
      \draw[fp turn] (\fpLeft,\fpQ) -- (\fpP,\fpNext);
    }
    \node[circle,fill=fpTeal,inner sep=1.5pt] at (\fpP,0.91) {};
    \node[fp equilibrium,fill=fpTeal] at (\fpP,\fpQ) {};
    \node[anchor=north west,inner sep=2pt,fill=fpGreenWash,
      text=fpTeal] at (0.68,0.30) {Interior NE};
    \draw[fp spiral arrow] (0.01,-0.21) -- (0.13,-0.21);
    \node[fp key] at (0.17,-0.21) {Trajectories spiral to the interior NE};
  \end{scope}
\end{tikzpicture}%
\endgroup%
        \caption{%
        Geometry of continuous-time classical FP in $2\times2$ games \citep{metrick1994fictitious}, with unique zeros $(\bar p,\bar q)$. \textbf{(a) Coordination-type:} Off the stable set, trajectories converge to pure equilibria. Points on the interior NE's stable set converge to the interior NE. \textbf{(b) Spiral-type:} Trajectories spiral toward the interior NE.}
        \label{fig:fp_geo}
\end{figure}

We first revisit the classical analysis of the convergence and geometry of FP in $2\times2$ games by \citet{metrick1994fictitious}, which motivates our introduction of the single-crossing property. In $2\times 2$ games, \citet{metrick1994fictitious} showed that continuous-time FP has two canonical geometric patterns, as illustrated in Figure~\ref{fig:fp_geo}. In both cases, the strategy square $[0,1]^2$ is partitioned by the \textit{switching lines} $q=\bar q$ and $p=\bar p$ into four regions, each with a constant pair of pure best responses. We call them \textit{best-response regions}. In the \emph{coordination-type} case (Figure~\ref{fig:fp_geo_coordination}), the interior NE is a saddle whose stable set separates the basins of the two attracting pure equilibria.  In the \emph{spiral-type} case (Figure~\ref{fig:fp_geo_spiral}), the vector field rotates across the four best-response regions and trajectories spiral toward the interior NE.

To understand these patterns, we observe that mean functional $\mu$
gives two \emph{affine} advantages:
\begin{equation}
\label{eq:affine_advantages}
\begin{aligned}
d^1_\mu(q)
=q(a-c)+(1-q)(b-d)
 =\lambda_1(q-\bar q), \quad
d^2_\mu(p)
=p(e-f)+(1-p)(g-h)
 =\lambda_2(p-\bar p),
\end{aligned}
\end{equation}
where
\[
\lambda_1
:=d^1_\mu(1)-d^1_\mu(0)=a-c-b+d,
\qquad
\lambda_2
:=d^2_\mu(1)-d^2_\mu(0)=e-f-g+h.
\]
Since both advantage functions have a single sign change,
$d^i_\mu(0)d^i_\mu(1)<0$ for each player. Consequently,
$\lambda_1,\lambda_2\neq0$, and the unique zeros are
\[
\bar q=\frac{d-b}{a-c-b+d},
\qquad
\bar p=\frac{h-g}{e-f-g+h},
\]
both lying in $(0,1)$ (see the dashed line in Figure~\ref{fig:scp_advantages}). Thus, each advantage changes sign exactly once.
As shown in Figure~\ref{fig:scp_regions}, the switching lines $q=\bar q$ and $p=\bar p$ partition the
strategy square $[0,1]^2$ into four regions,  each with a constant pure best-response pair $(x(q),y(p))$.  Within each best-response region, continuous-time FP in Equation~\eqref{eqt:cfp_2by2} follows a straight segment toward the corresponding corner.
If $\lambda_1\lambda_2>0$, the two advantages change sign in
the same direction, producing coordination-type geometry.
When both slopes are positive, the attracting pure equilibria
are $(0,0)$ and $(1,1)$. When both are negative, they are
$(1,0)$ and $(0,1)$, giving the same geometry after relabeling
actions. If $\lambda_1\lambda_2<0$, the advantages change sign
in opposite directions, producing spiral-type geometry.

Our key observation is that the essential feature is  the \textit{sign pattern} of the advantages. Their magnitudes and slopes do not enter the best-response rule. In particular, an advantage need
not be affine, or even monotone, to retain a single sign
change. Suppose a nonlinear evaluation functional $\rho$ produces advantages with unique zeros $q_\rho$ and $p_\rho$ satisfying
\[
\operatorname{sgn}d^1_\rho(q)
 =\sigma_1\operatorname{sgn}(q-q_\rho),
\qquad
\operatorname{sgn}d^2_\rho(p)
 =\sigma_2\operatorname{sgn}(p-p_\rho),
\qquad
\sigma_1,\sigma_2\in\{-1,1\}.
\]
These advantages generate exactly the same best-response
correspondences as the affine advantages
$\sigma_1(q-q_\rho)$ and $\sigma_2(p-p_\rho)$.
Consequently, under the same rule at indifference,
the continuous-time NL-FP dynamics have the corresponding classical FP geometry.

\begin{figure}[t]
        \centering
        \begingroup
\input{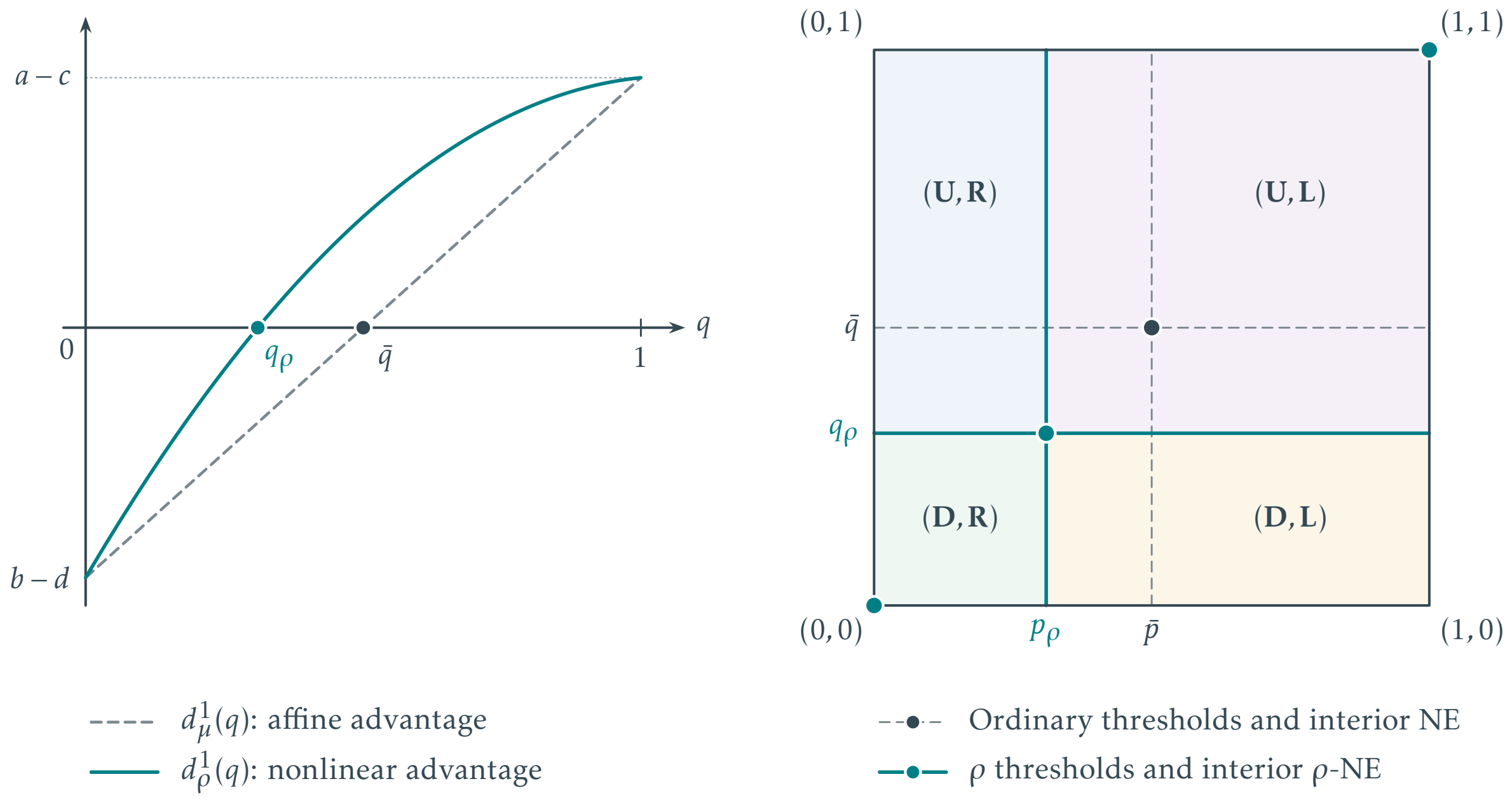}%
\def\scpR{0.31}%
\pgfmathsetmacro{\scpBend}{0.9*(0.5-\scpR)/(\scpR*(1-\scpR))}%
\begin{tikzpicture}[
  x=\fpUnit,y=\fpUnit,font=\footnotesize,text=fpInk,
  line cap=round,line join=round
]
  \begin{scope}
    \node[font=\small\bfseries] at (0.5,-0.43)
      {\panellabel{fig:scp_advantages} Single-crossing advantages};

    \draw[draw=fpFlow!70,line width=0.45pt,densely dotted]
      (0,0.95) -- (1,0.95);
    \draw[draw=fpInk,line width=0.75pt,-{Stealth[length=1.7mm,width=1.2mm]}]
      (0,0) -- (0,1.06);
    \draw[draw=fpInk,line width=0.75pt,-{Stealth[length=1.7mm,width=1.2mm]}]
      (-0.04,0.5) -- (1.08,0.5);
    \node[anchor=east,inner sep=4pt] at (0,0.95) {$a-c$};
    \node[anchor=east,inner sep=4pt] at (0,0.05) {$b-d$};
    \node[anchor=north east,inner sep=3pt] at (0,0.5) {$0$};
    \node[anchor=west,inner sep=3pt] at (1.08,0.5) {$q$};
    \draw[draw=fpInk,line width=0.55pt] (1,0.485) -- (1,0.515);
    \node[below,inner sep=3pt] at (1,0.485) {$1$};

    \draw[fp threshold,line width=0.85pt] (0,0.05) -- (1,0.95);
    \draw[fp spiral] plot[domain=0:1,samples=81]
      (\x,{0.05+0.9*\x+\scpBend*\x*(1-\x)});
    \node[fp equilibrium,inner sep=1.7pt] at (0.5,0.5) {};
    \node[fp equilibrium,fill=fpTeal,inner sep=1.7pt] at (\scpR,0.5) {};
    \node[below,inner sep=5pt] at (0.54,0.5) {$\bar q$};
    \node[below,text=fpTeal,inner sep=5pt] at (0.35,0.5) {$q_\rho$};

    \draw[fp threshold,line width=0.85pt] (0.01,-0.21) -- (0.13,-0.21);
    \node[fp key] at (0.17,-0.21) {$d_\mu^1(q)$: affine advantage};
    \draw[fp spiral] (0.01,-0.30) -- (0.13,-0.30);
    \node[fp key] at (0.17,-0.30) {$d_\rho^1(q)$: nonlinear advantage};
  \end{scope}

  \begin{scope}[shift={(1.42,0)}]
    \node[font=\small\bfseries] at (0.5,-0.43)
      {\panellabel{fig:scp_regions} Preserved response regions};
    \fill[fpBlueWash] (0,\scpR) rectangle (\scpR,1);
    \fill[fpPurpleWash] (\scpR,\scpR) rectangle (1,1);
    \fill[fpGreenWash] (0,0) rectangle (\scpR,\scpR);
    \fill[fpGoldWash] (\scpR,0) rectangle (1,\scpR);

    \draw[fp threshold] (0.5,0) -- (0.5,1);
    \draw[fp threshold] (0,0.5) -- (1,0.5);
    \draw[fp spiral] (\scpR,0) -- (\scpR,1);
    \draw[fp spiral] (0,\scpR) -- (1,\scpR);
    \fpCorners
    \node[below,inner sep=4pt] at (0.5,0) {$\bar p$};
    \node[left,inner sep=4pt] at (0,0.5) {$\bar q$};
    \node[below,text=fpTeal,inner sep=4pt] at (\scpR,0) {$p_\rho$};
    \node[left,text=fpTeal,inner sep=4pt] at (0,\scpR) {$q_\rho$};

    \node at (0.155,0.74) {$(\mathbf U,\mathbf R)$};
    \node at (0.75,0.74) {$(\mathbf U,\mathbf L)$};
    \node at (0.155,0.155) {$(\mathbf D,\mathbf R)$};
    \node at (0.75,0.155) {$(\mathbf D,\mathbf L)$};
    \node[fp equilibrium] at (0.5,0.5) {};
    \node[fp equilibrium,fill=fpTeal] at (\scpR,\scpR) {};
    \node[fp equilibrium,fill=fpTeal] at (0,0) {};
    \node[fp equilibrium,fill=fpTeal] at (1,1) {};

    \draw[fp threshold] (0.01,-0.21) -- (0.13,-0.21);
    \node[fp equilibrium,inner sep=1.6pt] at (0.07,-0.21) {};
    \node[fp key] at (0.17,-0.21) {Ordinary thresholds and interior NE};
    \draw[fp spiral] (0.01,-0.30) -- (0.13,-0.30);
    \node[fp equilibrium,fill=fpTeal,inner sep=1.6pt] at (0.07,-0.30) {};
    \node[fp key] at (0.17,-0.30) {$\rho$ thresholds and interior $\rho$-NE};
  \end{scope}
\end{tikzpicture}%
\endgroup%
        \caption{SSCP preserves the geometric structure of NL-FP, illustrated here for a certainty-normalized evaluation. \textbf{(a)} The nonlinear advantage $d^1_\rho$ has a unique zero $q_\rho$, shifted from the ordinary threshold $\bar q$ of $d^1_\mu$. \textbf{(b)} In the coordination-type case, the switching lines move from $(\bar p,\bar q)$ to $(p_\rho,q_\rho)$, preserving the four best-response regions and the two pure equilibria.}
        \label{fig:scp}
\end{figure}

Figure~\ref{fig:scp} illustrates this mechanism in the
coordination-type case. The nonlinear advantage retains a
single sign change, while the switching lines move from
$(\bar p,\bar q)$ to $(p_\rho,q_\rho)$. The interior $\rho$-NE
therefore shifts, but the arrangement of the four
best-response regions %
and the two  %
pure NE
is preserved. The equilibrium location and basin boundaries
may change without changing the qualitative geometric pattern. This observation motivates us to study  the \textit{single crossing property}  as a condition on the
sign pattern of the advantage functions. We give a formal definition of single crossing property in Definition \ref{def:scp}.

\begin{defi}[Single-crossing property]
\label{def:scp}
An advantage function $d^i:[0,1]\to\mathbb{R}$ satisfies
SCP if
$d^i(0)d^i(1)<0$ and there exist
$0<\underline r_i\le\bar r_i<1$ such that
\[
\begin{aligned}
d^i(s)d^i(0)>0,~
 \forall s\in[0,\underline r_i),\quad
d^i(s)=0,~
 \forall
 s\in[\underline r_i,\bar r_i],\quad
d^i(s)d^i(1)>0,~
 \forall
 s\in(\bar r_i,1].
\end{aligned}
\]
The property is \emph{strict} SCP (SSCP) if
$\underline r_i=\bar r_i$, equivalently, if $d^i$
has a unique zero.
\end{defi}

Note that SCP does not require continuity.
For a fixed game, we also say that a functional induces (S)SCP when its induced advantage satisfies (S)SCP. We next illustrate SCP and SSCP using two common risk measures in finance and decision theory: the entropic risk measure (ERM) and CVaR. These examples show how the zero set of an advantage function can be a single point or an interval.

\begin{exm}[SSCP of ERM]
We consider the entropic risk measure \citep{howard1972risk,follmer2004stochastic}
$\rho_\beta(X)=-\beta^{-1}\log\mathbb{E}[e^{-\beta X}]$,
with $\beta>0$. The advantage is
\[
d^1_\beta(q)
=\frac{1}{\beta}\log
\frac{q e^{-\beta c}+(1-q)e^{-\beta d}}
     {q e^{-\beta a}+(1-q)e^{-\beta b}}.
\]
It is obvious that $d^1_\beta$ is continuous on this interval.
If $(a-c)(b-d)<0$, its endpoint values have opposite
signs. Moreover, there is a unique interior zero
\[
q_\beta
=
\frac{e^{-\beta b}-e^{-\beta d}}
     {e^{-\beta b}-e^{-\beta d}
      +e^{-\beta c}-e^{-\beta a}}
\in(0,1).
\]
Since $d^1_\beta$ is continuous in $q$, the opposite endpoint signs and unique zero imply SSCP \citep{mohammadi2026entropic}. As $\beta\to0$, the threshold satisfies $q_\beta\to\bar q$, recovering the mean-evaluation threshold.
\end{exm}

\begin{exm}[SCP of CVaR]
We consider lower-tail CVaR %
under our reward convention \citep{rockafellar2002conditional,acerbi2002spectral}:
$\rho_\alpha(X)=\alpha^{-1}\int_0^\alpha F_X^{-1}(u)\,du$,
with $\alpha\in(0,1]$. For payoff rows $(1,0)$ and
$(0,1)$, the advantage is
\[
d^1_\alpha(q)
=\frac{(q+\alpha-1)_+-(\alpha-q)_+}{\alpha},
\]
where $(x)_+:=\max\{x,0\}$. When $0<\alpha<\frac{1}{2}$, this advantage is negative for
$q<\alpha$, zero on $[\alpha,1-\alpha]$, and positive
for $q>1-\alpha$. Thus, it satisfies SCP with
$\underline r_1=\alpha$ and $\bar r_1=1-\alpha$,
but not SSCP. When $\alpha\ge\frac{1}{2}$, the zero is uniquely
$q=\frac{1}{2}$, and SSCP holds.
\end{exm}

\begin{rmk}
Our terminology for SCP follows that of \citet[Definition~1]{kartik2024single}.
Under the continuity of $d^i$, Definition~\ref{def:scp} is equivalent to their single-crossing condition together with
$d^i(0)d^i(1)<0$.
The endpoint condition ensures a reversal of strict preference,
excluding cases without such a reversal that their broader
definition permits.
The zero set is then a nonempty closed interval
$[\underline r_i,\bar r_i]\subset(0,1)$, called
an \textit{indifference interval}.
SSCP requires this interval to reduce to a single point.
\end{rmk}
Thus, SCP allows an indifference interval $[\underline r_i,\bar r_i]$, while SSCP requires a unique indifference point $r_i$. Both directions of crossing are permitted. SSCP
preserves the classical four-region geometry,
while SCP permits an indifference interval between the two strict-sign regions. We show in Proposition \ref{prop:sscp} that
SSCP preserves the convergence and geometry. The key observation is that every advantage function $d^i$ satisfying SSCP has exactly the same sign pattern as an affine advantage $\bar d^i(s)=d^i(1)(s-r_i)$. The proof is provided in Appendix \ref{app:sscp_pf}.

\begin{figure}[t]
        \centering
        \begingroup
\input{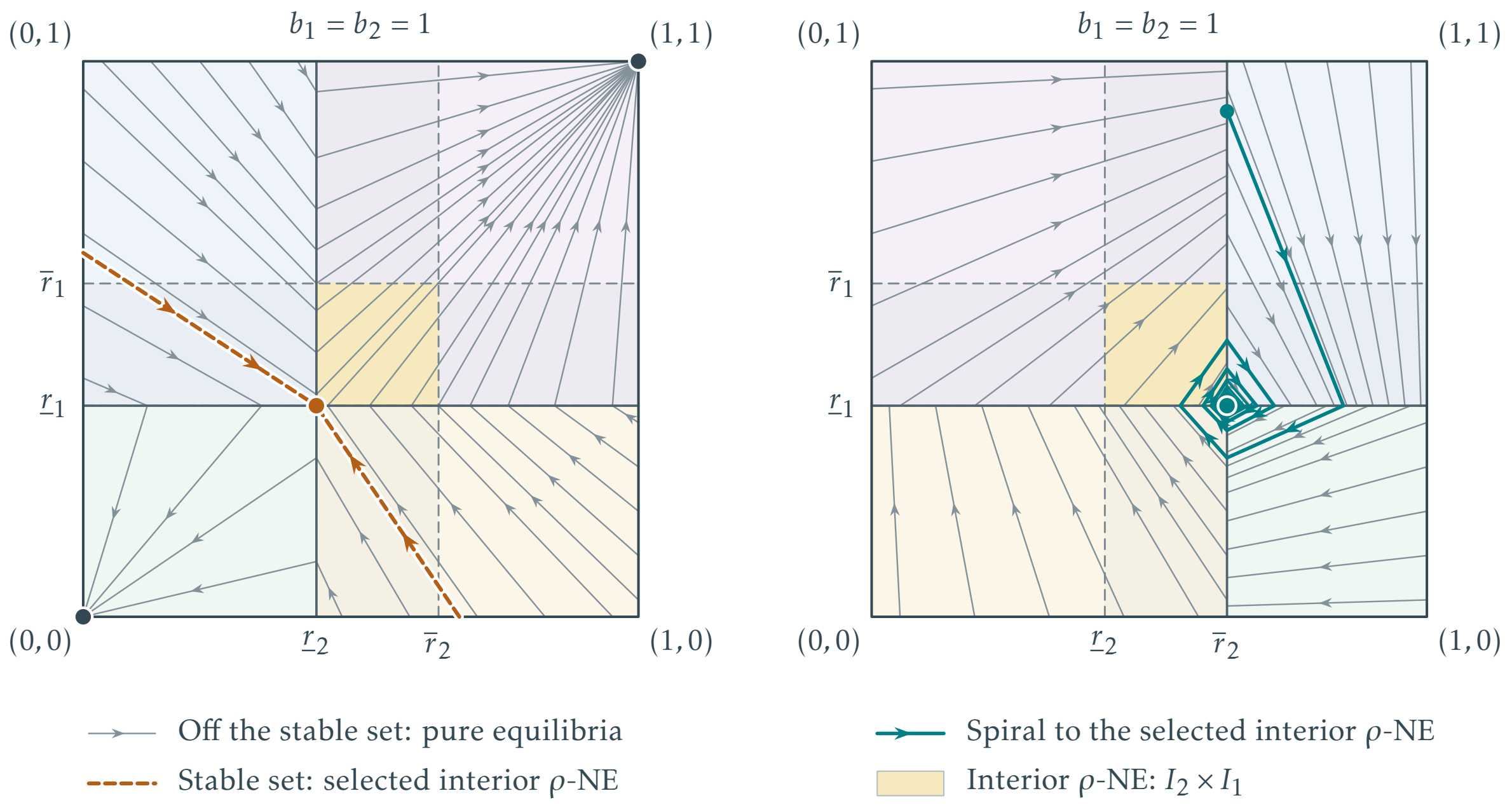}%
\def\fpLoP{0.42}%
\def\fpHiP{0.64}%
\def\fpLoQ{0.38}%
\def\fpHiQ{0.60}%
\def\fpWeakBands{%
  \fill[fpInk,opacity=0.035] (\fpLoP,0) rectangle (\fpHiP,1);
  \fill[fpInk,opacity=0.035] (0,\fpLoQ) rectangle (1,\fpHiQ);
  \fill[fpEquilibriumWash] (\fpLoP,\fpLoQ) rectangle (\fpHiP,\fpHiQ);
}
\def\fpWeakFrame{%
  \foreach \x in {\fpLoP,\fpHiP}
    \draw[fp threshold] (\x,0) -- (\x,1);
  \foreach \y in {\fpLoQ,\fpHiQ}
    \draw[fp threshold] (0,\y) -- (1,\y);
  \draw[draw=fpInk!85,line width=0.7pt] (\fpP,0) -- (\fpP,1);
  \draw[draw=fpInk!85,line width=0.7pt] (0,\fpQ) -- (1,\fpQ);
  \fpCorners
  \node[below,inner sep=4pt] at (\fpLoP,0) {$\underline r_2$};
  \node[below,inner sep=4pt] at (\fpHiP,0) {$\overline r_2$};
  \node[left,inner sep=4pt] at (0,\fpLoQ) {$\underline r_1$};
  \node[left,inner sep=4pt] at (0,\fpHiQ) {$\overline r_1$};
}
\begin{tikzpicture}[
  x=\fpUnit,y=\fpUnit,font=\footnotesize,text=fpInk,
  line cap=round,line join=round
]
  \begin{scope}
    \def\fpP{\fpLoP}\def\fpQ{\fpLoQ}
    \node[font=\small\bfseries] at (0.5,-0.43)
      {\panellabel{fig:weak_scp_coordination} Coordination-type};
    \node at (0.5,1.065) {$b_1=b_2=1$};
    \fill[fpBlueWash] (0,\fpQ) rectangle (\fpP,1);
    \fill[fpPurpleWash] (\fpP,\fpQ) rectangle (1,1);
    \fill[fpGreenWash] (0,0) rectangle (\fpP,\fpQ);
    \fill[fpGoldWash] (\fpP,0) rectangle (1,\fpQ);
    \fpWeakBands

    \foreach \t in {0.08,0.29,0.50,0.71,0.92} {
      \pgfmathsetmacro{\sx}{\fpP*\t}
      \pgfmathsetmacro{\sy}{\fpQ+(1-\fpQ)*\t}
      \fpRay{\sx}{1}{1}{0}{0}{\fpP}{\fpQ}{1}\fpToPure
      \fpRay{0}{\sy}{1}{0}{0}{\fpP}{\fpQ}{1}\fpToPure
      \pgfmathsetmacro{\sx}{\fpP+(1-\fpP)*\t}
      \pgfmathsetmacro{\sy}{\fpQ*\t}
      \fpRay{\sx}{0}{0}{1}{\fpP}{1}{0}{\fpQ}\fpToPure
      \fpRay{1}{\sy}{0}{1}{\fpP}{1}{0}{\fpQ}\fpToPure
    }
    \fpWeakFrame

    \pgfmathsetmacro{\fpStableY}{\fpQ/(1-\fpP)}
    \pgfmathsetmacro{\fpStableX}{\fpP/(1-\fpQ)}
    \draw[white,line width=2.7pt]
      (0,\fpStableY) -- (\fpP,\fpQ) -- (\fpStableX,0);
    \draw[fp stable] (0,\fpStableY) -- (\fpP,\fpQ);
    \draw[fp stable] (\fpStableX,0) -- (\fpP,\fpQ);
    \node[fp equilibrium] at (0,0) {};
    \node[fp equilibrium] at (1,1) {};
    \node[fp equilibrium,fill=fpStable] at (\fpP,\fpQ) {};

    \draw[fp flow] (0.01,-0.21) -- (0.13,-0.21);
    \node[fp key] at (0.17,-0.21) {Off the stable set: pure equilibria};
    \draw[draw=fpStable,line width=1.15pt,dash pattern=on 3pt off 1.8pt]
      (0.01,-0.30) -- (0.13,-0.30);
    \node[fp key] at (0.17,-0.30) {Stable set: selected interior $\rho$-NE};
  \end{scope}

  \begin{scope}[shift={(1.42,0)}]
    \def\fpP{\fpHiP}\def\fpQ{\fpLoQ}
    \node[font=\small\bfseries] at (0.5,-0.43)
      {\panellabel{fig:weak_scp_spiral} Spiral-type};
    \node at (0.5,1.065) {$b_1=b_2=1$};
    \fill[fpPurpleWash] (0,\fpQ) rectangle (\fpP,1);
    \fill[fpBlueWash] (\fpP,\fpQ) rectangle (1,1);
    \fill[fpGoldWash] (0,0) rectangle (\fpP,\fpQ);
    \fill[fpGreenWash] (\fpP,0) rectangle (1,\fpQ);
    \fpWeakBands

    \foreach \t in {0.08,0.29,0.50,0.71,0.92} {
      \pgfmathsetmacro{\lx}{\fpP*\t}
      \pgfmathsetmacro{\rx}{\fpP+(1-\fpP)*\t}
      \pgfmathsetmacro{\by}{\fpQ*\t}
      \pgfmathsetmacro{\ty}{\fpQ+(1-\fpQ)*\t}
      \fpRay{0}{\ty}{1}{1}{0}{\fpP}{\fpQ}{1}
      \fpRay{\lx}{\fpQ}{1}{1}{0}{\fpP}{\fpQ}{1}
      \fpRay{\rx}{1}{1}{0}{\fpP}{1}{\fpQ}{1}
      \fpRay{\fpP}{\ty}{1}{0}{\fpP}{1}{\fpQ}{1}
      \fpRay{1}{\by}{0}{0}{\fpP}{1}{0}{\fpQ}
      \fpRay{\rx}{\fpQ}{0}{0}{\fpP}{1}{0}{\fpQ}
      \fpRay{\lx}{0}{0}{1}{0}{\fpP}{0}{\fpQ}
      \fpRay{\fpP}{\by}{0}{1}{0}{\fpP}{0}{\fpQ}
    }
    \fpWeakFrame

    \foreach \turn [remember=\fpNext as \fpTop (initially 0.91)] in {1,...,9} {
      \pgfmathsetmacro{\fpRight}{1-(1-\fpP)*\fpQ/\fpTop}
      \pgfmathsetmacro{\fpBottom}{\fpP*\fpQ/\fpRight}
      \pgfmathsetmacro{\fpLeft}{\fpP*(1-\fpQ)/(1-\fpBottom)}
      \pgfmathsetmacro{\fpNext}{1-(1-\fpQ)*(1-\fpP)/(1-\fpLeft)}
      \ifnum\turn<4\relax
        \tikzset{fp turn/.style={fp spiral arrow}}
      \else
        \tikzset{fp turn/.style={fp spiral,line width=0.7pt}}
      \fi
      \draw[fp turn] (\fpP,\fpTop) -- (\fpRight,\fpQ);
      \draw[fp turn] (\fpRight,\fpQ) -- (\fpP,\fpBottom);
      \draw[fp turn] (\fpP,\fpBottom) -- (\fpLeft,\fpQ);
      \draw[fp turn] (\fpLeft,\fpQ) -- (\fpP,\fpNext);
    }
    \node[circle,fill=fpTeal,inner sep=1.5pt] at (\fpP,0.91) {};
    \node[fp equilibrium,fill=fpTeal] at (\fpP,\fpQ) {};
    \draw[fp spiral arrow] (0.01,-0.21) -- (0.13,-0.21);
    \node[fp key] at (0.17,-0.21) {Spiral to the selected interior $\rho$-NE};
    \filldraw[fill=fpEquilibriumWash,draw=fpInk!35,line width=0.4pt]
      (0.01,-0.322) rectangle (0.13,-0.278);
    \node[fp key] at (0.17,-0.30) {Interior $\rho$-NE: $I_2\times I_1$};
  \end{scope}
\end{tikzpicture}%
\endgroup%
        \caption{Continuous-time NL-FP geometry under SCP, with $b_1=b_2=1$ in both panels. The shaded rectangle $I_2\times I_1$ is the set of interior $\rho$-NE; solid lines mark the active thresholds. \textbf{(a) Coordination-type:} Off the stable set, trajectories converge to pure equilibria. On the stable set, they converge to $(\underline r_2,\underline r_1)$. \textbf{(b) Spiral-type:} Trajectories spiral toward $(\overline r_2,\underline r_1)$. The continuous-time selection stops at the selected interior NE of the reduced game.}
        \label{fig:weak_scp}
\end{figure}

\begin{prop}[Convergence under SSCP]
\label{prop:sscp}
Every $2\times2$ game $(G,\rho)$ in which each $d^i$ satisfies SSCP has the same best-response correspondence as a
$2\times2$ game $(\bar{G},\mu)$, whose advantage functions are $\bar d^1$ and $\bar d^2$, where $\bar d^i(s)=d^i(1)(s-r_i)$.
Under the same tie-breaking rule, continuous-time \rhoFP{} in $G$ has the same trajectories as continuous-time FP in $\bar{G}$, and discrete-time \rhoFP{} in $G$ has the same iterates as discrete-time FP in $\bar{G}$. These trajectories and iterates converge to a $\rho$-NE as $t\to\infty$ and $k\to\infty$, respectively. In particular, every $\rho$-NE of $(G,\rho)$ is an NE of $(\bar{G},\mu)$.
\end{prop}

In the case of (non-strict) SCP,  player~1 is indifferent between its two actions throughout the horizontal region $[0,1]\times[\underline r_1,\overline r_1]$, and player~2 throughout the vertical region $[\underline r_2,\overline r_2]\times[0,1]$ in the $(p,q)$ plane. Lemma \ref{lem:ne_scp} below shows that the set of interior $\rho$-NE %
under SCP expands from a singleton $(r_2,r_1)$ to a closed rectangle formed by the indifference intervals. The proof of Lemma \ref{lem:ne_scp} is given in Appendix \ref{app:scp}.
\begin{lem}[$\rho$-NE of $2\times 2$ games under SCP]
\label{lem:ne_scp}
Suppose $d^1,d^2$ satisfy SCP, with zero intervals
$I_i=[\underline r_i,\overline r_i]\subset(0,1)$,
where $\underline r_i<\overline r_i$.
Let $\sigma_i:=\operatorname{sgn}d^i(1)\in\{-1,1\}$.
Then the $\rho$-NE set is given by
\[
\mathcal E_\rho
=
(I_2\times I_1)
\;\cup\;
\begin{cases}
\{(0,0),(1,1)\},
    & \sigma_1=\sigma_2=1,\\
\{(0,1),(1,0)\},
    & \sigma_1=\sigma_2=-1,\\
\varnothing,
    & \sigma_1\sigma_2=-1.
\end{cases}
\]
In particular, the interior $\rho$-NE form the closed
rectangle $I_2\times I_1$, and any other $\rho$-NE is pure.
\end{lem}

With this equilibrium set characterized, we now establish convergence when each player uses a fixed action throughout its indifference interval.
\begin{prop}[Convergence under SCP]
\label{prop:scp}
Suppose $d^1,d^2$ satisfy SCP with $\underline r_i<\overline r_i$, and each player $i$ uses a fixed action throughout its indifference interval $I_i$. Then continuous-time \rhoFP{} converges to a $\rho$-NE as $t\to\infty$, and discrete-time \rhoFP{} converges to a $\rho$-NE as $k\to\infty$. In particular, for a continuous-time trajectory $\z_t:=(p_t,q_t)$ and discrete-time iterates $\z_k:=(p_k,q_k)$,
\[
\lim_{t\to\infty}\operatorname{dist}(\z_t,\mathcal E_\rho)=0,
\qquad
\lim_{k\to\infty}\operatorname{dist}(\z_k,\mathcal E_\rho)=0.
\]
The $\rho$-NE to which the continuous-time trajectory or the discrete-time iterates converge may depend on the tie-breaking rule.
\end{prop}
The proof is given in Appendix~\ref{app:scp}. Figure~\ref{fig:weak_scp} illustrates the geometry under SCP: the horizontal indifference region $[0,1]\times I_1$ and the vertical indifference region $I_2\times[0,1]$ intersect in the rectangle $I_2\times I_1$. By Lemma~\ref{lem:ne_scp}, this rectangle is  the set of interior $\rho$-NE.

\subsection{SCP identification}
\label{sec:sc_id}
We have established convergence and characterized the geometry of NL-FP {in $2\times 2$ games} under SCP. Building on the ERM and CVaR examples above, we now develop sufficient conditions for identifying SCP across broader classes of evaluation functionals. Table~\ref{tab:fun_scp} in Appendix \ref{app:fun} summarizes the conditions under which representative nonlinear evaluation functionals induce advantage functions satisfying SCP. A comprehensive identification of SCP for several common classes of risk measures can be found in Appendix \ref{app:fun_scp}.

Since SCP implies $d^i(0)d^i(1)<0$, we focus on the case $d^i(0)<0<d^i(1)$ for simplicity. The other case $d^i(0)>0>d^i(1)$ follows analogously by reversing the signs. We start with a fundamental sufficient condition (\textbf{SC1}), which applies to general evaluation functionals.

\vspace{4pt}
\noindent
\textbf{Sufficient Condition 1 (SC1).} $d^i$ satisfies SSCP if it is continuous and \textit{strictly} increasing in $(0,1)$. $d^i$ satisfies SCP if it is continuous and nondecreasing in $(0,1)$.
\par
\vspace{4pt}

\begin{proof}
Continuity and the endpoint inequalities imply the existence of a zero. Strict monotonicity implies the uniqueness of zero, and monotonicity implies the SCP.
\end{proof}
\textbf{SC1} is especially useful for evaluation functionals that need not be monetary risk measures, such as rank-dependent expected utility (RDEU) \citep{quiggin1982theory}. Positive homogeneity and translation invariance (Assumption~\ref{asp:rm}(e) and~(c), respectively) need not hold for RDEU. In particular, failure of translation invariance excludes a functional from the class of risk measures in Definition~\ref{def:rm}. We use \textbf{SC1} to prove the SCP of RDEU in Appendix \ref{app:id_scp}.

For risk measures, we exploit Assumption \ref{asp:rm} (a)-(c) to obtain additional sufficient conditions. We observe that the
Assumption \ref{asp:rm} (a) and (c)  imply $\rho(m)=m$ for every constant payoff $m$, thus
\[ d^1(0)=\rho(a,b;0)-\rho(c,d;0)=b-d,\quad d^1(1)=\rho(a,b;1)-\rho(c,d;1)=a-c. \]
Then $d^i(0)<0<d^i(1)$ is equivalent to $b-d<0<a-c$, i.e., $a>c$, $d>b$. Without loss of generality, we consider $a\ge d$. The other case $a>c, d>b, a<d$ is equivalent to it via relabeling actions. By the translation-invariance of $\rho$,
\begin{align*}
    d^1(q)=b+\rho(a-b,0;q)-(d+\rho(c-d,0;q))= \Psi(a-b,q) - \Psi(c-d,q) +(b-d),
\end{align*}
where $\Psi(\Delta,q):=\rho(\Delta,0;q)$. Since $b-d$ is independent of $q$, the monotonicity required by \textbf{SC1} is determined by $\Psi(a-b,q) - \Psi(c-d,q)$. Moreover, $b-d<0<a-c$ implies $a-b>c-d$. This representation motivates \textbf{SC2}, which gives a monotonicity condition directly on the family of two-point evaluations $\Psi$.
\vspace{6pt}

\noindent
\textbf{Sufficient Condition 2 (SC2).}
Let $\rho$ be a continuous RM. Then $\rho$ satisfies SSCP
(respectively, SCP) if, for every $\Delta_1>\Delta_2\ge 0$, the map
\(
q \longmapsto \Psi(\Delta_1,q)-\Psi(\Delta_2,q)
\)
is strictly increasing (respectively, nondecreasing) on $(0,1)$.
\par
\vspace{4pt}

The proof is provided in Appendix~\ref{app:sc_pf}.
\textbf{SC2} can be used to establish (S)SCP for common risk measures in finance and decision theory, including ERM, CVaR, lower-tail optimized certainty equivalents (OCEs), and coherent, spectral, and distortion risk measures, under the respective conditions detailed in Appendix~\ref{app:id_scp}.

Finally, we provide \textbf{SC3},  applicable to general functionals, including ones that are not RMs such as CPT. We establish the SSCP of CPT using \textbf{SC3} in Appendix \ref{app:cpt}.

\vspace{4pt}
\noindent
\textbf{Sufficient Condition 3 (SC3).} $d^i$ satisfies SSCP if it is continuous and for $0< q_1<q_2<1$, the function $H_{q_1,q_2}(x,y)
:=
\rho(x,y;q_2)
-
\rho(x,y;q_1)$ is strictly increasing in $x$ and strictly decreasing in $y$. $d^i$ satisfies SCP if it is continuous and for $0< q_1<q_2<1$, the function $H_{q_1,q_2}(x,y)
$ is non-decreasing in $x$ and non-increasing in $y$.

\subsection{SCP can be violated}
The preceding results identify conditions under which broad classes of evaluation functionals preserve SCP. We now show that some functionals commonly used in decision theory, including mean-semivariance (MSV) \citep{estrada2007mean,estrada2007opt}, certain concave risk measures, and criteria based on higher-order moments \citep{pham2011efficient,zhou2021solving}, can induce advantage functions that violate SCP. We illustrate this with
MSV, which is common in finance, especially in portfolio theory, and defer the concave RMs and higher order moments to Appendix \ref{app:scp_vio}. The (reward-side) MSV is defined as
\[
\rho_\lambda(X):=
\mathbb{E}[X]
-
\lambda\,\mathrm{LSV}(X),
\qquad
\lambda>0,
\]
where $\mathrm{LSV}(X)
:=
\mathbb{E}
\!\left[
\bigl(\mathbb{E}[X]-X\bigr)_+^2
\right]$ is the lower semivariance. Unlike variance, MSV penalizes only below-mean deviations.
It is not a monetary RM %
as it violates monotonicity (Assumption \ref{asp:rm} (b)).

For a two-point distribution $(\Delta,0;q)$ with $\Delta>0$,  $\mathrm{LSV}(\Delta,0;q)=(1-q)q^2\Delta^2$, and thus  $\rho_\lambda(\Delta,0;q)
=
q\Delta
-
\lambda q^2(1-q)\Delta^2.$ Taking  $\Delta_1=a-b=1, \Delta_2=c-d=0.2,$ we have
\begin{align*}
    d^1(q)=b-d+0.8q-0.96\lambda q^2(1-q).
\end{align*}
Choosing $\lambda=5$ and $b-d=-0.02$ gives $d^1(q)=4.8q^3-4.8q^2+0.8q-0.02.$ This cubic function is not monotone. It has three zeros $0<q_1<q_2<q_3<1$, with $d^1(q)<0$ for $q\in(0,q_1)\cup(q_2,q_3)$ and $d^1(q)>0$ for $q\in(q_1,q_2)\cup(q_3,1)$. Therefore, these repeated sign changes violate SCP.

 The violation of SCP, however, does not necessarily imply failure of convergence of NL-FP. In fact, we will show that NL-FP converges to $\rho$-NE when both advantages have an arbitrary finite number of crossing points.

\section{Geometric Analysis of Continuous-time NL-FP in \texorpdfstring{$2 \times 2$}{2 x 2} Games}
\label{sec:ct}

When SCP fails, the advantage functions may change sign multiple times, introducing additional switching lines %
and a richer phase portrait than the classical FP geometry shown in Figure~\ref{fig:scp_regions}. Nevertheless, this increased geometric complexity need not prevent convergence. In this section, we study the dynamics of continuous-time NL-FP. We identify and exploit a \textit{finite-grid structure} underlying the dynamics in $2\times 2$ games. Under the regularity assumptions stated below, the switching lines partition the strategy square into rectangular cells, within each of which the pure best-response pair is constant and trajectories follow straight line segments toward a corner. This structure allows us to establish the convergence of every trajectory of the selected continuous-time dynamics to a single $\rho$-NE and to characterize the landscape. %

We impose a \textit{finite-crossing} assumption below (Assumption \ref{asp:2by2}), under which the convergence of continuous-time NL-FP in $2 \times 2$ games is established. Assumption \ref{asp:2by2} requires finitely many interior zeros %
with a sign change at each zero. It does not require continuity of the evaluation functionals. Consequently, each $d^i$ has a constant nonzero sign on every
connected component of $[0,1]\setminus Z^i$, where
$Z^i$ is the zero set of $d^i$, and these signs alternate across
successive zeros.
\begin{asp}[Finite-crossing]
\label{asp:2by2}
For each $i\in[2]$,
$d^i:[0,1]\to\mathbb{R}$ has finitely many zeros, changes sign
at every zero, and satisfies $d^i(0)d^i(1)\neq 0$. Moreover,
for every $0\le s<t\le 1$, $d^i(s)d^i(t)<0$ implies that $d^i(r)=0$ for some $r\in(s,t)$.
\end{asp}

Let $n\ge0$ and $m\ge0$ be the numbers of zeros of $d^1$
and $d^2$, respectively. Thus, the interior zero sets are finite sets, denoted by $\cP=\{p_1,p_2,\cdots,p_m\},\ \cQ=\{q_1,q_2,\cdots,q_n\}$.
We list each nonempty zero set in increasing order $0<q_1<\cdots<q_n<1,\
0<p_1<\cdots<p_m<1.$
We set $p_0=q_0=0$, $p_{m+1}=q_{n+1}=1$, and interpret a
zero set with no elements as empty. We note that no restriction is imposed on the parity of $m$ or $n$, and we allow the empty zero set ($m=0$ or $n=0$). Therefore,  Assumption \ref{asp:2by2} specializes to SCP for $m=n=1$, i.e., the case of single zero.

\subsection{Finite-grid geometry}
\label{sec:grid_2by2}
The zeros partition the strategy square $[0,1]^2$ into $(m+1)\times(n+1)$ grid cells (see Figure~\ref{fig:2by2_pf_a}). We use $ \mathcal V:=\cP\times\cQ$
to denote the set of interior (cell) vertices, and let $\mathcal C:=\{0,1\}^2$ be the set of all four corners. On an open cell $R_{ij}:=(p_{i-1},p_i) \times (q_{j-1},q_j)$,
both advantage functions have constant nonzero signs. Consequently, the pure best-response pair is the same for every $(p,q)\in R_{ij}$, and we denote it by $\vc_{ij}:=(x_j,y_i)\in\C$.  Within an open cell $R_{ij}$, the correspondence is the singleton $\{\vc_{ij}\}$, so it follows from Equation~\eqref{eqt:cfp_2by2} that the continuous-time dynamics of NL-FP satisfy
\[
\dot \z_t=\vc_{ij}-\z_t.
\]
Let $t_0$ be a time at which the trajectory starts at $\z_{t_0} \in R_{ij}$. Solving the linear differential equation above gives, for $t\ge t_0$ as long as the trajectory remains in $R_{ij}$,
\[
\z_t=\vc_{ij}+\bigl(\z_{t_0}-\vc_{ij}\bigr)e^{-(t-t_0)},
\]
and hence it follows a straight line segment toward the corresponding corner $\vc_{ij}$ (see Figure~\ref{fig:2by2_pf_a}).
A trajectory may reach a cell boundary and switch into an adjacent cell, so we must specify how it continues at the switching points.
We call the common boundary segment of two adjacent cells, with both endpoints excluded, an \textit{open switching segment}. A point on such a segment is called a \textit{regular switching point}. A trajectory crossing at such a point is called a \textit{regular switch}. A finite itinerary is called \textit{regular} if it avoids vertices. For example, the open switching segment between $R_{ij}$ and $R_{i+1,j}$ is $S_{i,i+1;j}:=\{p_i\}\times(q_{j-1},q_j),
\ i\in[m],\ j\in[n+1].$

\begin{figure}[t]
    \centering
    \begingroup
\input{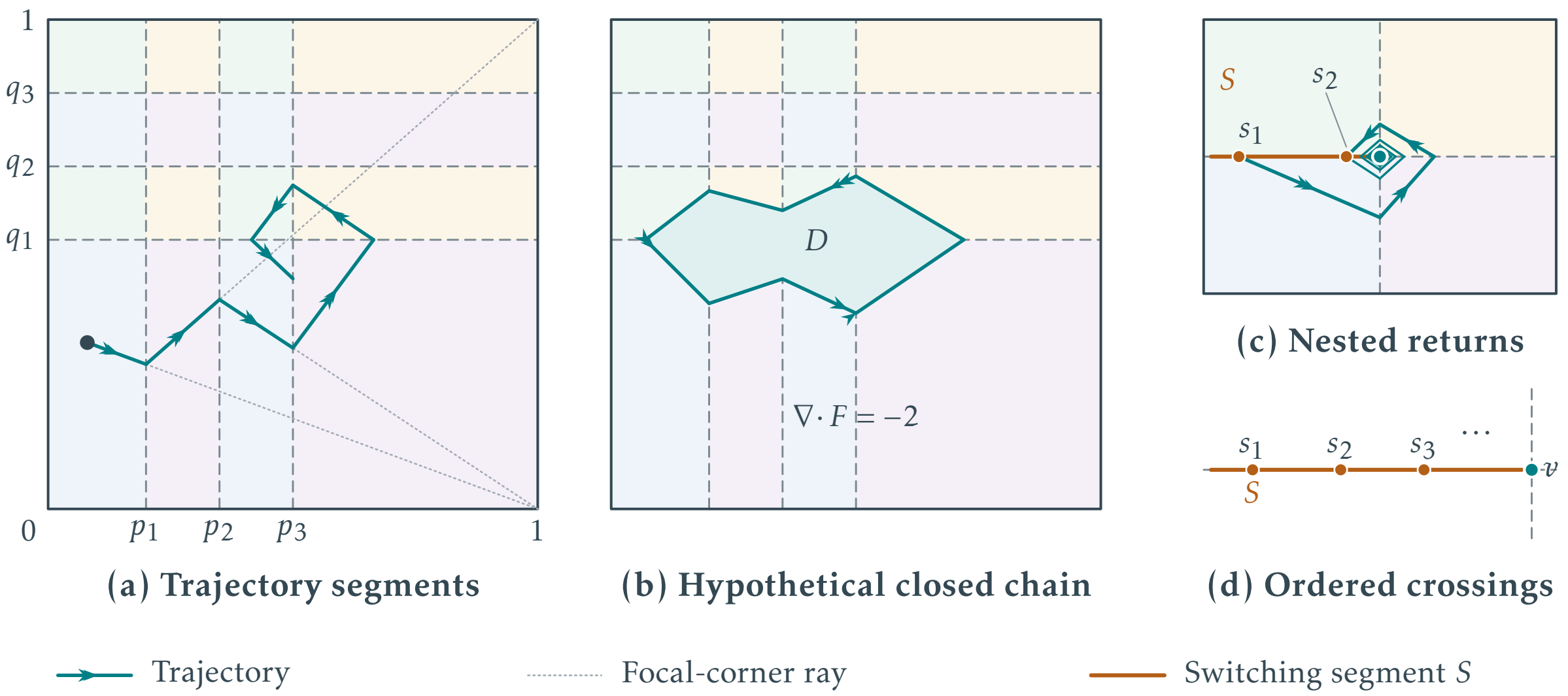}%
\pgfmathsetlengthmacro{\proofUnit}{0.30\textwidth}%
\tikzset{
  proof path/.style={fp spiral arrow,line width=1pt,
    decoration={markings,
      mark=at position .55 with {\arrow{Stealth[length=1.5mm,width=1.1mm]}}}},
  proof guide/.style={draw=fpInk!45,line width=.5pt,densely dotted},
  proof section/.style={draw=fpStable,line width=1.15pt},
  proof crossing/.style={circle,draw=white,line width=.5pt,
    fill=fpStable,inner sep=1.3pt}
}%
\def\proofGrid{%
  \foreach \i/\xmin/\xmax in {0/0/.20,1/.20/.35,2/.35/.50,3/.50/1} {
    \foreach \j/\ymin/\ymax in {0/0/.55,1/.55/.70,2/.70/.85,3/.85/1} {
      \pgfmathtruncatemacro{\proofX}{1-mod(\j,2)}
      \pgfmathtruncatemacro{\proofY}{mod(\i,2)}
      \ifnum\proofX=0\relax
        \ifnum\proofY=0\relax\def\proofWash{fpGreenWash}
        \else\def\proofWash{fpGoldWash}\fi
      \else
        \ifnum\proofY=0\relax\def\proofWash{fpBlueWash}
        \else\def\proofWash{fpPurpleWash}\fi
      \fi
      \fill[\proofWash] (\xmin,\ymin) rectangle (\xmax,\ymax);
    }
  }
  \foreach \x in {.20,.35,.50}
    \draw[fp threshold] (\x,0) -- (\x,1);
  \foreach \y in {.55,.70,.85}
    \draw[fp threshold] (0,\y) -- (1,\y);
  \draw[fpInk,line width=.75pt] (0,0) rectangle (1,1);
}%
\begin{tikzpicture}[
  x=\proofUnit,y=\proofUnit,font=\footnotesize,text=fpInk,
  line cap=round,line join=round
]
  \node[font=\small\bfseries] at (.5,-.16)
    {\panellabel{fig:2by2_pf_a} Trajectory segments};
  \node[font=\small\bfseries] at (1.65,-.16)
    {\panellabel{fig:2by2_pf_b} Hypothetical closed chain};
  \node[font=\small\bfseries] at (2.72,.34)
    {\panellabel{fig:phase_left} Nested returns};

  \proofGrid
  \foreach \x/\i in {.20/1,.35/2,.50/3}
    \node[below,inner sep=3pt] at (\x,0) {$p_\i$};
  \foreach \y/\j in {.55/1,.70/2,.85/3}
    \node[left,inner sep=3pt] at (0,\y) {$q_\j$};
  \node[below left,inner sep=3pt] at (0,0) {$0$};
  \node[below,inner sep=3pt] at (1,0) {$1$};
  \node[left,inner sep=3pt] at (0,1) {$1$};
  \draw[proof guide] (.08,.34) -- (1,0);
  \draw[proof guide] (.20,.29565217) -- (1,1);
  \draw[proof guide] (.35,.42771739) -- (1,0);
  \foreach \xa/\ya/\xb/\yb in {
      .08/.34/.20/.29565217,
      .20/.29565217/.35/.42771739,
      .35/.42771739/.50/.32901338,
      .50/.32901338/.66467290/.55,
      .66467290/.55/.50/.66148763,
      .50/.66148763/.41572962/.55,
      .41572962/.55/.50/.47067250}
    \draw[proof path] (\xa,\ya) -- (\xb,\yb);
  \node[circle,fill=fpInk,inner sep=1.5pt] at (.08,.34) {};

  \begin{scope}[shift={(1.15,0)}]
    \proofGrid
    \path[fill=fpTeal!12] (.20,.42) -- (.35,.47) -- (.50,.40)
      -- (.72,.55) -- (.50,.68) -- (.35,.61) -- (.20,.65)
      -- (.07,.55) -- cycle;
    \draw[proof path] (.20,.42) -- (.35,.47) -- (.50,.40) -- (.72,.55);
    \draw[proof path] (.72,.55) -- (.50,.68) -- (.35,.61) -- (.20,.65);
    \draw[proof path] (.20,.65) -- (.07,.55) -- (.20,.42);
    \node[fill=fpTeal!12,inner sep=1.5pt] at (.42,.55) {$D$};
    \node at (.5,.19) {$\nabla\!\cdot F=-2$};
  \end{scope}

  \begin{scope}[shift={(2.36,0)}]
    \fill[fpGreenWash] (0,.72) rectangle (.36,1);
    \fill[fpGoldWash] (.36,.72) rectangle (.72,1);
    \fill[fpBlueWash] (0,.44) rectangle (.36,.72);
    \fill[fpPurpleWash] (.36,.44) rectangle (.72,.72);
    \draw[fp threshold] (0,.72) -- (.72,.72);
    \draw[fp threshold] (.36,.44) -- (.36,1);
    \draw[fpInk,line width=.75pt] (0,.44) rectangle (.72,1);
    \draw[proof section] (.015,.72) -- (.345,.72);
    \node[text=fpStable,anchor=south west,inner sep=2pt] at (.015,.84) {$S$};
    \foreach \k in {0,...,15} {
      \pgfmathsetmacro{\proofRadius}{.4/(1+.8*\k)}
      \pgfmathsetmacro{\proofNext}{.4/(1+.8*(\k+1))}
      \pgfmathsetmacro{\proofAngle}{180+90*\k}
      \ifnum\k<4\relax
        \tikzset{proof turn/.style={proof path}}
      \else
        \tikzset{proof turn/.style={fp spiral,line width=.65pt}}
      \fi
      \draw[proof turn]
        ({.36+.72*\proofRadius*cos(\proofAngle)},
         {.72+.56*\proofRadius*sin(\proofAngle)}) --
        ({.36+.72*\proofNext*cos(\proofAngle+90)},
         {.72+.56*\proofNext*sin(\proofAngle+90)});
    }
    \node[proof crossing] at (.072,.72) {};
    \node[proof crossing] at ({.72*17/42},.72) {};
    \node[anchor=south,inner sep=2pt] at (.10,.73) {$s_1$};
    \draw[fpInk!60,line width=.4pt] ({.72*17/42},.74) -- (.25,.85);
    \node[anchor=south,inner sep=1pt] at (.25,.85) {$s_2$};
    \node[fp equilibrium,fill=fpTeal,inner sep=1.5pt] at (.36,.72) {};

    \node[font=\small\bfseries] at (.36,-.16)
      {\panellabel{fig:phase_right} Ordered crossings};
    \draw[fp threshold] (0,.08) -- (.72,.08);
    \draw[fp threshold] (.67,-.06) -- (.67,.25);
    \draw[proof section] (.015,.08) -- (.65,.08);
    \foreach \x/\i in {.10/1,.28/2,.45/3} {
      \node[proof crossing] at (\x,.08) {};
      \node[above,inner sep=3pt] at (\x,.08) {$s_\i$};
    }
    \node[text=fpStable,below,inner sep=3pt] at (.10,.08) {$S$};
    \node at (.56,.15) {$\cdots$};
    \node[fp equilibrium,fill=fpTeal,inner sep=1.5pt] at (.67,.08) {};
    \node[right,inner sep=3pt] at (.67,.08) {$v$};
  \end{scope}

  \draw[proof path] (.02,-.34) -- (.17,-.34);
  \node[fp key] at (.21,-.34) {Trajectory};
  \draw[proof guide] (.98,-.34) -- (1.13,-.34);
  \node[fp key] at (1.17,-.34) {Focal-corner ray};
  \draw[proof section] (2.13,-.34) -- (2.28,-.34);
  \node[fp key] at (2.32,-.34) {Switching segment $S$};
\end{tikzpicture}%
\endgroup%
    \caption{Geometry of the continuous-time convergence proof.
        \textbf{(a)} Each trajectory segment points toward its corresponding
        corner; dotted lines extend representative rays.
        \textbf{(b)} A hypothetical loop enclosing $D$.
        \textbf{(c)} A trajectory returns to the same open switching
        segment $S$ (orange), with crossings $s_1,s_2$ in the same direction.
        \textbf{(d)} Successive crossings are strictly ordered.
        Gray dashed lines mark crossings, and teal circles mark
        equilibria. %
        }
    \label{fig:all_geometric_panels}
\end{figure}

\paragraph{Selection convention.}
\label{par:selection_2by2}
At a regular switching point, the trajectory continues across the switching segment into the adjacent cell.
At a point of a vertical open switching segment $\{p_i\}\times(q_{j-1},q_j)$, the two adjacent cells prescribe the same horizontal velocity $\dot p_t=x_j-p_i\neq0$. Analogously, the two adjacent cells prescribe the same vertical velocity $\dot q_t=y_i-q_j\neq0$. Since these components are nonzero, a trajectory reaching an open switching segment crosses it and cannot remain on it. The  crossing direction is determined by the sign of $y_i-q_j\ne0$ or $x_j-p_i$.

At an interior vertex $v$, both players are indifferent,
so their current mixed strategies are best responses and $v$ is
a $\rho$-NE. Choosing these responses keeps the trajectory at $v$ in
Equation \eqref{eqt:cfp_2by2}. However, the inclusion may also allow
a trajectory to leave. To specify a unique continuation, we require
$\z_t=\z_\tau$ for all $t\ge\tau$ whenever $\z_\tau$ is a
$\rho$-NE, including when $\tau=0$. We call the trajectories satisfying this rule the selected continuous-time NL-FP trajectories.
This convention affects only trajectories that reach a saddle
crossing. The initial states that reach such a crossing
form a set of measure zero.

\subsection{Convergence to \texorpdfstring{$\rho$}{rho}-NE: %
 A geometric proof}

\begin{thm}[Continuous-time   convergence in $2\times2$ games]
\label{thm:ct_conv_2by2}
Under Assumption \ref{asp:2by2}, every selected continuous-time $\rho$-FP trajectory converges to a $\rho$-NE,
\[ \lim_{t\to\infty} \z_t =\z^\star \in \mathcal E_\rho=(\cP\times \cQ)\cup \mathcal C_\rho.\]
\end{thm}
We provide a proof sketch below.
The complete proof is deferred to Appendix \ref{app:pf_ct_conv_2by2}.

\begin{proof}[Proof sketch of Theorem \ref{thm:ct_conv_2by2}]
If $d^1$ has no zeros ($n=0$), the row player's pure best response
is independent of $q$. Thus $p_t$ moves monotonically toward that
action and crosses each vertical switching line at most once.
The column player's best response is eventually constant
as well, and the trajectory converges to a pure $\rho$-NE.
The case $m=0$ follows by interchanging the players. We may thus
assume $m,n\ge1$, so the grid has both vertical and horizontal
switching lines, and each intersection is an interior $\rho$-NE.

Figure~\ref{fig:2by2_pf_a} shows how a trajectory moves
through the grid. In each cell it follows a straight line toward the corresponding corner. When it crosses a switching
segment, it continues toward the corner for the next cell. If there
are only finitely many switches, the trajectory either reaches an
interior vertex and stops or stays in the closure of its last cell
and approaches its corresponding corner. In either case, the limit
is a $\rho$-NE.

With infinitely many switches, the trajectory must cross some
segment $S$ repeatedly, as in Figure~\ref{fig:phase_left}.
Returning to the same point would form a loop. To see why this is
impossible, consider the hypothetical loop around $D$ in
Figure~\ref{fig:2by2_pf_b}. The velocity follows the
loop, so the net flow across it is zero. Yet applying Green's theorem cell by cell to the region $D$ yields
\[
0=\sum_{i,j}\iint_{D\cap R_{ij}}\nabla\!\cdot(\vc_{ij}-\z)\,dA
=-2\operatorname{Area}(D)<0,
\] since the divergence is
$-2$ in each cell and the contributions across shared edges cancel.
Moreover, the next crossings of $S$ preserve the order of the
starting points on $S$ (Lemma~\ref{lem:order}). Thus, if $s_2$ is
to the right of $s_1$ as drawn in
Figure~\ref{fig:phase_right}, every later crossing is
farther to the right. The reverse order gives the reverse direction.
In either case, the crossing points approach a limit on the
closure of $S$.

The limits on different switching segments must also agree.
Every transition that occurs infinitely often lies on a cycle of
such transitions. Taking limits around that cycle joins the
limiting crossing points by straight pieces. If any two consecutive
limits differed, these pieces would form a closed chain of positive
length, which the argument in Figure~\ref{fig:2by2_pf_b}
rules out. Both horizontal and vertical crossings must recur.
Otherwise, one player's best response would eventually stay fixed,
and the corresponding belief would move monotonically, leaving
only finitely many crossings. The common limit is therefore an
interior vertex. Between two consecutive switches, the trajectory
follows the straight line segment joining the two crossing points.
Every point on this segment is no farther from the vertex than the
farther endpoint. Since both endpoints converge to the vertex and
the switching times tend to infinity (Lemma~\ref{lem:nonaccumulation}),
the whole trajectory converges to the vertex, which is a $\rho$-NE.
\end{proof}

\begin{figure}[t]
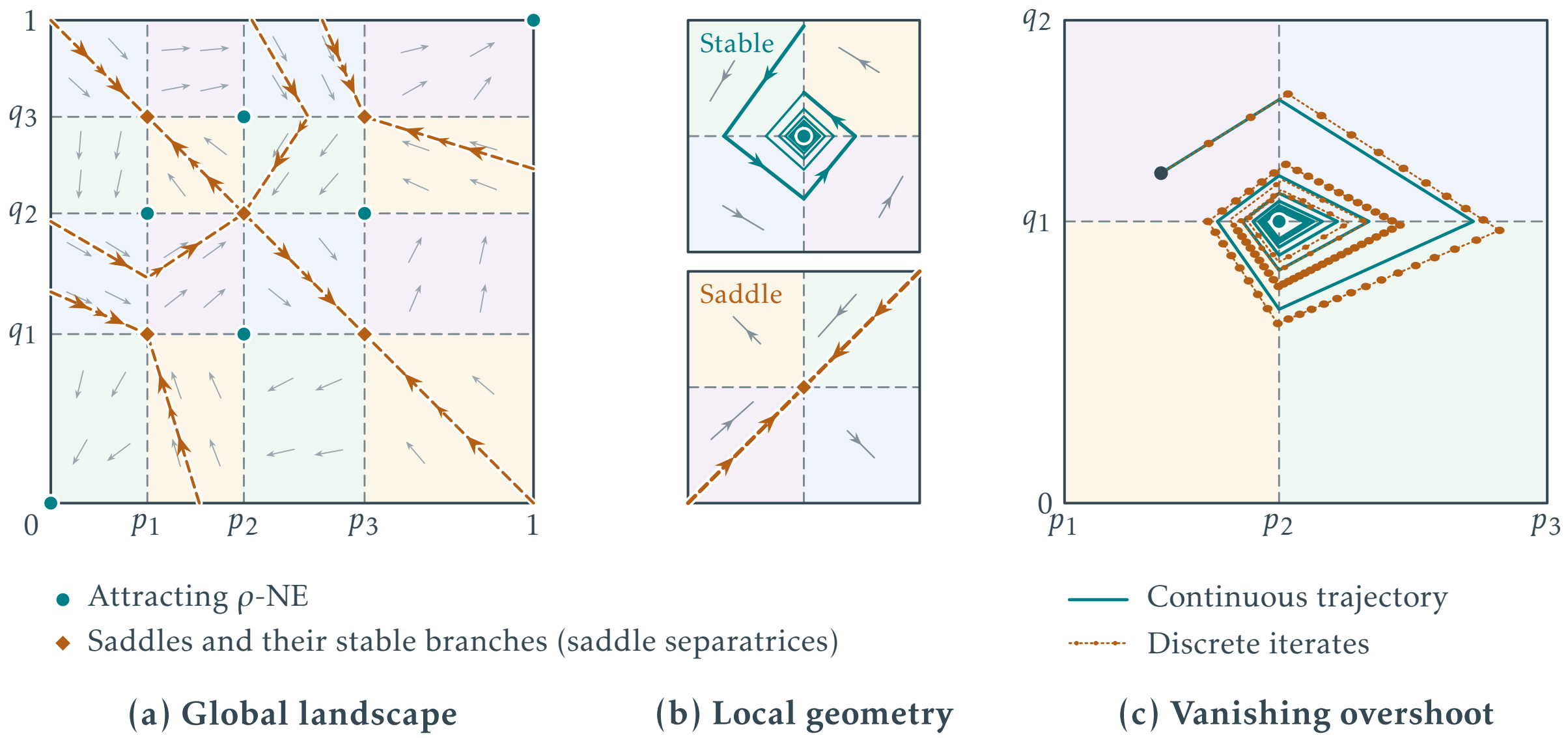

    \centering
    \begingroup
\input{fig/fp-geometry-common}%
\input{fig/rho-fp-paths}%
\pgfmathsetlengthmacro{\rhoUnit}{0.297\textwidth}%
\def\rhoDiamond#1#2{%
  \path[draw=white,line width=.7pt,fill=fpStable]
    (#1,{#2+.019}) -- ({#1+.019},#2) --
    (#1,{#2-.019}) -- ({#1-.019},#2) -- cycle;
}
\begin{tikzpicture}[
  x=\rhoUnit,y=\rhoUnit,font=\footnotesize,text=fpInk,
  line cap=round,line join=round,
  rho attractor/.style={fp equilibrium,fill=fpTeal,inner sep=1.6pt},
  rho branch/.style={fp stable,line width=.85pt,
    decoration={markings,
      mark=at position .58 with {\arrow{Stealth[length=1.35mm,width=1mm]}}}},
  rho continuous/.style={draw=fpTeal,line width=.9pt},
  rho discrete/.style={draw=fpStable,line width=.55pt,densely dotted}
]
  \node[font=\small\bfseries] at (.5,-.44)
    {\panellabel{fig:rho_fp_global} Global landscape};
  \node[font=\small\bfseries] at (1.56,-.44)
    {\panellabel{fig:rho_fp_local} Local geometry};
  \node[font=\small\bfseries] at (2.60,-.44)
    {\panellabel{fig:rho_fp_overshoot} Vanishing overshoot};

  \foreach \i/\xmin/\xmax in {0/0/.20,1/.20/.40,2/.40/.65,3/.65/1} {
    \foreach \j/\ymin/\ymax in {0/0/.35,1/.35/.60,2/.60/.80,3/.80/1} {
      \pgfmathtruncatemacro{\rhoX}{mod(\j,2)}
      \pgfmathtruncatemacro{\rhoY}{mod(\i,2)}
      \ifnum\rhoX=0\relax
        \ifnum\rhoY=0\relax\def\rhoWash{fpGreenWash}
        \else\def\rhoWash{fpGoldWash}\fi
      \else
        \ifnum\rhoY=0\relax\def\rhoWash{fpBlueWash}
        \else\def\rhoWash{fpPurpleWash}\fi
      \fi
      \fill[\rhoWash] (\xmin,\ymin) rectangle (\xmax,\ymax);
      \foreach \u in {.30,.70} {
        \foreach \v in {.30,.70} {
          \pgfmathsetmacro{\sx}{\xmin+\u*(\xmax-\xmin)}
          \pgfmathsetmacro{\sy}{\ymin+\v*(\ymax-\ymin)}
          \pgfmathsetmacro{\dx}{\rhoX-\sx}
          \pgfmathsetmacro{\dy}{\rhoY-\sy}
          \pgfmathsetmacro{\rhoNorm}{sqrt(\dx*\dx+\dy*\dy)}
          \draw[draw=fpFlow!80,line width=.35pt,
            -{Stealth[length=.9mm,width=.65mm]}]
            ({\sx-.030*\dx/\rhoNorm},{\sy-.030*\dy/\rhoNorm}) --
            ({\sx+.030*\dx/\rhoNorm},{\sy+.030*\dy/\rhoNorm});
        }
      }
    }
  }
  \foreach \x/\i in {.20/1,.40/2,.65/3} {
    \draw[fp threshold] (\x,0) -- (\x,1);
    \node[below,inner sep=3pt] at (\x,0) {$p_\i$};
  }
  \foreach \y/\j in {.35/1,.60/2,.80/3} {
    \draw[fp threshold] (0,\y) -- (1,\y);
    \node[left,inner sep=3pt] at (0,\y) {$q_\j$};
  }
  \draw[draw=fpInk,line width=.75pt] (0,0) rectangle (1,1);
  \node[below left,inner sep=3pt] at (0,0) {$0$};
  \node[below,inner sep=3pt] at (1,0) {$1$};
  \node[left,inner sep=3pt] at (0,1) {$1$};

  \foreach \branch in {\rhoSepOneA,\rhoSepOneB,\rhoSepTwoA,\rhoSepTwoB,
      \rhoSepThreeA,\rhoSepThreeB,\rhoSepFourA,\rhoSepFourB,
      \rhoSepFiveA,\rhoSepFiveB} {
    \draw[white,line width=2pt] \branch;
    \draw[rho branch] \branch;
  }
  \foreach \x/\y in {.20/.60,.40/.35,.40/.80,.65/.60,0/0,1/1}
    \node[rho attractor] at (\x,\y) {};
  \foreach \x/\y in {.20/.35,.20/.80,.40/.60,.65/.35,.65/.80}
    \rhoDiamond{\x}{\y}
  \node[rho attractor] at (.025,-.20) {};
  \node[fp key] at (.075,-.20) {Attracting $\rho$-NE};
  \rhoDiamond{.025}{-.29}
  \node[fp key] at (.075,-.29) {Saddles and their stable branches (saddle separatrices)};

  \begin{scope}[shift={(1.32,0)}]
    \input{fig/rho-fp-local}
  \end{scope}

  \begin{scope}[shift={(2.10,0)}]
    \fill[fpPurpleWash] (0,{.35/.60}) rectangle ({.20/.45},1);
    \fill[fpBlueWash] ({.20/.45},{.35/.60}) rectangle (1,1);
    \fill[fpGoldWash] (0,0) rectangle ({.20/.45},{.35/.60});
    \fill[fpGreenWash] ({.20/.45},0) rectangle (1,{.35/.60});
    \draw[fp threshold] ({.20/.45},0) -- ({.20/.45},1);
    \draw[fp threshold] (0,{.35/.60}) -- (1,{.35/.60});
    \draw[draw=fpInk,line width=.75pt] (0,0) rectangle (1,1);
    \foreach \x/\label in {0/p_1,{.20/.45}/p_2,1/p_3}
      \node[below,inner sep=3pt] at (\x,0) {$\label$};
    \foreach \y/\label in {0/0,{.35/.60}/q_1,1/q_2}
      \node[left,inner sep=3pt] at (0,\y) {$\label$};
    \begin{scope}
      \clip (0,0) rectangle (1,1);
      \begin{scope}[cm={2.222222,0,0,1.666667,(-.444444,0)}]
        \draw[rho continuous] \rhoContinuousPath;
        \draw[rho discrete] \rhoDiscretePath;
        \foreach \point in \rhoDiscretePoints
          \fill[fpStable] \point circle[radius=.65pt];
        \foreach \point in \rhoDiscreteLatePoints
          \fill[fpStable] \point circle[radius=.45pt];
        \node[circle,fill=fpInk,inner sep=1.4pt] at \rhoCommonStart {};
        \node[rho attractor] at \rhoAttractor {};
      \end{scope}
    \end{scope}
    \draw[rho continuous] (.01,-.20) -- (.13,-.20);
    \node[fp key] at (.17,-.20) {Continuous trajectory};
    \draw[rho discrete] (.01,-.29) -- (.13,-.29);
    \foreach \x in {.025,.065,.105}
      \fill[fpStable] (\x,-.29) circle[radius=.65pt];
    \node[fp key] at (.17,-.29) {Discrete iterates};
  \end{scope}
\end{tikzpicture}%
\endgroup%
    \caption{Geometry underlying Theorems~\ref{thm:ct_conv_2by2}, \ref{thm:2by2_lsc},  and~\ref{thm:dt_conv_2by2}. \textbf{(a)} Gray dashed lines  partition the strategy square into best-response cells. Teal circles mark stable equilibria; orange diamonds mark saddle vertices, whose stable branches are shown as orange dashed curves. \textbf{(b)} Top: a stable vertex with an inward spiral. Bottom: a saddle vertex with two incoming stable branches. \textbf{(c)} An enlarged neighborhood of $(p_2,q_1)$ compares a continuous-time trajectory (teal) and discrete-time iterates (orange dots) from the same initial belief (dark dot). A discrete step can overshoot a switching line, but each overshoot  is at most the current step size $\gamma_{k+1}\to0$.}
    \label{fig:rho_fp}
\end{figure}

\subsection{Global landscape and local geometry}
We now classify the possible limits and characterize their attraction basins.
Let $\phi_t(\z)$ denote the selected continuous-time trajectory
starting at $\z$ under the convention in Section~\ref{sec:grid_2by2}.
\begin{defi}
   An interior vertex $v=(p_i,q_j)$ is called a \textit{stable vertex} (equilibrium) if $x_j\ne y_i$ (Figure~\ref{fig:rho_fp_local}, top). Otherwise, it is called a \textit{saddle vertex} (equilibrium) (Figure~\ref{fig:rho_fp_local}, bottom).
\end{defi}

Let $\cX$ denote the strategy square. We introduce the following notation for interior vertices:
\[
\mathcal V_{\mathrm{st}}
:=\{(p_i,q_j)\in\mathcal V:x_j\ne y_i\},\qquad
\mathcal V_{\mathrm{sad}}
:=\{(p_i,q_j)\in\mathcal V:x_j=y_i\}.
\]
Let $\mathcal E_{\mathrm{st}}:=\mathcal C_\rho\cup\mathcal V_{\mathrm{st}}$ be the stable equilibria. For
$a\in\mathcal E_{\mathrm{st}}$ and $b\in\mathcal V_{\mathrm{sad}}$, we define the \textit{(attraction) basin} of $a$ and \textit{stable set} of saddle $b$ by
\[
B(a):=\{\z\in \cX:\lim_{t\to\infty}\phi_t(\z)=a\},
\qquad
W^s(b):=\{\z\in \cX:\lim_{t\to\infty}\phi_t(\z)=b\}.
\]
\begin{thm}[Landscape and local geometry]
   \label{thm:2by2_lsc}
Under Assumption \ref{asp:2by2}, every selected continuous-time $\rho$-FP trajectory converges as $t\to\infty$ to one of the following:
\begin{itemize}[leftmargin=*]
    \item  a corner  $c \in \mathcal{C}_\rho$
    with finitely many cell switches %
    \item  a saddle vertex $b\in\mathcal V_{\mathrm{sad}}$ with finitely many cell switches,
    \item  a stable interior vertex $a\in\V_{\mathrm{st}}$ with infinitely many cell switches unless initialized at that vertex.
\end{itemize}
A stable interior vertex $a$ is locally asymptotically stable and nearby trajectories spiral inward. The stable set $W^s(b)$ of a saddle vertex $b$ consists of two one-dimensional stable branches. %
The stable sets of saddle vertices form the boundaries of basins (Figure~\ref{fig:rho_fp_global}).
\end{thm}
The proof of Theorem \ref{thm:2by2_lsc} is deferred to Appendix \ref{app:pf_2by2_lsc}. Theorem~\ref{thm:2by2_lsc} shows how the classical geometry in $2\times2$ games
shown in Figure~\ref{fig:fp_geo} reappears locally when the advantage functions have multiple crossing points. The saddle geometry in Figure~\ref{fig:rho_fp_local}, bottom, corresponds to the coordination-type geometry in Figure~\ref{fig:fp_geo_coordination}, while the inward spiral in Figure~\ref{fig:rho_fp_local}, top, corresponds to the spiral-type geometry in Figure~\ref{fig:fp_geo_spiral}.
In the classical FP geometry illustrated in Figure~\ref{fig:fp_geo}, each advantage has one crossing, yielding four best-response regions and a unique interior NE.
Multiple crossings introduce additional switching lines and vertices, at which these local patterns recur. Saddle and spiral
geometries can therefore coexist within a single game,
producing the richer global landscape in
Figure~\ref{fig:rho_fp_global}.

Figure~\ref{fig:rho_fp_global} shows how these \textit{local} geometric patterns combine into the \textit{global} landscape. The teal points and the orange points represent the stable vertices and   saddle vertices, respectively. The saddle stable sets
form the basin boundaries. We call each such stable set a \textit{saddle separatrix}.
Each connected component remaining after removing the saddle separatrices contains one stable vertex and equals its attraction basin. %
Initial beliefs determine which equilibrium is reached, and convergence to a saddle vertex occurs for initial beliefs in its stable set. %
These saddle separatrices have measure zero, so almost every initial belief converges to a stable vertex. %
Theorem~\ref{thm:2by2_lsc} complements the convergence result of Theorem~\ref{thm:ct_conv_2by2} by characterizing the local geometry of $\rho$-NE, their attraction basins, and the saddle separatrices forming the basin boundaries.

\section{Convergence of Discrete-time NL-FP in \texorpdfstring{\(2\times 2\)}{2 x 2} Games}
\label{sec:dt}

In this section, we leverage the continuous-time convergence result
in Theorem~\ref{thm:ct_conv_2by2} and the global landscape
in Theorem~\ref{thm:2by2_lsc} to establish the convergence of
discrete-time NL-FP (Theorem~\ref{thm:dt_conv_2by2}). We recall that the recursion for the discrete-time NL-FP iterates
$\{\z_k\}_{k\ge0}$:
\[
\z_{k+1}
=
\z_k+\gamma_{k+1}\bigl(\vc_k-\z_k\bigr),
\qquad
\gamma_{k+1}=\frac{1}{k+2},
\qquad k=0,1,\ldots.
\]
Here $\z_0\in\cX$ is the initial belief, and
$\vc_k\in\BR^1_\rho(q_k)\times\BR^2_\rho(p_k)\subseteq\C$
is the selected pure best-response pair at iteration $k$.
At indifference, the choice may depend on $k$ and the preceding history.

In continuous time, the best-response pair changes when the trajectory
crosses a switching segment into an adjacent cell. In discrete time,
the pair $\vc_k$ is selected at $\z_k$ and held fixed for the
entire update to $\z_{k+1}$. %
The displacement from the crossing point to
$\z_{k+1}$ is called an \emph{overshoot} in classical FP
\cite{metrick1994fictitious}. In particular, if the line segment $[\z_k,\z_{k+1}]$ corresponding to the $k$-th iteration intersects a switching segment at $\widehat\xi_k \in [\z_k,\z_{k+1}]$, its overshoot is $\z_{k+1}-\widehat\xi_k$,
which satisfies
\[
\|\z_{k+1}-\widehat\xi_k\|_\infty
\le \|\z_{k+1}-\z_k\|_\infty
\le \gamma_{k+1}\longrightarrow0\qquad\text{as }k\to\infty.
\]
Figure~\ref{fig:rho_fp_overshoot} illustrates this overshoot phenomenon. However, vanishing overshoots alone do not guarantee convergence. The main challenge is that the discrete-time iterates can move across switching segments between cells and across the stable sets of saddle vertices that separate continuous-time attraction basins.
Our key geometric observation is that every continuous
trajectory converging neither to a corner nor to a saddle
eventually enters a neighborhood of a stable vertex.
Within this neighborhood, the continuous dynamics recover
the classical spiral geometry and return map
of \citet{metrick1994fictitious}.
We combine comparison along fixed finite regular itineraries
with local capture near stable equilibria and a global
analysis of saddle passages. These arguments show that
overshoots and possible basin changes cannot prevent
convergence, while allowing convergence to a saddle.

\begin{thm}[Discrete-time convergence in $2\times 2$ games]
\label{thm:dt_conv_2by2}
Suppose Assumption \ref{asp:2by2} holds. Under arbitrary pure-action tie-breaking rules at each iteration, every sequence of iterates $\{\z_k\}_{k\geq 0}$
converges to a $\rho$-NE,
\[ \lim_{k\to\infty} \z_k =\z^\star \in \mathcal E_\rho=(\cP\times \cQ)\cup \mathcal C_\rho.\]
\end{thm}

We provide a proof sketch of Theorem \ref{thm:dt_conv_2by2} below. The complete proof appears in Appendix~\ref{app:dt}.

\begin{proof}[Proof sketch of Theorem \ref{thm:dt_conv_2by2}]

Figure~\ref{fig:rho_fp_overshoot} illustrates how discrete updates differ
from the continuous trajectory. Starting from the same point,
both follow the same straight line within a cell, but a discrete
update may cross a switching line before changing direction.
The overshoot is at most $\gamma_{k+1}\to0$, so sufficiently late
discrete iterates follow any fixed finite portion of a continuous
trajectory that avoids vertices
(Lemma~\ref{lem:uni_comp}). Near a stable vertex,
we establish local contraction to ensure that the iterates remain nearby and converge (Lemma~\ref{lem:local_att}). Combining these results shows that a subsequential limit in a  basin shown in
Figure~\ref{fig:rho_fp_global} forces convergence to its equilibrium.

The remaining possibility is an approach along an incoming saddle
branch, as illustrated in Figure~\ref{fig:rho_fp_local}, bottom. The  iterates then either converge to the saddle or leave close to one
of its outgoing branches (Lemma~\ref{lem:passage}). Following that
branch leads into an attraction basin or toward another saddle.
These connections between saddles cannot form a cycle, since
a cycle would form a loop of continuous trajectory, excluded by the continuous-time proof
(Lemma~\ref{lem:acyclic}). Because there are finitely many saddles,
repeating this argument eventually gives convergence to a saddle
or entry into an attraction basin.

These arguments place every subsequential limit among the finitely
many corners and interior vertices. Boundedness and the vanishing
increments $\|\z_{k+1}-\z_k\|_\infty\le\gamma_{k+1}\to0$ imply
that the set of subsequential limits is nonempty and connected,
so it consists of a single point. A corner that is not a
$\rho$-NE cannot be the limit, since a strict best response would
drive one coordinate toward the opposite pure action. Hence the
iterates converge to a single $\rho$-NE.
\end{proof}

\section{When Does NL-FP Converge in \texorpdfstring{\(2\times n\)}{2 x n} Games?}
\label{sec:2byn}

Having established convergence of continuous- and discrete-time NL-FP
in $2\times2$ games, we now examine whether
these results extend to $2\times n$ games. For classical FP,
\citet{berger2005fictitious} established convergence to the NE set under the nondegeneracy condition that each player has a unique best response to every pure action of the opponent. We show  that  nonlinear evaluation can prevent convergence even in a
$2\times3$ game with a CVaR row player and a risk-neutral column player in Section~\ref{sec:non_conv_2byn}.
To identify conditions that restore convergence, we
develop a geometric analysis of classical FP using a planar projection in Section~\ref{sec:rev_fp},
and extend this approach to NL-FP under
finite-crossing and linear-projection conditions in Section~\ref{sec:conv_2byn}.

\paragraph{NL-FP in $2 \times n$ games.} The mixed-strategy profile is specified by $p\in[0,1]$ and $\q\in\Delta_n$ in $2 \times n$ games.
We encode the first and second row actions by $1$ and $0$, respectively, while column actions retain their indices in $[n]$.
The empirical beliefs evolve according to
\begin{equation}
\label{eqt:dt_2byn}
p_{k+1}=p_k+\gamma_{k+1}(x_k-p_k),\qquad
\q_{k+1}=\q_k+\gamma_{k+1}(\e_{J_k}-\q_k),
\qquad
\gamma_{k+1}=\frac{1}{k+2}.
\end{equation}
Here $k\ge0$, and the initial belief
$(p_0,\q_0)\in[0,1]\times\Delta_n$ is arbitrary. Let $\z_k:=(p_k,\q_k)$.
At each iteration, $x_k\in\BR^1_\rho(\q_k)$ and $J_k\in\BR^2_\rho(p_k)$ are
\textit{arbitrary} pure best responses at $\z_k$. In particular,
choices at indifference may depend on the iteration and the
preceding history.
We define the advantage function $d^1:\Delta_n\to \R$ for player 1 as $d^1(\q):=\rho^1(A_{1\cdot},\q)-\rho^1(A_{2\cdot},\q).$
Because player~1 has two actions, its unique pure best response is action $1$ when $d^1(\q)>0$ and action $0$ when $d^1(\q)<0$. Both actions are best responses when $d^1(\q)=0$. This is the sign-based rule in Equation~\eqref{eqt:br_2by2}, now applied to the vector belief $\q\in\Delta_n$.
The pure-action best-response correspondence for player 2 is
\begin{equation}
    \BR^2_\rho(p)=\arg\max_{j\in[n]} \rho^2(B_{j1},B_{j2};p).
\end{equation}
For pure-response notation, let
$x(\q)\in\BR^1_\rho(\q)$ and $y(p)\in\BR^2_\rho(p)$ denote the selected pure best responses. When a best-response set is not a singleton, a fixed deterministic tie-breaking rule selects one element. With the encoding above, $x(\q)\in\{0,1\}$ and $y(p)\in[n]$. The discrete choices in Equation~\eqref{eqt:dt_2byn} need not
be generated by these rules.
The continuous-time NL-FP inclusion, up to a rescaling of time, is
\begin{equation}
\label{eqt:cfp_2byn}
    \dot{p}_t\in\operatorname{BR}^1(\q_t)-p_t,\quad
    \dot{\q}_t\in\operatorname{BR}^2(p_t)-\q_t.
\end{equation}
Under the pure selections above, this takes the form
\[
    \dot{p}_t=x(\q_t)-p_t,\quad \dot{\q}_t=\mathbf{e}_{y(p_t)}-\q_t.
\]
The continuous-time convergence and landscape results below use
the mixed selections at projected equilibria specified in
Sections~\ref{sec:rev_fp} and~\ref{sec:grid_2byn}.

We introduce some useful notation for classical FP. Let $\z_t:=(p_t,\q_t)$. Let $a_j:=A_{1j}-A_{2j}$ for $j\in[n]$. For each $p\in[0,1]$, $J(p)=\operatorname{BR}^2_\rho(p)$ is the set of pure column best-response indices, and $\operatorname{BR}^2(p)=\Delta(J(p))$ is the set of column mixed strategies $\q\in\Delta_n$ satisfying $\operatorname{supp}(\q)\subseteq J(p)$.

\subsection{Nonconvergence when the row player is nonlinear}
\label{sec:non_conv_2byn}
We first show in Proposition \ref{prop:cvar_nonconvergence} that the convergence of continuous-time NL-FP in $2 \times n$ games can fail even when $n=3$. Proposition \ref{prop:cvar_nonconvergence} provides a counterexample in which the column player evaluates the payoffs by expectation, whereas the row player uses CVaR. Further details of the counterexample are provided in Appendix~\ref{app:2by3}.
\begin{prop}[Nonconvergence of (CVaR, mean)-FP in a $2\times3$ game]
\label{prop:cvar_nonconvergence}
There exists a $2\times3$ game where the row player uses
 CVaR with  $\alpha=1/5$ and the column
player uses expected payoff, such that continuous-time (CVaR, mean)-FP admits a nonconstant periodic trajectory. In particular, over one period of length $\log 4$ in rescaled time, the trajectory $(p_t,\q_t)$ traces eight line segments in $[0,1]\times\Delta_3\subset\mathbb R^4$.
\end{prop}
Hence, continuous-time NL-FP can cycle and need not converge
in $2\times3$ games, in contrast to our convergence result
for $2\times2$ games. This failure occurs even when the column
player is risk-neutral. While the column player keeps choosing action~1, the row player's belief $\q_t$ moves along a line segment toward $\e_1$. Along this segment, the row player's CVaR advantage changes sign three times, whereas an expected-payoff advantage is affine and can change sign at most once. The row player's best response therefore switches repeatedly, making $p_t$ move back and forth. As $p_t$ crosses the column player's crossing points, the column player changes its best response and redirects $\q_t$. These switches return $(p_t,\q_t)$ to its starting point.

The roles of the two players are therefore \textit{asymmetric}. As we will show in Section \ref{sec:conv_2byn}, when the two-action (row) player is risk-neutral, NL-FP converges in $2\times n$ games under the assumptions of Theorem~\ref{thm:conv_2byn}, even when the column
player uses a nonlinear evaluation functional.

\subsection{Revisiting the convergence of classical FP}
\label{sec:rev_fp}

We first revisit the convergence of classical FP, which provides the basis for our convergence analysis of NL-FP in $2\times n$ games.
\citet{berger2005fictitious} used the \textit{linear} projection $(p,\q)\mapsto(p,y)$, where $\mathbf a=(a_j)_{j\in[n]}$ is the (expected-payoff) advantage vector, and $y:=\mathbf a^\top\q=d^1(\q)$ is the row player's expected-payoff advantage, and proved that $\operatorname{dist}(\z_t,\mathcal E_{\mathrm{NE}})\to0$ as $t\to\infty$. The \textit{nondegeneracy} condition in that result requires each player to have a unique best response to every pure action of the opponent: $a_j\ne0$ for every $j\in[n]$, and unique column best responses at $p=0,1$. It allows more than two column actions to tie at an interior mixed strategy $0<p<1$ of row player.

Building on our analysis of $2\times2$ games, we use Berger's projection to give a new geometric proof of convergence to a \textit{single} NE, i.e., $\z_t\to\z^\star \in\mathcal E_{\mathrm{NE}}$ as $t\to\infty$, and to characterize the attraction basins  and saddle separatrices. These stronger results require a condition of \textit{simple crossing}, which is stated in Assumption~\ref{asp:fp_2byn}. %

We call a column action \textit{active} if it is a best
response to at least one mixed row strategy $p\in[0,1]$.
Suppose there are $K+1 \leq n$ active column actions. Without loss of
generality, we relabel them as $1,\ldots,K+1$.
 $\q\in\Delta_n$ still has one coordinate for
each of the $n$ column actions, since even inactive actions may
have positive probability in the initial belief.

\begin{asp}[Simple crossing]
\label{asp:fp_2byn}
There exist $0=x_0<x_1<\cdots<x_K<x_{K+1}=1$ such that
\[
\operatorname{BR}^2(p)=\{\e_j\},\qquad \forall p\in I_j:=(x_{j-1},x_j),\quad j\in[K+1].
\]
We call each $x_i$ ($i\in[K]$) a \textit{crossing point}.
At every crossing point $x_i$, it holds that $\operatorname{BR}^2(x_i)=\operatorname{co}\{\e_i,\e_{i+1}\}$. At the endpoints,
$\operatorname{BR}^2(x_0)=\{\e_1\}$ and
$\operatorname{BR}^2(x_{K+1})=\{\e_{K+1}\}$.
Moreover, $a_j\neq0$ for $j\in[K+1]$.
\end{asp}
This condition allows us to recover convergence of
the mixed-strategy profile $(p_t,\q_t)$ from that of the projected one $(p_t,y_t)$.
Suppose $(p_t,y_t)\to(x_i,0)$ as $t\to\infty$, with
$a_i a_{i+1}<0$. For all sufficiently large $t$, $\operatorname{BR}^2(x_i)=\operatorname{co}\{\e_i,\e_{i+1}\}$ implies that only column
actions $i$ and $i+1$ can be selected as best responses.
All other coordinates satisfy $\dot q_j=-q_j$ and tend to zero.
Consequently,
\[
q_i(t)+q_{i+1}(t)\to1,\qquad
a_iq_i(t)+a_{i+1}q_{i+1}(t)\to0
\quad\text{as }t\to\infty.
\]
Since $a_i\ne a_{i+1}$, these relations determine the
limits of $q_i(t)$ and $q_{i+1}(t)$, yielding the column mixed
strategy $\q^{i,*}$ in Proposition~\ref{prop:fp_2byn}.
At a projected pure-NE limit, the unique column best response
at endpoints similarly gives $\q_t\to\e_1$ or
$\q_t\to\e_{K+1}$, respectively. When the condition is not satisfied,
there might be three or more tied column actions, we cannot determine a unique column mixed strategy. Appendix~\ref{app:exm_fp_2byn}
gives an example for the unrestricted inclusion satisfying Berger's nondegeneracy condition
in which the projected trajectory is constant but
$(p_t,\q_t)$ is periodic when the simple-crossing condition is not imposed.

\paragraph{Selection convention.}
\label{par:selection_fp_2byn}
For the continuous-time results in this subsection, whenever
$(p_\tau,\mathbf a^\top\q_\tau)=(x_i,0)$ with $a_i a_{i+1}<0$
(including when $\tau=0$), we select the mixed responses $(x_i,\q^{i,*})$
defined in Proposition~\ref{prop:fp_2byn} for all $t\ge\tau$.
The projection stays fixed, while $\dot\q_t=\q^{i,*}-\q_t$
gives $\q_t\to\q^{i,*}$. Any NE reached remains fixed.
We call these selected continuous-time FP trajectories.

\begin{prop}[Convergence of continuous-time FP in $2\times n$ games]
\label{prop:fp_2byn}
Under Assumption \ref{asp:fp_2byn}, every
selected continuous-time FP trajectory $(p_t,\q_t)$ converges to a NE as $t\to\infty$. In particular, its limit is one of the following:
i) a pure NE $(0,\e_1)$ if $a_1<0$; (ii) a pure NE $(1,\e_{K+1})$ if $a_{K+1}>0$; (iii) a mixed NE $(x_i, \q^{i,*})$ if $a_i a_{i+1}<0$, where $\q^{i,*}=\frac{-a_{i+1}}{a_i-a_{i+1}}\, \e_i+\frac{a_i}{a_i-a_{i+1}}\, \e_{i+1}.$
\end{prop}
The proof of Proposition \ref{prop:fp_2byn} is deferred to Appendix \ref{app:pf_fp_2byn}.

\begin{figure}[!tb]
        \centering
        \begingroup
\input{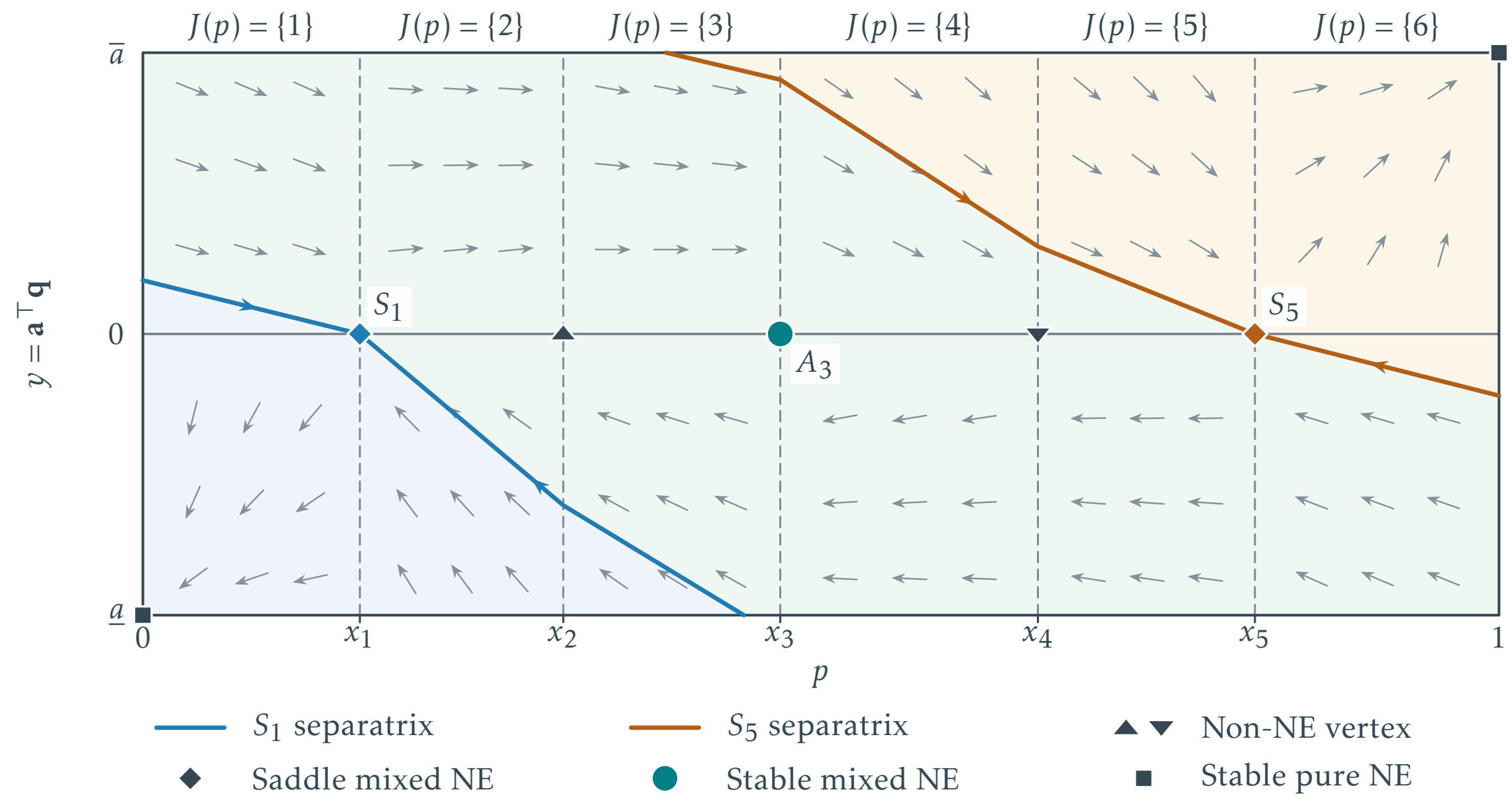}%
\definecolor{fpSixBlue}{HTML}{207DB3}%
\pgfmathsetlengthmacro{\fpSixXUnit}{0.82\textwidth}%
\pgfmathsetlengthmacro{\fpSixYUnit}{0.17\textwidth}%
\def\fpSixLeftPath{(0,{4/21}) -- (.16,0) -- (.31,{-39/64})
  -- ({3224/7275},-1)}%
\def\fpSixRightPath{({3566.2/9242.5},1) -- (.47,{5533/6120})
  -- (.66,{14/45}) -- (.82,0) -- (1,{-9/41})}%
\begin{tikzpicture}[
  x=\fpSixXUnit,y=\fpSixYUnit,font=\footnotesize,text=fpInk,
  line cap=round,line join=round,
  six flow/.style={draw=fpFlow,line width=.45pt,
    -{Stealth[length=1.35mm,width=1mm]}},
  six separator/.style={line width=1.25pt},
  six toward saddle/.style={six separator,postaction={decorate},
    decoration={markings,
      mark=at position .53 with {\arrow{Stealth[length=1.8mm,width=1.3mm]}}}},
  six saddle/.style={rectangle,rotate=45,draw=white,line width=.65pt,
    inner sep=2.5pt},
  six pure/.style={rectangle,fill=fpInk,draw=white,line width=.65pt,
    inner sep=2.5pt},
  six mixed/.style={circle,fill=fpTeal,draw=white,line width=.65pt,
    inner sep=2.7pt},
  six label/.style={fill=white,fill opacity=.9,text opacity=1,inner sep=1.5pt}
]
  \fill[fpGreenWash] (0,-1) rectangle (1,1);
  \path[fill=fpBlueWash] \fpSixLeftPath -- (0,-1) -- cycle;
  \path[fill=fpGoldWash] \fpSixRightPath -- (1,1) -- cycle;

  \foreach \j/\left/\right/\height in {
    1/0/.16/-1, 2/.16/.31/.65,
    3/.31/.47/.30, 4/.47/.66/-.75,
    5/.66/.82/-.35, 6/.82/1/1} {
    \node at ({(\left+\right)/2},1.09) {$J(p)=\{\j\}$};
    \foreach \u in {.23,.50,.77} {
      \foreach \sy in {-.87,-.60,-.30,.30,.60,.87} {
        \pgfmathsetmacro{\sx}{\left+\u*(\right-\left)}
        \pgfmathsetmacro{\dx}{ifthenelse(\sy>0,1,0)-\sx}
        \pgfmathsetmacro{\dy}{\height-\sy}
        \pgfmathsetmacro{\scale}{.013/sqrt(\dx*\dx+(.17/.82*\dy)^2)}
        \draw[six flow]
          ({\sx-\scale*\dx},{\sy-\scale*\dy}) --
          ({\sx+\scale*\dx},{\sy+\scale*\dy});
      }
    }
  }

  \foreach \i/\switch in {1/.16,2/.31,3/.47,4/.66,5/.82} {
    \draw[fp threshold] (\switch,-1) -- (\switch,1);
    \draw[fpInk,line width=.6pt] (\switch,-1) -- (\switch,-1.025);
    \node[below,inner sep=3pt] at (\switch,-1) {$x_{\i}$};
  }
  \draw[fpInk!70,line width=.65pt] (0,0) -- (1,0);
  \draw[fpInk,line width=.75pt] (0,-1) rectangle (1,1);

  \draw[six toward saddle,fpSixBlue] (0,{4/21}) -- (.16,0);
  \draw[six toward saddle,fpSixBlue]
    ({3224/7275},-1) -- (.31,{-39/64}) -- (.16,0);
  \draw[six toward saddle,fpStable]
    ({3566.2/9242.5},1) -- (.47,{5533/6120}) -- (.66,{14/45}) -- (.82,0);
  \draw[six toward saddle,fpStable] (1,{-9/41}) -- (.82,0);

  \node[six saddle,fill=fpSixBlue] at (.16,0) {};
  \node[six saddle,fill=fpStable] at (.82,0) {};
  \node[six mixed] at (.47,0) {};
  \node[six pure] at (0,-1) {};
  \node[six pure] at (1,1) {};
  \path[fill=fpInk,draw=white,line width=.6pt]
    (.31,.036) -- (.300,-.023) -- (.320,-.023) -- cycle;
  \path[fill=fpInk,draw=white,line width=.6pt]
    (.66,-.036) -- (.650,.023) -- (.670,.023) -- cycle;
  \node[six label,anchor=south west] at (.165,.025) {$S_1$};
  \node[six label,anchor=north west] at (.477,-.035) {$A_3$};
  \node[six label,anchor=south west] at (.825,.025) {$S_5$};

  \node[left,inner sep=5pt] at (0,-1) {$\underline a$};
  \node[left,inner sep=5pt] at (0,0) {$0$};
  \node[left,inner sep=5pt] at (0,1) {$\overline a$};
  \node[rotate=90] at (-.080,0) {$y=\mathbf a^\top\q$};
  \node[below,inner sep=3pt] at (0,-1) {$0$};
  \node[below,inner sep=3pt] at (1,-1) {$1$};
  \node at (.5,-1.22) {$p$};

  \draw[six separator,fpSixBlue] (.01,-1.40) -- (.06,-1.40);
  \node[fp key] at (.08,-1.40) {$S_1$ separatrix};
  \draw[six separator,fpStable] (.36,-1.40) -- (.41,-1.40);
  \node[fp key] at (.43,-1.40) {$S_5$ separatrix};
  \path[fill=fpInk]
    (.725,-1.371) -- (.716,-1.423) -- (.734,-1.423) -- cycle;
  \path[fill=fpInk]
    (.751,-1.429) -- (.742,-1.377) -- (.760,-1.377) -- cycle;
  \node[fp key] at (.78,-1.40) {Non-NE vertex};
  \node[six saddle,fill=fpInk] at (.035,-1.58) {};
  \node[fp key] at (.08,-1.58) {Saddle mixed NE};
  \node[six mixed] at (.385,-1.58) {};
  \node[fp key] at (.43,-1.58) {Stable mixed NE};
  \node[six pure] at (.738,-1.58) {};
  \node[fp key] at (.78,-1.58) {Stable pure NE};
\end{tikzpicture}%
\endgroup%
        \caption{Projected landscape of classical FP in a $2\times6$ game, with vertical coordinate $y=\mathbf a^\top\q\in[\underline a,\overline a]=[\min_j a_j,\max_j a_j]$. Each crossing point $x_i$ defines the vertical switching lines, and $y=0$ defines the row player's indifference line. The stable sets of the saddle vertices $S_1$ and $S_5$ are the blue and orange separatrices. They separate three attraction basins, corresponding to two pure NE and the mixed NE whose projection is $A_3$. The triangles mark non-NE vertices.}
        \label{fig:2byn_lsc}
\end{figure}

To identify which initial beliefs lead to each NE, we examine
the global landscape and local geometry of classical FP in the projected space.

\paragraph{Global landscape and local geometry.}
For the selected projected dynamics $(p_t,y_t)\in \R^2$,
the lines $p=x_i$ divide the plane into $K+1$ vertical strips (Figure \ref{fig:2byn_lsc}),
and the indifference line for row player $y=0$  divides each strip into an upper and a lower cell. These cells form the
$(K+1)\times2$ grid. Only their intersections with
$[0,1]\times[\min_j a_j,\max_j a_j]$ correspond to strategy profiles.
We call $v_i:=(x_i,0)$ a \textit{(crossing) vertex}.
Under Assumption~\ref{asp:fp_2byn}, a crossing vertex with
$a_i a_{i+1}<0$ is the projection of $(x_i,\q^{i,*})$,
the unique mixed NE with row strategy $x_i$.

The local geometry of each crossing vertex $v_i$ has three types:
\begin{enumerate}[label=(\roman*),leftmargin=*]
\item a stable projected equilibrium if $a_i>0>a_{i+1}$,
and nearby trajectories spiral inward.
\item a saddle projected equilibrium if $a_i<0<a_{i+1}$,
which has two stable branches.
\item  not an NE if $a_i a_{i+1}>0$. Near this vertex, $\dot y$ has the common strict sign of
$a_i$ and $a_{i+1}$, so trajectories pass upward or downward
rather than converging to it.
\end{enumerate}
Figure~\ref{fig:2byn_lsc} illustrates these cases:
$S_1$ and $S_5$ are saddle vertices, $A_3$ is a stable vertex,
and the triangles at $v_2$ and $v_4$ mark vertices that do not
correspond to NE.

Under Assumption~\ref{asp:fp_2byn}, the pure-NE projections
are $(0,a_1)$ when $a_1<0$ and $(1,a_{K+1})$ when
$a_{K+1}>0$. These points, when present, and the stable
mixed-NE projections have attraction basins. The stable sets of saddle
vertices form piecewise-linear separatrices between these basins.
Proposition~\ref{prop:lsc_2byn} in
Appendix~\ref{app:cls_fp_lsc_2byn} establishes this structure.
The attraction basins in the original space $[0,1]\times\Delta_n$ are
the inverse images of the projected basins under the mapping
$(p,\q)\mapsto(p,\mathbf a^\top\q)$.

\subsection{Finite-grid geometry}
\label{sec:grid_2byn}
We now extend the planar analysis of classical FP to NL-FP.
Our assumptions reflect the asymmetric roles of the two players
in $2\times n$ games. Both players may use nonlinear evaluation
functionals, subject to Assumption~\ref{asp:mean_rho}
for the $n$-action column player and
Assumption~\ref{asp:fin_sw} for the two-action row player.

\paragraph{Assumption on column player.}
Assumption~\ref{asp:mean_rho} extends the finite-crossing structure  in Assumption~\ref{asp:2by2} from two column actions to $n$.

\begin{asp}[Finite crossings]
\label{asp:mean_rho}
There exist points $0=r_0<r_1<\cdots<r_K<r_{K+1}=1$
and column actions $j_1,\ldots,j_{K+1}\in[n]$ such that
$J(p)=\{j_k\}$ for all $p\in I_k:=(r_{k-1},r_k)$ and
$k\in[K+1]$. At the endpoints, $J(0)=\{j_1\}$ and
$J(1)=\{j_{K+1}\}$.
The points $r_1,\ldots,r_K$ are called \textit{crossing points}.
Each crossing point is \textit{simple}, meaning that
$j_i\ne j_{i+1}$ and $J(r_i)=\{j_i,j_{i+1}\}$ for every $i\in[K]$.
\end{asp}

Assumption~\ref{asp:mean_rho} imposes a condition of finite crossings. The row player's mixed-strategy space $[0,1]$ is partitioned by $K$ crossing points, the column player has a unique constant pure best response between successive crossings.  In the classical FP case, the endpoint requirements in Assumption~\ref{asp:mean_rho} reduces to the nondegeneracy condition
of \citet{berger2005fictitious} on the column player. The additional simple crossing requirement rules out intervals with multiple column best responses and ties among three or more actions at a crossing point.

Note that the indices $j_1,\ldots,j_{K+1}$ need not be distinct.
Under nonlinear evaluation, the same column action may be
uniquely optimal on several disjoint intervals.

\paragraph{Assumption on row player.}
As with the column player, we impose a finite-crossing
condition on the row player (Assumption \ref{asp:fin_sw}).
The column player's best response depends on the scalar
$p\in[0,1]$, whereas the row player faces a vector $\q\in\Delta_n$.
To obtain planar dynamics, we additionally require the row
player's best response to depend on $\q$ only through a
 linear projection $s:=\alpha^\top\q$.

\begin{asp}[Finite crossings through a linear projection]
\label{asp:fin_sw}
There exist $\alpha\in\mathbb{R}^n$, an integer $L\ge0$,
points satisfying $\min_j\alpha_j=c_0<c_1<\cdots<c_L<c_{L+1}=\max_j\alpha_j$
and a correspondence
$\chi:I\to\{\{0\},\{1\},[0,1]\}$ on $I:=[c_0,c_{L+1}]$
such that the following holds. For every $\q\in\Delta_n$,
$\operatorname{BR}^1(\q)=\chi(\alpha^\top\q)$.
The correspondence $\chi$ has the constant singleton value
$\chi(s)=\{\chi_\ell\}$ on each $(c_\ell,c_{\ell+1})$,
where $\chi_\ell\in\{0,1\}$ and $\ell=0,\ldots,L$.
At the endpoints, $\chi(c_0)=\{\chi_0\}$ and
$\chi(c_{L+1})=\{\chi_L\}$.
At each \textit{crossing level} $c_\ell$, $\ell\in[L]$, we require
$\chi(c_\ell)=[0,1]$ and $\chi_{\ell-1}\ne\chi_\ell$. In addition, for the column actions $j_k$ in Assumption~\ref{asp:mean_rho},
$\alpha_{j_k}\ne c_\ell$ for every $k\in[K+1]$ and $\ell\in[L]$.
\end{asp}

Here $\chi(s)$ is the mixed best responses set for row player,
represented by the probability of choosing action~1.
This assumption also requires the value $\alpha_{j_k}$
 to differ from every crossing level. Note that $\alpha$ need not equal the advantage vector $\mathbf a$.

When $L=0$, there are no crossing levels and
$\chi(s)=\{\chi_0\}$ for all $s\in I$, with $\chi_0\in\{0,1\}$.
We use the convention $[0]=\varnothing$, so the conditions
indexed by $\ell\in[L]$ impose no requirements in this case.

\paragraph{Examples of finite crossings.}
The condition on the row player covers the ERM \citep{howard1972risk,follmer2004stochastic}. For a row player using ERM with parameter $\beta>0$, we set
$\alpha_j=(e^{-\beta A_{2j}}-e^{-\beta A_{1j}})/\beta$.
The row player's best response is determined by the sign
of $s=\alpha^\top\q$. If $A_{1j}\ne A_{2j}$ for every $j$,
this correspondence has either a single crossing level
$c_1=0$ or a constant and pure best response, and thus satisfies
Assumption~\ref{asp:fin_sw}.
Moreover, $\alpha_j\to a_j$ as $\beta\to0$, so the scalar
$s=\alpha^\top\q$ approaches the expected-payoff advantage
$\mathbf a^\top\q$. On the other hand, the column player may use more general nonlinear functional. For instance,  mean-variance and  CVaR satisfy Assumption~\ref{asp:mean_rho}. %

\paragraph{Projected grid.} Together, Assumptions~\ref{asp:mean_rho} and~\ref{asp:fin_sw} yield a planar system in $(p,s)$
with a $(K+1)\times(L+1)$ grid of best-response regions. Within a cell where the row player's pure best response is $x\in\{0,1\}$
and the column player's is action $j$, the projected dynamics
satisfy $\dot p=x-p$ and $\dot s=\alpha_j-s$.
Thus, the trajectory traces a straight line segment toward
$(x,\alpha_j)$ whenever it remains in that cell.
Figure~\ref{fig:2byn_lsc} illustrates these directions and
crossings in the classical case, where $s=\mathbf a^\top\q$
and the row indifference line $s=0$ gives a
$(K+1)\times2$ grid. For NL-FP in $2\times n$ games,
the $L$ horizontal switching lines $s=c_\ell$ extend this structure to a $(K+1)\times(L+1)$ grid.
This structure allows us to build on the geometric analysis
of classical FP in Section~\ref{sec:rev_fp} and extend the
convergence arguments developed for $2\times2$ games to
the $2\times n$ setting, as shown in Theorem \ref{thm:conv_2byn}.

\paragraph{Selection convention.}
\label{par:selection_2byn}
We use the definitions of open switching segments and regular
switches from Section~\ref{sec:grid_2by2} for the projected grid.
On a vertical open switching segment $p=r_i$, the adjacent cells
have the same nonzero horizontal velocity $x-r_i$.
On a horizontal open switching segment $s=c_\ell$ within $I_k$,
they have the same nonzero vertical velocity $\alpha_{j_k}-c_\ell$.
Thus, a trajectory reaching either segment crosses into the
adjacent cell.

As in Section~\ref{sec:grid_2by2}, we require
$\z_t=\z_\tau$ for all $t\ge\tau$ whenever
$\z_\tau=(p_\tau,\q_\tau)$ is a $\rho$-NE, including when $\tau=0$.
At a projected vertex $v=(r_i,c_\ell)$, however, indifference
alone does not imply that the full strategy profile is a $\rho$-NE.
The vertex is the projection of a mixed $\rho$-NE precisely when
$(\alpha_{j_i}-c_\ell)(\alpha_{j_{i+1}}-c_\ell)<0$.
If the projected trajectory reaches such a vertex at time $\tau$,
including when $\tau=0$, we choose the mixed row response $r_i$
and the mixed column response $\q^{i,\ell}$ given in
Theorem~\ref{thm:conv_2byn} for all $t\ge\tau$.
These responses give $\dot p_t=\dot s_t=0$ and
$\dot\q_t=\q^{i,\ell}-\q_t$.
Thus, the projection remains at $v$, while the full strategy
profile converges to $(r_i,\q^{i,\ell})$; if it is already that
$\rho$-NE, it remains fixed.
At every other interior vertex, the two adjacent values
$\alpha_{j_i}-c_\ell$ and $\alpha_{j_{i+1}}-c_\ell$ have the same
strict sign, so the projected trajectory passes through.
We call trajectories satisfying this convention the selected
continuous-time NL-FP trajectories. For $\alpha=\mathbf a$ and
crossing level $0$, this recovers the classical selection in
Section~\ref{sec:rev_fp}.

\subsection{Convergence of both continuous- and discrete-time NL-FP}
\label{sec:conv_2byn}

\begin{thm}[Convergence of continuous-time $\rho$-FP in $2\times n$ games]
\label{thm:conv_2byn}
    Assume that the column player satisfies Assumption \ref{asp:mean_rho} and the row player satisfies Assumption \ref{asp:fin_sw}. Then every selected continuous-time $\rho$-FP trajectory in every $2\times n$ game converges to a $\rho$-NE as $t\to\infty$. In particular, its limit is one of the following:
i) a pure $\rho$-NE $(0,\e_{j_1})$ if $0\in\chi(\alpha_{j_1})$; (ii) a pure $\rho$-NE $(1,\e_{j_{K+1}})$ if $1\in\chi(\alpha_{j_{K+1}})$; (iii) a mixed $\rho$-NE $(r_i,\q^{i,\ell})$ if $(\alpha_{j_i}-c_\ell)(\alpha_{j_{i+1}}-c_\ell)<0$, where %
\[ \q^{i,\ell}
=
\frac{\alpha_{j_{i+1}}-c_\ell}
{\alpha_{j_{i+1}}-\alpha_{j_i}}\,\e_{j_i}
+
\frac{c_\ell-\alpha_{j_i}}
{\alpha_{j_{i+1}}-\alpha_{j_i}}\,\e_{j_{i+1}}. \]
\end{thm}

The proof of Theorem~\ref{thm:conv_2byn} is deferred to
Appendix~\ref{app:pf_ct_2byn}.
The asymmetric assumptions reflect the different dimensions of
the opponents' beliefs. The column player responds to the scalar
$p$, so its finite-crossing condition gives vertical switching
lines. The row player faces the vector $\q$, thus assuming its best
response to depend only on $s=\alpha^\top\q$ allows us to
describe the projected dynamics using $(p,s)$ alone:
for $p\in I_k$, $\dot s=\alpha_{j_k}-s$.
The finite crossing levels supply horizontal switching
lines. Together, these conditions yield the finite planar grid
underlying the convergence analysis and guarantee that
every selected continuous-time NL-FP trajectory converges to a
$\rho$-NE.

When the row player is risk-neutral, the relevant scalar reduces to
$s=\mathbf a^\top\q$, and the best response is determined by its sign. %
Its only possible indifference level is $s=0$:
there is one crossing level ($L=1$) when
$\min_j a_j<0<\max_j a_j$, and no crossings ($L=0$)
when all $a_j$ have the same strict sign.
We refer to the resulting dynamics as $(\mu,\rho)$-FP,
where $\mu(X)=\mathbb{E}[X]$ is the row player's evaluation
and $\rho$ is the column player's evaluation.

\paragraph{Convergence of continuous-time $(\mu,\rho)$-FP.}
For a risk-neutral row player, we relax Assumption~\ref{asp:fin_sw} to the nondegeneracy condition in Assumption~\ref{asp:mean_rho_2}.
It is the specialization of the nondegeneracy requirement in Assumption~\ref{asp:fin_sw}
to $\alpha=\mathbf a$ and row indifference level $0$.
This condition requires $a_j\ne0$ for active column actions only, while the nondegeneracy condition of \citet{berger2005fictitious} applies to every column action.

\begin{asp}[Nondegeneracy]
\label{asp:mean_rho_2}
For every $k\in[K+1]$, $a_{j_k}=A_{1j_k}-A_{2j_k}\ne0$ for the column actions $j_k$ in Assumption~\ref{asp:mean_rho}.
\end{asp}

\begin{coro}
    \label{coro:conv_mean_rho}
Under Assumptions~\ref{asp:mean_rho} and~\ref{asp:mean_rho_2}, every selected continuous-time $(\mu,\rho)$-FP trajectory in every $2\times n$ game converges to a $\rho$-NE as $t\to\infty$. In particular, its limit is one of the following: i) a pure $\rho$-NE $(0,\e_{j_1})$ if $a_{j_1}<0$; (ii) a pure $\rho$-NE $(1,\e_{j_{K+1}})$ if $a_{j_{K+1}}>0$; (iii) a mixed $\rho$-NE $(r_i,\q^{i,*})$ if $a_{j_i} a_{j_{i+1}}<0$, where $\q^{i,*}=\frac{-a_{j_{i+1}}}{a_{j_i}-a_{j_{i+1}}}\,\e_{j_i}+\frac{a_{j_i}}{a_{j_i}-a_{j_{i+1}}}\,\e_{j_{i+1}}.$
\end{coro}

\paragraph{Convergence of discrete-time NL-FP in $2\times n$ games.}
Returning to nonlinear evaluation by both players, we now establish the convergence of discrete-time NL-FP  under Assumptions~\ref{asp:mean_rho} and~\ref{asp:fin_sw}.

\begin{thm}[Convergence of discrete-time $\rho$-FP in $2\times n$ games]
\label{thm:conv_dt_2byn}
    Assume that the column player satisfies Assumption \ref{asp:mean_rho} and the row player satisfies Assumption \ref{asp:fin_sw}. Under arbitrary pure-action tie-breaking rules at each iteration,
    every sequence of iterates $\{\z_k\}_{k\geq 0}$ generated by Equation~\eqref{eqt:dt_2byn}
converges to a $\rho$-NE,
\[ \lim_{k\to\infty} \z_k =\z^\star \in \mathcal E_\rho.\]
\end{thm}
The proof of Theorem~\ref{thm:conv_dt_2byn} is deferred to
Appendix~\ref{app:pf_dt_2byn}.  The proof extends the comparison with continuous-time trajectories
used in the $2\times2$ case. The local arguments near stable
and saddle vertices can be reused.
The main difficulty is that the projected grid can
contain non-NE vertices that trajectories pass through.
Near such a vertex, the discrete iterates and a continuous
trajectory may cross the two switching lines in different
orders. We show that the additional error caused by these
different orders tends to zero with the step size.
This allows the comparison to continue through non-NE
vertices and the arguments based on continuous-time
attraction basins and saddle vertices to apply.

\section{Concluding Remarks}
\label{sec:con}

We study nonlinear fictitious play, in which players evaluate each pure action through a general nonlinear functional of the payoff distribution induced by the opponent's empirical play. Our analysis identifies the \emph{finite crossing property} {of advantage functions} as a key to convergence. In $2\times2$ games, the single-crossing property preserves the classical FP geometry \citep{metrick1994fictitious}. Beyond SCP, the finite-crossing assumption yields the convergence of both the selected continuous-time dynamics and the discrete-time iterates to a $\rho$-NE. %
We establish a novel geometric analysis that characterizes the attraction basins of $\rho$-NE and the saddle separatrices that form  their basins.

In $2\times n$ games, however, nonlinear evaluation can prevent the convergence of NL-FP.  We show that a simple $2\times3$ game instance with a CVaR row player and a risk-neutral column player admits a nonconstant periodic orbit in continuous time. Nevertheless, we establish convergence of the selected continuous-time dynamics and the discrete-time iterates under the stated finite-crossing assumptions, when the two-action player's best response depends on the opponent's strategy only through a linear projection. %
We build on our geometric analysis to describe the global landscape and local geometry of classical FP in $2\times n$ games under the corresponding assumptions.

We believe these results motivate the extension of the analysis beyond the finite crossing condition,  and the identification of convergence conditions that do not require a linear projection, beyond two-action games. Another promising direction is to study nonlinear FP with \textit{finite memory}, where players form beliefs using only a fixed number of the opponent's most recent actions. It would also be valuable to quantify the convergence rates of nonlinear FP to a $\rho$-NE.

\section*{Acknowledgment}
The authors acknowledge the support from the Army
Research Office (ARO) grant W911NF-24-1-0085, the
NSF CAREER Award \#2443704, the AFOSR YIP Award FA9550-25-1-0258, an AI Safety Research Award from
Coefficient Giving, a Cisco Faculty Research Award, a JP Morgan Faculty Research Award, and an Amazon Research Award.

\paragraph{AI-use Disclosure.}
The proofs of Proposition \ref{prop:cvar_nonconvergence}, Theorem  \ref{thm:conv_dt_2byn}, and Theorem  \ref{thm:lsc_2byn}  were AI-assisted. For Theorem \ref{thm:conv_dt_2byn},  GPT-6.0 Astra helped refine the convergence argument and formalize intermediate steps, and also helped construct the counterexample in Proposition \ref{prop:cvar_nonconvergence}. For Theorem  \ref{thm:lsc_2byn}, we established the convergence and the geometric structure of NL-FP in $2\times2$ games and identified sufficient conditions for convergence in $2\times n$ games, formulated in Assumptions \ref{asp:mean_rho}-\ref{asp:fin_sw}; with these results as input, GPT-6.0 Astra extended our finite-grid analysis for $2\times2$ games to the two-dimensional projected dynamics.
We additionally used AI assistance to convert figures into TikZ code to improve their visual appearance and polish the manuscript for readability. The authors carefully checked and substantially rewrote all AI-assisted results, and filled in nontrivial logical gaps. The authors take full responsibility for all the content in the manuscript.

\bibliographystyle{IEEEtranN}
\bibliography{ref}

\clearpage

\appendix

\section{Related Work}
\label{app:lit}
\paragraph{Fictitious play.}

Fictitious play was introduced by \citet{brown1951iterative} and first analyzed in zero-sum games by \citet{robinson1951iterative}. Although FP may not converge in general games \citep{shapley1963some}, strong convergence results are available for certain games. \citet{miyasawa2010convergence} established convergence in nondegenerate $2\times2$ games under a particular tie-breaking rule, and \citet{metrick1994fictitious} subsequently gave a geometric proof based on the structure of the best-response correspondence in both continuous and discrete time. \citet{berger2005fictitious} later extended the geometric approach to nondegenerate $2\times n$ games. More recently, such convergence of FP was further extended to stochastic games \cite{sayin2022fictitious,ozdaglar2023independent,park2023multi,sayin2022fictitiousb,unlu2026fictitious}.
These results exclusively rely on \emph{expected-utility best responses}, unlike our setting, in which each pure action is evaluated by applying a \emph{nonlinear functional} to the payoff distribution induced by the opponent's mixed strategy. The difference between the evaluations of two actions defines the advantage function, which may have one or several sign-changing thresholds. We recover the classical $2\times 2$ geometry when there is a single crossing, and develop a different geometric analysis when multiple crossings are present.

\paragraph{Equilibrium under nonlinear payoff evaluation.}

When preferences over payoff distributions cannot be represented by expected utility, the definition and existence of mixed-strategy equilibrium depend critically on how a player evaluates randomization. Early work showed that nonlinear preferences can invalidate standard equilibrium-existence and dynamic-consistency arguments \cite{crawford1990equilibrium,dekel1991existence,ritzberger1996games}.
\citet{fiat2010players} studied games in which players are not expectation maximizers and demonstrated that risk-adjusted games need not possess mixed equilibria when the nonlinear criterion is applied to all sources of randomness.

A separate body of literature studied strategic interaction under rank-dependent utility and CPT. \citet{keskin2016equilibrium} introduced Nash and correlated equilibrium notions for CPT players and studied their existence. \citet{phade2019geometry} showed that the geometry of CPT Nash and correlated equilibria can differ substantially from its expected-utility counterpart.

\paragraph{Learning under nonlinear payoff evaluation.}
Recent work studied learning in games under specific nonlinear payoff evaluations, but with different learning dynamics and equilibrium targets. For convex risk measures under the minimization convention,
\citet{slumbers2023game} defined a mean-variance risk-averse equilibrium (RAE) and proposed stochastic fictitious play as a computational method, but established only that any convergent sequence of observed strategies has an RAE as its limit; the convergence of the learning process itself is not guaranteed. \citet{phade2023learning} studied calibrated learning in repeated games with cumulative-prospect-theory preferences, and proved the convergence of empirical play to mediated CPT correlated equilibria, rather than Nash equilibrium.
\citet{mazumdar2025tractable} established the convergence of no-regret algorithms to risk-averse quantal-response coarse correlated equilibria in matrix and Markov games; \citet{zhang2026provably} extended this framework to actor-critic learning. In contrast, we analyze exact fictitious play and its convergence to an actionwise $\rho$-\emph{Nash} equilibrium. \citet{mohammadi2026entropic} studied entropic risk-sensitive learning and its equilibrium selection in coordination games.
In summary, existing results typically pair a particular nonlinear criterion with a specialized learning procedure and equilibrium concept. By contrast, we allow a broad class of actionwise nonlinear functionals,  and focus on the natural dynamics of fictitious play, as well as convergence to the induced Nash equilibrium.
\paragraph{Risk functionals and single crossing.}

The nonlinear evaluations covered by our analysis draw on several established models of decision-making under risk,  including coherent and convex/concave risk measures \citep{artzner1999coherent,follmer2002convex}, optimized certainty equivalents  \citep{ben2007old}, conditional value-at-risk \citep{rockafellar2000optimization}, mean-semivariance and other downside-risk criteria \citep{estrada2007mean}, spectral risk measures \citep{acerbi2002spectral}, rank-dependent expected utility \citep{quiggin1982theory}, and CPT \citep{tversky1992advances}. Rather than revisiting their axiomatic foundations, we study how these functionals alter the geometry of best-response and fictitious-play dynamics in a \(2\times2\) game.

\section{Proofs for Section \ref{sec:pre}}
\label{app:form_pf}

\begin{proof}[Proof of Proposition \ref{prop:ne_2by2}]
By Definition \ref{def:rho_ne}, every pure action used with positive
probability must be a best response. For player~1, this is
equivalent to
\[
p>0\implies d^1(q)\ge0,
\qquad
1-p>0\implies d^1(q)\le0.
\]
The analogous conditions for player~2 are
\[
q>0\implies d^2(p)\ge0,
\qquad
1-q>0\implies d^2(p)\le0.
\]
These four implications characterize equilibrium for every $(p,q)\in[0,1]^2$.
In the interior, both players use both actions, so equilibrium
is equivalent to $d^1(q)=d^2(p)=0$, giving $\cP\times \cQ$.
At a corner, the conditions reduce to those defining $\C_\rho$.
On an open edge, the mixing player must be indifferent, while
the other player's chosen pure action must be optimal; these
are exactly the conditions defining $\cE_\rho^{\mathrm{edge}}$.
This proves the decomposition.
Finally, nonzero endpoint advantages rule out indifference
for the mixing player on every open edge.
\end{proof}

\begin{proof}[Proof of Proposition \ref{pro:rho-ne-existence}]
If a pure $\rho$-NE exists, i.e., $\C_\rho\neq\varnothing$, the conclusion is immediate.
Suppose therefore that none of the four pure profiles
is a $\rho$-NE. We show that the following holds
\[
d^1(0)d^1(1)<0,
\qquad
d^2(0)d^2(1)<0.
\]

First, suppose $d^1(0)\ge0$.
Since $(1,0)$ is not an equilibrium, we must have $d^2(1)>0$.
Since $(1,1)$ is not an equilibrium, it follows that $d^1(1)<0$.
Similarly, excluding $(0,1)$ gives $d^2(0)<0$, and excluding
$(0,0)$ then gives $d^1(0)>0$.
Thus, both endpoint products are strictly negative. If instead $d^1(0)<0$, successively excluding
$(0,0)$, $(0,1)$, and $(1,1)$ gives $d^2(0)>0, d^1(1)>0, d^2(1)<0.$
Again, both endpoint products are strictly negative.

By the assumed condition, there exist $q^\star,p^\star\in(0,1)$
such that $d^1(q^\star)=0$ and $d^2(p^\star)=0.$ At $(p^\star,q^\star)$, each player is indifferent between
its two pure actions. Hence, every action in the support of
each player's strategy is a best response, so
$(p^\star,q^\star)$ is a $\rho$-NE by Definition \ref{def:rho_ne}.

Finally, continuity of each $d^i$ implies the condition
by the intermediate value theorem.
\end{proof}

\begin{proof}[Proof of Proposition \ref{prop:ne_2byn}]
Let $F(p,\mathbf{q})
:=
\operatorname{BR}^1(\mathbf{q})
\times
\operatorname{BR}^2(p)$ on the compact convex set $[0,1]\times\Delta_n$. Lemma \ref{lem:con} implies the continuity of $d^1$ and every $G_j$. By Berge's maximum theorem \citep[Chapter~VI, Section~3]{berge1963topological} and the convex-hull construction in Equation~\eqref{eqt:br}, both mixed correspondences $\operatorname{BR}^1$ and $\operatorname{BR}^2$ are upper-hemicontinuous with nonempty, compact, and convex values. Thus, $F$ satisfies the hypotheses of Kakutani's fixed-point theorem \citep{kakutani1941generalization}, so there exists $(p^{\star},\mathbf{q}^{\star})$ such that $p^{\star}\in\operatorname{BR}^1(\mathbf{q}^{\star}), \mathbf{q}^{\star}\in\operatorname{BR}^2(p^{\star}).$
If $p^{\star}=0$, then $\mathbf{q}^{\star}\in\Delta(J(0)), \
d^1(\mathbf{q}^{\star})\le0,$
so $(p^{\star},\mathbf{q}^{\star})\in\left\{(0,\mathbf{q}):\mathbf{q}\in\Delta(J(0)),\ d^1(\mathbf{q})\le0\right\}$. If $p^{\star}=1$, then $\mathbf{q}^{\star}\in\Delta(J(1)),
\ d^1(\mathbf{q}^{\star})\ge0,$
so $(p^{\star},\mathbf{q}^{\star})\in\left\{(1,\mathbf{q}):\mathbf{q}\in\Delta(J(1)),\ d^1(\mathbf{q})\ge0\right\}$. Finally, if
$0<p^{\star}<1$, then both row actions are used, which requires $\mathbf{q}^{\star}\in\Delta(J(p^\star)),\ d^1(\mathbf{q}^{\star})=0.$
Hence, $(p^{\star},\mathbf{q}^{\star})\in\left\{(p,\mathbf{q}):0<p<1,\ \mathbf{q}\in\Delta(J(p)),\
d^1(\mathbf{q})=0\right\}$. Conversely, every profile in the displayed union satisfies Definition \ref{def:rho_ne}.
\end{proof}

\section{Functionals Preserving or Violating SCP}
\label{app:fun_scp}

We first provide the convergence proofs for games with a dominant action and games satisfying (S)SCP. We then prove the sufficient conditions from Section~\ref{sec:sc_id}, apply them to representative functionals, and present counterexamples to SCP.

\subsection{Convergence in games with a dominant action}
\label{app:dom}

We consider the easy cases in $2 \times 2$ games where at least one player has a dominant action (pure strategy). %
After relabeling actions if necessary, this means that
there exists $i\in[2]$ such that $d^i(s)\ge0,\ \forall s\in[0,1].$ The action is strongly dominant if the inequality is strict for every $s\in[0,1]$. Without loss of generality, suppose player~1 has a dominant pure action $\bar a^1\in\{0,1\}$, which satisfies $\bar a^1\in\operatorname{BR}^1_\rho(s),\ \forall s\in[0,1].$
We establish convergence under the regularity and tie-breaking conditions stated below.
\begin{prop}[Convergence of $\rho$-FP in $2\times2$ games with a dominant action]
\label{prop:dominant}
Assume that $d^{2}$ is continuous and has finitely many zeros, and both players use fixed pure-action tie-breaking rules such that \(\bar a^1\) is chosen whenever player~$1$ is
indifferent. Then every continuous-time $\rho$-FP trajectory converges as $t\to\infty$, and every sequence of discrete-time $\rho$-FP iterates converges as $k\to\infty$, to a pure $\rho$-NE $(\bar a^1,\bar a^2)$, where \(\bar a^2\in\operatorname{BR}^{2}_\rho(\bar a^1)\).
\end{prop}

\begin{proof}[Proof of Proposition \ref{prop:dominant}]
We recall that  $x(q)$ and $y(p)$ denote the selected pure best responses of players~1
and~2 (see Equation~\eqref{eqt:rho_fp_2by2}).  By assumption, $x(q)=\bar a^1$ for all $q$.  In continuous time,
therefore,
\[
p_t=\bar a^1+\bigl(p_0-\bar a^1\bigr)e^{-t}.
\]
If $p_0\ne\bar a^1$, then $p_t$ approaches \(\bar a^1\) monotonically.
Because $d^2$ has only finitely many zeros, the trajectory can meet only finitely
many of player~2's switching points.  Consequently, there are a time $T$
and an action \(\bar a^2\in\{0,1\}\) such that
$y(p_t)=\bar a^2$ for all $t\ge T$.  If $p_0=\bar a^1$, the same conclusion
holds immediately by the fixed tie-breaking rule.  It follows that
\[
q(t)=\bar a^2+\bigl(q(T)-\bar a^2\bigr)e^{-(t-T)},
\qquad t\ge T,
\]
and hence $(p_t,q(t))\to(\bar a^1,\bar a^2)$.

For discrete time, $p_{k+1}-\bar a^1=(1-\gamma_{k+1})(p_k-\bar a^1).$ Thus, $p_k$ also approaches \(\bar a^1\) monotonically (or reaches it in one
step).  The finiteness of the zero set again makes $y(p_k)$ eventually
constant, say equal to \(\bar a^2\).  Thereafter $q_{k+1}-\bar a^2=(1-\gamma_{k+1})(q_k-\bar a^2),$ which converges to zero because
\(\sum_k\gamma_{k+1}=\infty\). Hence, the discrete iterates also converge to $(\bar a^1,\bar a^2)$ as $k\to\infty$.

Finally, \(\bar a^1\) is a best response to every $q$, and continuity implies
that the eventual action \(\bar a^2\) is a best response at $p=\bar a^1$. We note that if
\(d^2(\bar a^1)=0\), both actions are best responses.  Therefore, \((\bar a^1,\bar a^2)\) is a \(\rho\)-NE by Definition~\ref{def:rho_ne}.
\end{proof}

\subsection{Proof of Proposition \ref{prop:sscp}}
\label{app:sscp_pf}
\begin{proof}[Proof of Proposition \ref{prop:sscp}]
Let $r_1,r_2\in(0,1)$ be the SSCP thresholds of $d^1$ and $d^2$,
respectively. By SSCP and the assumption $d^i(0)d^i(1)<0$,
\[
\operatorname{sgn}d^i(s)
=\operatorname{sgn}\bigl(d^i(1)(s-r_i)\bigr),
\qquad s\in[0,1],\quad i=1,2,
\]
including at $s=r_i$, where both sides vanish.
Construct a bimatrix game $\bar G$ with payoff matrices
\[
\bar A:=d^1(1)
\begin{pmatrix}
1&0\\
r_1&r_1
\end{pmatrix},
\qquad
\bar B:=d^2(1)
\begin{pmatrix}
1&r_2\\
0&r_2
\end{pmatrix},
\]
where both matrices use rows $U,D$ and columns $L,R$.
Its advantage functions under mean evaluation are
\[
\begin{aligned}
\bar d^1(q)
&=q(\bar A_{11}-\bar A_{21})
 +(1-q)(\bar A_{12}-\bar A_{22})=d^1(1)\bigl[q(1-r_1)-(1-q)r_1\bigr]
 =d^1(1)(q-r_1),\\
\bar d^2(p)
&=p(\bar B_{11}-\bar B_{12})
 +(1-p)(\bar B_{21}-\bar B_{22})=d^2(1)\bigl[p(1-r_2)-(1-p)r_2\bigr]
 =d^2(1)(p-r_2).
\end{aligned}
\]
Thus, the advantage function in $(G,\rho)$ and its affine counterpart in $(\bar{G},\mu)$ have identical signs and zero sets. In both games, $U$ is the unique
row best response when $d^1(q)>0$, $D$ is the unique row best
response when $d^1(q)<0$, and both actions are optimal when
$d^1(q)=0$. The same statement holds for $L,R$ with $d^2(p)$.
Consequently, the pure best responses and their
mixed best-response correspondences coincide,
and the Nash equilibria of $\bar G$ are exactly the $\rho$-NE of $G$.

Starting from identical initial beliefs, the two discrete
processes have identical updates when their step sizes and
fixed deterministic tie-breaking rules are matched. Their
continuous best-response inclusions also coincide.
Since $d^i(1)\ne0$ and $r_i\in(0,1)$, the game $\bar{G}$ has
strict best responses to every pure opponent action and is
nondegenerate. The classical convergence result for FP in nondegenerate
$2\times2$ games under these conventions \citep{metrick1994fictitious} therefore yields a
limiting NE of $\bar G$, which is a $\rho$-NE of $G$.
\end{proof}

\subsection{Proof of Proposition \ref{prop:scp}}
\label{app:scp}

\begin{proof}[Proof of Lemma \ref{lem:ne_scp}]
By the definition of $\rho$-NE, a profile
$(p,q)\in(0,1)^2$ is an interior $\rho$-NE if and only if
both players are indifferent between their two actions: $d^1(q)=0$ and $d^2(p)=0.$ Under SCP, these conditions are equivalent to
$q\in I_1$ and $p\in I_2$. Hence, the interior $\rho$-NE
are exactly $I_2\times I_1$.

There are no equilibria on the open edges of the strategy
square. Indeed, a player mixing against a pure opponent
would require $d^i(0)=0$ or $d^i(1)=0$, contradicting SCP.
It therefore remains to check the four corners. At $(0,0)$, the equilibrium conditions are
$d^1(0)<0$ and $d^2(0)<0$; at $(1,1)$, they are
$d^1(1)>0$ and $d^2(1)>0$.
Since SCP gives
$\operatorname{sgn}d^i(0)=-\sigma_i$,
both corners are equilibria precisely when
$\sigma_1=\sigma_2=1$.
Similarly, $(0,1)$ and $(1,0)$ are equilibria precisely
when $\sigma_1=\sigma_2=-1$.
If $\sigma_1\sigma_2=-1$, no corner is an equilibrium.
This proves the characterization.
\end{proof}

\begin{proof}[Proof of Proposition \ref{prop:scp}]
We fix an arbitrary deterministic tie-breaking rule.
Let $b_i\in\{0,1\}$ be player $i$'s action when indifferent $s\in[\underline r_i, \overline r_i]$, and set
\[
r_i:=
\begin{cases}
\underline r_i,
& (b_i=1 \text{ and } d^i(1)>0)\text{ or }(b_i=0 \text{ and }d^i(1)<0),\\
\overline r_i, & (b_i=0 \text{ and } d^i(1)>0)\text{ or }(b_i=1 \text{ and }d^i(1)<0).
\end{cases}
\]
The selected best response under SCP is the same as the best response induced by the affine advantage $\bar d^i(s):=d^i(1)(s-r_i)$, with matched
tie-breaking at $r_i$. Hence, the continuous-time trajectories and discrete-time iterates
coincide with their respective counterparts under ordinary FP in the
reduced nondegenerate $2\times2$ game and converge by Proposition~\ref{prop:sscp}.
If the limit is interior, it is $(r_2,r_1)$, where both original advantage
functions vanish. If it is pure, the reduced and original strict best responses
coincide. Thus, the limit is a $\rho$-NE.
\end{proof}

\subsection{A summary of SCP identification}
\label{app:fun}

Table~\ref{tab:fun_scp} summarizes the conditions under which representative
nonlinear evaluation functionals induce advantage functions satisfying SCP.

\begin{table}[!htb]
\centering
\caption{Conditional SCP properties of representative evaluation functionals.}
\label{tab:fun_scp}
\small
\renewcommand{\arraystretch}{1.1}
\begin{minipage}{\textwidth}
All conclusions concern advantage functions in $2\times2$ games under
Assumption~\ref{asp:cont} and $d^i(0)d^i(1)<0$.
\end{minipage}
\smallskip

\begin{tabularx}{\textwidth}{@{}>{\raggedright\arraybackslash}p{.21\textwidth}
>{\raggedright\arraybackslash}p{.19\textwidth}
>{\raggedright\arraybackslash}X@{}}
\toprule
Functional class & SCP conclusion & Conditions and qualifications \\
\midrule
ERM
& SSCP
& Risk-averse parameter $\theta>0$; satisfies SC2. Concave, but not coherent. \\
\addlinespace
Lower-tail OCE
& SSCP under smooth sufficient conditions
& $\ell\in C^2(\mathbb R)$, $\ell(0)=0$, $\ell'(0)=1$, and
  $\ell',\ell''>0$ everywhere. Other admissible losses require separate
  verification; OCE membership alone does not imply SSCP. \\
\addlinespace
Coherent RM
& SSCP under strictness;  SCP may occur
& Strict increase of $q\mapsto\rho(1,0;q)$ suffices via SC1.
  Coherence alone does not ensure a unique zero or probability continuity. \\
\addlinespace
Lower-tail CVaR
& SCP; SSCP when the zero is unique
& $0<\alpha\le1$. For payoff rows $(1,0)$ and $(0,1)$ with
  $\alpha<\tfrac12$, the advantage vanishes on $[\alpha,1-\alpha]$.
  CVaR is both an OCE and a coherent RM. \\
\addlinespace
Spectral RM
& SCP; SSCP under strictness
& $\phi\ge0$, $\int_0^1\phi(u)\,du=1$.
  Positive integral on every subinterval of positive length gives SSCP via SC1.
  Coherence additionally requires $\phi$ nonincreasing. \\
\addlinespace
Distortion RM
& SCP; SSCP under strictness
& Continuous, nondecreasing $g$ with $g(0)=0$, $g(1)=1$.
  Strict increase gives SSCP via SC1.
  Coherence additionally requires convex $g$ in the reward convention. \\
\addlinespace
RDEU / CPT
& SSCP under strictness and continuity
& Strictly increasing utility $v$ with $v(0)=0$ and continuous,
  strictly increasing probability weights normalized at $0$ and $1$;
  SC1 for RDEU, SC3 for CPT. These classes overlap with RMs. \\
\addlinespace
General concave RM
& SCP can fail
& Concavity alone is insufficient; the example below has three crossings. \\
\addlinespace
MSV / higher-moment criteria
& SCP can fail
& The examples below have three crossings. Such criteria need not be
  monotone and hence need not be monetary RMs. \\
\bottomrule
\end{tabularx}

\end{table}

\subsection{Proof of sufficient conditions enabling SCP}
\label{app:sc_pf}

\begin{proof}[Proof of SC2]
 We first prove the case of SSCP. We fix $\Delta_1> \Delta_2\ge 0$. We define $G_{\Delta_1,\Delta_2}(q)
:=
\Psi(\Delta_1,q)-\Psi(\Delta_2,q).$
By \textbf{SC2}, $G_{\Delta_1,\Delta_2}$ is strictly increasing on $(0,1)$. We extend this property from nonnegative payoff gaps
to arbitrary $\Delta_1>\Delta_2$ with $\Delta_1\neq0$. If
$\Delta_1>0\ge\Delta_2$,
\[
G_{\Delta_1,\Delta_2}= \Psi(\Delta_1,q)-\Psi(\Delta_2,q)=\Psi(\Delta_1,q)-\Psi(0,q)+\Psi(0,q)-\Psi(\Delta_2,q)
=G_{\Delta_1,0}+G_{0,\Delta_2}.
\]
The first term is strictly increasing by \textbf{SC2}, while the second term $G_{0,\Delta_2}=\rho(0,0;q)-\rho(\Delta_2,0;q)=-\rho(\Delta_2,0;q)$ is nondecreasing by monotonicity of $\rho$. Hence, their sum $G_{\Delta_1,\Delta_2}$ is strictly increasing. %
Finally, suppose $0>\Delta_1>\Delta_2$. For every
$\Delta<0$,
\[
(\Delta,0;q)
\stackrel{d}{=}
(0,\Delta;1-q)
\stackrel{d}{=}
\Delta+(-\Delta,0;1-q).
\]
Translation invariance therefore implies $\Psi(\Delta,q)=\rho(\Delta,0;q)=
\Delta+\Psi(-\Delta,1-q).$ Since $-\Delta_2>-\Delta_1>0$,
\begin{align*}
\Psi(\Delta_1,q)-\Psi(\Delta_2,q)
&=
\Delta_1-\Delta_2
-\Bigl(
\Psi(-\Delta_2,1-q)-\Psi(-\Delta_1,1-q)
\Bigr).
\end{align*}
By \textbf{SC2}, the expression in parentheses is strictly increasing in
$1-q$. It is therefore strictly decreasing in $q$, so
$q\mapsto\Psi(\Delta_1,q)-\Psi(\Delta_2,q)$ is strictly increasing.  Thus, $d^1(q)=\Psi(a-b,q) - \Psi(c-d,q) +(b-d)$
is strictly increasing since $a-b>c-d, a-b\neq0$. Since $d^1(0)=b-d<0, d^1(1)=a-c>0$, continuity and strict monotonicity of $d^1$ therefore imply SSCP. For the SCP case, the same payoff-gap decomposition yields nondecreasing differences under the non-strict version of \textbf{SC2}. Consequently, $d^1$ is nondecreasing, so continuity and the endpoint inequalities imply SCP by \textbf{SC1}.
\end{proof}

\begin{proof}[Proof of SC3]
We observe that for any $0< q_1<q_2<1$,
    \begin{align*}
        d^1(q_2)-d^1(q_1) &= \rho(a,b;q_2) - \rho(c,d;q_2)-(\rho(a,b;q_1) - \rho(c,d;q_1)) \\
        &=H_{q_1,q_2}(a,b)-H_{q_1,q_2}(c,d)\\
        &=H_{q_1,q_2}(a,b)-H_{q_1,q_2}(c,b)+H_{q_1,q_2}(c,b)-H_{q_1,q_2}(c,d)\\
        &>0.
    \end{align*}
This proves that $d^1$ is strictly increasing. Together with continuity and the endpoint conditions, this gives a unique zero and hence SSCP. Under the non-strict assumptions in \textbf{SC3}, the two $H_{q_1,q_2}$-differences in the displayed decomposition are nonnegative. Thus, $d^1(q_2)-d^1(q_1)\ge0$, so $d^1$ is nondecreasing; continuity and the endpoint inequalities then give SCP by \textbf{SC1}.
\end{proof}

\subsection{Identification of SCP}
\label{app:id_scp}

In this section, we identify the (S)SCP of the following evaluation functionals using \textbf{SC1}-\textbf{SC3}.

\subsubsection{Lower-tail OCE}

For a proper, lower semicontinuous, convex, nondecreasing loss $\ell$,
the lower-tail OCE is $\underline{\rho}_\ell(X):=
\sup_{\eta\in\mathbb R}
\left\{\eta-\mathbb E[\ell(\eta-X)]\right\}.$
We assume finiteness for the two-point distributions considered and
the normalization $\sup_u\{u-\ell(u)\}=0$, which ensures
$\underline{\rho}_\ell(m)=m$ for constant payoffs. We observe that
\[
\Psi_\ell(\Delta,q)
=
\underline{\rho}_\ell(\Delta,0;q)
=
\sup_{\eta\in\mathbb R}
\left\{\eta-q\ell(\eta-\Delta)-(1-q)\ell(\eta)\right\}.
\]
Translation invariance gives
$\underline{\rho}_\ell(x,y;q)=y+\Psi_\ell(x-y,q)$, so the
advantage function is
\[
d^1(q)
=
b-d+\Psi_\ell(a-b,q)-\Psi_\ell(c-d,q).
\]
The following lemma establishes strictly increasing differences,
and hence \textbf{SC2}, for a smooth subclass of OCE losses.

\begin{lem}
Suppose $\ell\in C^2(\mathbb R),\
\ell(0)=0,\ \ell'(0)=1,\
\ell'(u)>0,\ \ell''(u)>0$ for all $u\in\mathbb R.$
Then $\Psi_\ell$ has strictly increasing differences on
$\mathbb R\times[0,1]$.
\end{lem}

\begin{proof}
Convexity gives $\ell(u)\ge u$, with equality at zero, so the
normalization holds. We write $F(\eta;\Delta,q)$ for the objective
defining $\Psi_\ell$. It is strictly concave in $\eta$, and $F_\eta(\min\{0,\Delta\};\Delta,q)\ge0
\ge F_\eta(\max\{0,\Delta\};\Delta,q).$ Thus, it has a unique global maximizer
$\eta=\eta(\Delta,q)\in[\min\{0,\Delta\},\max\{0,\Delta\}]$,
characterized by
\[
q\ell'(\eta-\Delta)+(1-q)\ell'(\eta)=1.
\]
For a fixed $\Delta$, this common compact maximization interval
ensures continuity of $\Psi_\ell(\Delta,\cdot)$ on $[0,1]$, with
$\Psi_\ell(\Delta,0)=0$ and $\Psi_\ell(\Delta,1)=\Delta$. For $q\in(0,1)$, the implicit-function theorem yields
\[
D:=q\ell''(\eta-\Delta)+(1-q)\ell''(\eta)>0,
\qquad
\lambda:=\frac{\partial\eta}{\partial\Delta}
=\frac{q\ell''(\eta-\Delta)}{D}\in(0,1).
\]
Using the first-order condition, differentiation in $q$ and then in $\Delta$ gives
\[
\begin{aligned}
\frac{\partial\Psi_\ell}{\partial q}
=\ell(\eta)-\ell(\eta-\Delta),\quad
\frac{\partial^2\Psi_\ell}{\partial\Delta\,\partial q}
=\lambda\ell'(\eta)+(1-\lambda)\ell'(\eta-\Delta)>0.
\end{aligned}
\]
Consequently, for every $\Delta_1>\Delta_2$,
$\Psi_\ell(\Delta_1,q)-\Psi_\ell(\Delta_2,q)$ is strictly
increasing in $q$. Continuity at $q=0,1$ extends this property
to $[0,1]$.
\end{proof}

The smooth assumptions above are sufficient but not necessary. The linear loss $\ell(u)=u$ gives
$\Psi_\ell(\Delta,q)=q\Delta$ and satisfies \textbf{SC2} directly,
while the hinge losses $\ell_\alpha(u)=u^+/\alpha$,
$0<\alpha\le1$, yield lower-tail CVaR and are covered by the calculation in Appendix~\ref{app:cvar}. Other losses may require separate verification of \textbf{SC2} or SCP. %

\subsubsection{ERM}
\label{app:erm}
We consider ERM $\rho_\beta(X)
=
-\frac{1}{\beta}
\log
\mathbb{E}[e^{-\beta X}]$, with $\beta>0$.
Then
\[
\Psi_\beta(\Delta,q)
=
-\frac{1}{\beta}
\log
\bigl(
1-q+q e^{-\beta\Delta}
\bigr).
\]
We have
\[
\Psi_\beta(\Delta_1,q)-\Psi_\beta(\Delta_2,q)=
\frac{1}{\beta}
\log
\left(
\frac{
1-q+q e^{-\beta\Delta_2}
}{
1-q+q e^{-\beta\Delta_1}
}
\right).
\]
For $\Delta_1>\Delta_2$, the ratio is strictly increasing in $q$ because its derivative is
$(e^{-\beta\Delta_2}-e^{-\beta\Delta_1})/(1-q+q e^{-\beta\Delta_1})^2>0$.
The SSCP of ERM follows from \textbf{SC2.}

\subsubsection{CVaR}
\label{app:cvar}
Let $\rho_\alpha$ denote lower-tail CVaR at level
$\alpha\in(0,1]$.
We have
\[
\Psi_\alpha(\Delta,q)
=
\begin{cases}
\dfrac{\Delta}{\alpha}(q+\alpha-1)_+,
& \Delta\ge 0,
\\[2ex]
\Delta\cdot\min\left\{\dfrac{q}{\alpha},1\right\},
& \Delta<0.
\end{cases}
\]
Thus, for $\Delta_1>\Delta_2\ge0$,
\[ \Psi_\alpha(\Delta_1,q)-\Psi_\alpha(\Delta_2,q)=\dfrac{\Delta_1-\Delta_2}{\alpha}(q+\alpha-1)_+, \]
which is nondecreasing for every $\alpha\in(0,1]$, and is strictly increasing when $\alpha=1$. Thus, \textbf{SC2} gives SCP for all levels and SSCP for $\alpha=1$.

\subsubsection{Coherent RMs}
Coherent RMs \citep{artzner1999coherent} include lower-tail CVaR / expected shortfall, worst-case value, robust lower expectations. However, ERM is not coherent because it violates positive homogeneity. A coherent risk measure is also a concave risk measure in the sense of Definition~\ref{def:rm}, since superadditivity and positive homogeneity in Assumption~\ref{asp:rm}(d)--(e) imply concavity.

The advantage functions induced by a coherent risk measure satisfy SCP under the additional condition that $q\mapsto\rho(1,0;q)$ is strictly increasing.
We note that $(\Delta,0;q)\overset{d}{=}\Delta\cdot (1,0;q)$.
    For any $\Delta_1>\Delta_2\ge0$, using the positive homogeneity of $\rho$, we have
    \[ \Psi(\Delta_1,q)-\Psi(\Delta_2,q) =(\Delta_1-\Delta_2)\rho(1,0;q). \]
    Monotonicity of $\rho$ makes $q\mapsto \rho(1,0;q)$ nondecreasing; the stated strictness condition makes it strictly increasing. This verifies \textbf{SC2} and hence yields SSCP.  If strictness of the advantage function fails only through a  zero interval, the advantage instead
satisfies SCP, and Proposition~\ref{prop:scp} still gives convergence.
In the proof, we do not need superadditivity.

\subsubsection{Spectral RMs}
A reward-side spectral RM is
\[
\rho_{\phi}(X):=\int_0^1 \phi(u)F_X^{-1}(u)\,du,
\qquad
\Phi(t):=\int_0^t\phi(u)\,du,
\]
where $\phi:[0,1]\to[0,\infty)$ satisfies
$\int_0^1\phi(u)\,du=1$. The coherent case
additionally requires $\phi$ to be nonincreasing. Let $X_{\Delta,q}$ take the value $\Delta\geq0$ with probability
$q$ and $0$ with probability $1-q$. Then
\[
\rho_{\phi}(X_{\Delta,q})
=\Delta\int_{1-q}^1\phi(u)\,du
=\Delta\bigl(1-\Phi(1-q)\bigr).
\]
Consequently, if $\Delta_1>\Delta_2\geq0$,
\[
\Psi_\phi(\Delta_1,q)-\Psi_\phi(\Delta_2,q)
=(\Delta_1-\Delta_2)\bigl(1-\Phi(1-q)\bigr),
\]
which is strictly increasing in $q$ if $\phi$ has positive
integral on every subinterval of positive length in $[0,1]$.
For example, this holds if $\phi>0$ almost everywhere.
Under this condition, \textbf{SC2} and SSCP hold.
Nonnegativity of $\phi$ alone yields only a nondecreasing
difference and is not sufficient for the strict SCP conclusion. If both
advantages have nonempty closed zero intervals and satisfy $d^i(0)d^i(1)<0$,
they satisfy SCP. Proposition~\ref{prop:scp} then ensures convergence when
both zero intervals have positive length, and each player uses one fixed
action throughout its indifference interval.

\subsubsection{Distortion RMs}
We define the reward-side distortion RM by
\[
\rho_g(X)
=
\int_0^\infty g\bigl(P(X>x)\bigr)\,dx
+
\int_{-\infty}^0
\left[g\bigl(P(X>x)\bigr)-1\right]dx,
\]
where $g:[0,1]\to[0,1]$ is continuous and nondecreasing and satisfies
$g(0)=0$ and $g(1)=1$. Convexity of $g$ gives the coherent case. For $X_{\Delta,q}$ with $\Delta\geq0$, $\rho_g(X_{\Delta,q})=\Delta g(q).$
Thus, if $\Delta_1>\Delta_2\geq0$, $\Psi_g(\Delta_1,q)-\Psi_g(\Delta_2,q)
=(\Delta_1-\Delta_2)g(q),$ which is strictly increasing in $q$ when $g$ is strictly increasing.
Under this condition \textbf{SC2}, and hence SCP, holds. If $g$ is merely
nondecreasing, the zero set may be an interval; this is SCP, and convergence
follows from Proposition~\ref{prop:scp}.

For an absolutely continuous distortion $g$ with $g(0)=0$ and
$g(1)=1$, the spectral and distortion representations coincide via
\[
g(q)=1-\Phi(1-q)=\int_{1-q}^1\phi(u)\,du,
\qquad
\Phi(t):=\int_0^t\phi(u)\,du,
\]
where $\phi(u)=g'(1-u)$ almost everywhere and
$\int_0^1\phi(u)\,du=1$. A general continuous nondecreasing
distortion need not admit an integrable spectral density. For a nonnegative, nonincreasing spectral density $\phi$,
or equivalently a nondecreasing, convex distortion $g$ within
this absolutely continuous class, the resulting functional is
monotone, cash-additive, positively homogeneous, and superadditive,
and hence a coherent RM under our reward convention. ERM with positive  parameter is concave but
not coherent, since it is not positively homogeneous.

\subsubsection{RDEU}
An RDEU is defined by
    \[ \rho_{v,w}(F):= \int_{-\infty}^{\infty}v(x)dw(F(x)),\]
where $w:[0,1]\rightarrow[0,1]$ is a strictly increasing weight function with $w(0)=0$ and $w(1)=1$; $v:\mathbb{R}\rightarrow\mathbb{R}$ is an (unbounded) strictly increasing differentiable utility function with $v(0)=0$.  RDEU is generally not positively homogeneous due to the nonlinearity of $v$, and it is not  even translation-invariant $v(x+c)\neq v(x)+c$. Therefore, RDEU does not belong to RMs as it violates Assumption \ref{asp:rm}. \textbf{SC1} is still applicable.

We fix $a>b$. We show that $d^1(q)=\rho_{v,w}(a,b;q)-\rho_{v,w}(c,d;q)$ is strictly increasing. We observe that
\[ \rho_{v,w}(a,b;q)=v(b)w(1-q)+v(a)(1-w(1-q)), \]
which is strictly increasing in $q$. The value of $\rho_{v,w}(c,d;q)$ depends on the ordering of $c$ and $d$. There are two cases: $c\leq d$ or $c>d$. When $c\leq d$, $\rho_{v,w}(a,b;q)$ is strictly increasing and $\rho_{v,w}(c,d;q)$ is non-increasing, so $d^1(q)$ is strictly increasing. When $c>d$, we have
\[ d^1(q)=(v(a)-v(c))(1-w(1-q))+(v(b)-v(d))w(1-q).\]
Because $v(a)-v(c)+v(d)-v(b)>0$ and $w$ is strictly increasing, $d^1$ is strictly increasing in $q$. The endpoint inequalities also follow from the strict monotonicity of $v$. Therefore, the SSCP of RDEU follows from \textbf{SC1}.

\subsubsection{CPT}
\label{app:cpt}
    CPT splits outcomes into gains and losses relative to a reference point, usually 0. CPT uses an RDEU-type formula over the gain domain ($x>0$) and loss domain ($x<0$) with different  weighting functions $w^+/w^-$. The CPT is defined as
\[ \rho_{v,w}(F):= \int_{-\infty}^{0}v(x)dw^-(F(x))+\int^{\infty}_{0}v(x)d(1-w^+(1-F(x))),\]
where $w^{+/-}:[0,1]\rightarrow[0,1]$ are strictly increasing continuous weight functions with $w^{+/-}(0)=0$ and $w^{+/-}(1)=1$; $v:\mathbb{R}\rightarrow\mathbb{R}$ is a strictly increasing differentiable utility function with $v(0)=0$.
To verify \textbf{SC3}, we compute $H_{q_1,q_2}(x,y)$ in each of the six possible payoff orderings: (i) $x\ge y\ge0$: $\rho_{v,w}(x,y;q)
=v(y)+w^{+}(q)\bigl[v(x)-v(y)\bigr].$ So $H_{q_1,q_2}(x,y)
=(w^+(q_2)-w^+(q_1))\bigl[v(x)-v(y)\bigr].$ (ii) $y\ge x\ge0$: $\rho_{v,w}(x,y;q)=v(x)+w^{+}(1-q)\bigl[v(y)-v(x)\bigr].$ So $H_{q_1,q_2}(x,y)
=(w^+(1-q_1)-w^+(1-q_2))\bigl[v(x)-v(y)\bigr].$ (iii) $0\ge x\ge y$: $\rho_{v,w}(x,y;q)=v(x)+w^{-}(1-q)\bigl[v(y)-v(x)\bigr].$ So $H_{q_1,q_2}(x,y)
=(w^-(1-q_1)-w^-(1-q_2))\bigl[v(x)-v(y)\bigr].$ (iv) $0\ge y\ge x$: $\rho_{v,w}(x,y;q)=v(y)+w^{-}(q)\bigl(v(x)-v(y)\bigr).$ So $H_{q_1,q_2}(x,y)
=(w^-(q_2)-w^-(q_1))\bigl[v(x)-v(y)\bigr].$ (v) $x\ge0\ge y$: $\rho_{v,w}(x,y;q)=w^{+}(q)\,v(x)+w^{-}(1-q)\,v(y).$ So $H_{q_1,q_2}(x,y)
=(w^+(q_2)-w^+(q_1))v(x)-(w^-(1-q_1)-w^-(1-q_2))v(y).$ (vi) $y\ge0\ge x$: $\rho_{v,w}(x,y;q)=w^{-}(q)\,v(x)+w^{+}(1-q)\,v(y).$ So $H_{q_1,q_2}(x,y)
=(w^-(q_2)-w^-(q_1))v(x)-(w^+(1-q_1)-w^+(1-q_2))v(y).$  In each of the six regions, $H_{q_1,q_2}(x,y)$ can be written in the separable form $H_{q_1,q_2}(x,y)=\Delta^+ \cdot v(x)-\Delta^- \cdot v(y)$ for strictly positive $\Delta^{+/-}$, so $H_{q_1,q_2}(x,y)$ is strictly increasing in $x$ and strictly decreasing in $y$. Thus, CPT satisfies \textbf{SC3}.

\subsection{Functionals that violate SCP}
\label{app:scp_vio}
\subsubsection{Concave RM}
Concavity alone does not imply SCP, as the following example shows. We consider
\[
\rho(X):=\min\left\{\mathbb{E}[X],\,
\mathrm{CVaR}_{0.5}(X)+2\right\},
\]
where $\mathrm{CVaR}_{0.5}$ is lower-tail CVaR. This is a concave RM: it is normalized, monotone, and translation invariant, and pointwise minima preserve concavity. For $X=(\Delta,0;q)$ with $\Delta\ge0$,
\[
\Psi(\Delta,q)
=\min\left\{q\Delta,\,
2\Delta\left(q-\frac12\right)_++2\right\}.
\]
We now take payoff rows $U=(10,0)$ and $D=(5,1)$. Translation invariance gives
\[
d^1(q)=-1+\Psi(10,q)-\Psi(4,q)
=
\begin{cases}
6q-1,&0\le q\le\frac15,\\
1-4q,&\frac15\le q\le\frac12,\\
16q-9,&\frac12\le q\le\frac45,\\
6q-1,&\frac45\le q\le1.
\end{cases}
\]
Hence, $d^1$ has three zeros, $\tfrac16$, $\tfrac14$, and $\tfrac9{16}$, and SCP fails.

\subsubsection{Higher order moments}
Following the mean--variance--skewness--kurtosis (MVSK) portfolio literature \citep{pham2011efficient,zhou2021solving}, we consider the reward-side evaluation functional
\[
\rho_{\boldsymbol{\lambda}}(X)
=
\lambda_1\mu_1(X)
-\lambda_2\mu_2(X)
+\lambda_3\mu_3(X)
-\lambda_4\mu_4(X),
\qquad
\lambda_k\geq 0,
\]
where
\[
\mu_1(X)=\mathbb{E}[X],\
\mu_k(X)=
\mathbb{E}\!\left[
\bigl(X-\mathbb{E}[X]\bigr)^k
\right],
\qquad k=2,3,4.
\]
Thus, the functional rewards the mean and third central moment while penalizing the variance and fourth central moment.

To show that this functional can violate SCP, we consider the payoff matrix
$$
A=
\begin{pmatrix}
1 & 0\\
9/625 & 9/625
\end{pmatrix},
$$
and let $\rho(X)
=
\mu_1(X)
-\frac{5}{3}\mu_2(X)
+2\mu_3(X)
-\frac{4}{3}\mu_4(X)$. If the opponent chooses the first column with probability \(q\), the first action generates a Bernoulli payoff \(X_q\) equal to \(1\) with probability \(q\) and \(0\) otherwise, whereas the second action yields the constant payoff \(9/625\). For \(X_q\),
\(
\mu_1(X_q)=q,\
\mu_2(X_q)=q(1-q),\
\mu_3(X_q)=q(1-q)(1-2q),\
\mu_4(X_q)=
q(1-q)\bigl[(1-q)^3+q^3\bigr].
\)
Consequently, $\rho(X_q)=q^2(2q-1)^2$,
and the advantage of the first action over the second is
\(
d^1(q)
=
\rho(1,0;q)
-
\rho\left(\frac{9}{625},\frac{9}{625};q\right)\
=
q^2(2q-1)^2-\frac{9}{625}\
=
\frac{(5q-3)(5q-1)(10q-3)(10q+1)}{625}.
\)
The advantage function has three zeros in \((0,1)\): $\frac{1}{5}, \frac{3}{10}, \frac{3}{5}$. Moreover, \(d^1(q)<0\) on
\(
\left(0,\frac{1}{5}\right)
\cup
\left(\frac{3}{10},\frac{3}{5}\right),
\)
whereas \(d^1(q)>0\) on
\(
\left(\frac{1}{5},\frac{3}{10}\right)
\cup
\left(\frac{3}{5},1\right).
\)
Hence, the advantage function does not satisfy SCP.

\section{Proofs for Section \ref{sec:ct}}
\label{app:ct}
\subsection{Proof of Theorem \ref{thm:ct_conv_2by2}}
\label{app:pf_ct_conv_2by2}

\paragraph{Step 1: piecewise straight trajectory.} We recall that $p_0=q_0=0$ and $p_{m+1}=q_{n+1}=1$. We define
\[
P_i:=(p_{i-1},p_i),\quad i\in[m+1],
\qquad
Q_j:=(q_{j-1},q_j),\quad j\in[n+1],
\]
so that $R_{ij}=P_i\times Q_j$. Its closure is the closed cell $\overline{R_{ij}}=[p_{i-1},p_i]\times[q_{j-1},q_j]$. The open vertical and horizontal
switching segments are, respectively,
\[
S_{i,i+1;j}:=\{p_i\}\times Q_j,
\ i\in[m],\ j\in[n+1];\qquad
S_{i;j,j+1}:=P_i\times\{q_j\},
\ i\in[m+1],\ j\in[n].
\]
Regular switches occur in these open segments;
their endpoints, including the cell vertices, are excluded.

We fix a rectangle \(R_{ij}\) ($1\leq i\leq m+1, 1\leq j \leq n+1$). Since \(x_j,y_i\in\{0,1\}\) are constant on
\(R_{ij}\), the solution from an entry time $t_0$ is, for $t\ge t_0$
while the trajectory remains in \(R_{ij}\),
\begin{equation}
\label{eqt:seg}
    p_t=x_j+(p_{t_0}-x_j)e^{-(t-t_0)},
\qquad
q_t=y_i+(q_{t_0}-y_i)e^{-(t-t_0)}.
\end{equation}
Inside \(R_{ij}\), the trajectory traces a straight segment toward
the corner $(x_j,y_i)$ with an exponential time parametrization.
Every regular switch is transverse. On a vertical
switching segment $S_{i,i+1;j}$, the normal velocity has the same
nonzero value on both sides:
\[
x_j-p_i=
\begin{cases}
1-p_i>0,
& x_j=1,
\\[1ex]
-p_i<0,
& x_j=0,
\end{cases}
\]
and on a horizontal switching segment $S_{i;j,j+1}$, it is likewise
the same and nonzero on both sides:
\[
y_i-q_j=
\begin{cases}
1-q_j>0,
& y_i=1,
\\[1ex]
-q_j<0,
& y_i=0.
\end{cases}
\]
Consequently, there is no sliding along an open switching segment.
At every regular switch, the trajectory passes from one open
rectangle to an adjacent open rectangle. This transversality statement
does not apply at cell vertices.

\begin{lem}[Nonaccumulation of switching times]
\label{lem:nonaccumulation}
Every bounded time interval contains only finitely many regular switches
of the selected trajectory. Consequently, if there are infinitely many
switching times $t_k$, then $t_k\to\infty$ as $k\to\infty$.
\end{lem}
\begin{proof}
Suppose that infinitely many switches accumulate at a finite time $T$.
Since $|\dot p|,|\dot q|\le1$, we have
$\|\z_t-\z_s\|_\infty\le|t-s|$ for $s,t<T$.
Thus, $\z_t$ is Cauchy as $t\uparrow T<\infty$ and has a unique
limit $\z_*\in[0,1]^2$. The finite union of closed switching segments
contains $\z_*$. If $\z_*$ is not an interior cell vertex, a sufficiently
small neighborhood meets just one switching line. Its normal velocity
has a fixed nonzero sign on both sides, so the normal coordinate is
strictly monotone and that line can be crossed at most once. This also
applies at an endpoint on the outer square boundary. Thus,
$\z_*=(p_i,q_j)$ for some $i,j$. We choose a small neighborhood $U$ of
$\z_*$ meeting only the four incident cells. Since $\z_t\to \z_*$,
there exists $t_0<T$ such that $\z_t\in U$ for every $t_0<t<T$.

The sign changes at the thresholds give
$x_{j+1}=1-x_j$ and $y_{i+1}=1-y_i$. If $x_j=y_i$, the vertex has
the saddle geometry: two opposite cells are incoming and the other
two are outgoing. A trajectory in an incoming cell either reaches
the vertex and stops, or crosses one switching segment into an
outgoing cell, where both coordinates move away from the vertex.
Hence, a trajectory remaining in this neighborhood has at most one
regular switch, contradicting accumulation.

If $x_j\ne y_i$, the four cell directions circulate around the vertex,
so an infinitely switching tail confined to this neighborhood makes
complete turns. For $(x_j,y_i)=(1,0)$, we start on the lower vertical
half-section $\z(r):=(p_i,q_j-r)$, with $r>0$ sufficiently small.
The four ray intersections, obtained from Equation~\eqref{eqt:seg}
(equivalently, by similar triangles \citep{metrick1994fictitious}),
give offsets
\begin{equation}
\label{eqt:local_spiral_offsets}
s=\frac{(1-p_i)r}{1-q_j+r},\qquad
u=\frac{(1-q_j)s}{p_i+s},\qquad
w=\frac{p_i u}{q_j+u},\qquad
r^+=\frac{q_j w}{1-p_i+w}.
\end{equation}
These maps are continuous and vanish at $r=0$, so all four legs lie
in the incident cells when $r$ is small enough. Composing them gives
the local return map below; reflection in $p\mapsto1-p$ gives the
other orientation:
\begin{equation}
\label{eqt:local_spiral_return}
r^+=\frac{Cr}{C+r},\qquad
C:=\begin{cases}
p_iq_j(1-q_j),& (x_j,y_i)=(1,0),\\
(1-p_i)q_j(1-q_j),& (x_j,y_i)=(0,1).
\end{cases}
\end{equation}
In either case $C>0$. Successive lower-section offsets therefore satisfy
\begin{equation}
\label{eqt:local_spiral_radii}
\frac1{r_{k+1}}=\frac1{r_k}+\frac1C,
\qquad r_k=\frac{Cr_0}{C+kr_0},\qquad k\ge0.
\end{equation}
The first segment of the $k$-th turn changes $q$ by $r_k$ and
$|\dot q|\le1$, so its duration is at least $r_k$.
Since $C,r_0>0$, the explicit formula gives
$r_k=C/(k+C/r_0)\ge C/(2k)$ for all $k\ge C/r_0$.
Thus, $\sum_k r_k=\infty$ by comparison with the harmonic series,
so these turns cannot fit into a finite time interval.
This excludes the remaining case and proves the lemma.
\end{proof}

For every trajectory, there are two cases: finitely many switches and infinitely many switches.

\paragraph{Step 2: case of finitely many switches.} We will show that the trajectory converges to a corner or to a zero-line intersection in this case.  We consider an interior trajectory with finitely many regular switches, possibly zero. Let $(p,q)$ start its final trajectory segment within a cell in $\overline{R_{ij}}$, after the last crossing or at the initial time if there are none. Inside that rectangle, the vector field points toward the corner \((x_j,y_i)\).

If \((x_j,y_i)\) lies in the closure of the rectangle, then the trajectory remains in the closure and converges exponentially to \((x_j,y_i)\) as \(t\to\infty\).

If $(x_j,y_i)\not\in\overline{R_{ij}}$, the cell solution must leave $\overline{R_{ij}}$ in finite time. Let $(p',q')$ be its first exit point. By Equation~\eqref{eqt:seg}, a trajectory segment in the interior cannot reach the outer square boundary in finite time. Thus, the exit lies on the closure of an internal switching segment $S'$; for example,
\[
\overline{S_{i,i+1;j}}=S_{i,i+1;j}\cup\{(p_i,q_{j-1})\}\cup\{(p_i,q_j)\}.
\]
If $(p',q')\in S'$, transversality gives another regular switch, a contradiction. Hence, $(p',q')\in\overline{S'}\setminus S'$ is an interior cell vertex $(p_k,q_l)\in\partial R_{ij}$. The trajectory reaches this vertex in finite time and remains there by the equilibrium-stopping convention.

Thus, the only remaining case is the case of infinitely many switches.
\paragraph{Step 3: no nonconstant closed chains of forward cell arcs.}
\begin{lem}[Green's theorem, flux form]
\label{lem:green}
    Let \(C\) be a positively oriented, \textbf{piecewise smooth}, simple closed curve
in the plane, and let \(D\) be the region bounded by \(C\). Let $\boldsymbol{F}(p,q)=(L(p,q),M(p,q))$ be a vector field defined on an open region containing \(\overline D\). If \(L\) and \(M\)
have \textbf{continuous partial derivatives} on that open region, then
\[
\oint_C \left(-M\,dp + L\,dq\right)
=
\iint_D
\left(
\frac{\partial L}{\partial p}
+
\frac{\partial M}{\partial q}
\right)\,dA,
\]
where the path of integration along \(C\) is counterclockwise. Equivalently,
\[ \int_{C} \boldsymbol{F}\cdot \boldsymbol{n}\,ds
=
\iint_{D} \nabla\cdot \boldsymbol{F}\,dA, \]
where $\boldsymbol{n}$ is the outward-pointing unit normal vector on $C$ \citep[Section~11.19]{apostol1969calculus}.
\end{lem}
\begin{lem}
\label{lem:no_orbit}
    There is no nonconstant closed chain formed by finitely many forward arcs of the cell equations $\dot z=\vc_{ij}-z$, each contained in $\overline{R_{ij}}$. Arcs may be joined at cell vertices, and an arc may start at a vertex and move away from it even when the selected flow would remain there. The chain need not itself be a trajectory of the selected flow.
\end{lem}
\begin{proof}[Proof of Lemma~\ref{lem:no_orbit}]
We fix a rectangle $R_{ij}$ and let \(\boldsymbol{F_{ij}}(p,q)\) denote the vector field in $R_{ij}$:
\[
\boldsymbol{F_{ij}}(p,q):=(L_{ij},M_{ij}):=(x_j-p,\ y_i-q).
\]
On every rectangle $R_{ij}$, we have
\[
\nabla\cdot \boldsymbol{F_{ij}}
=
\frac{\partial}{\partial p}(x_j-p)
+
\frac{\partial}{\partial q}(y_i-q)
=
-1-1
=
-2.
\]
We extend each affine field $\boldsymbol{F_{ij}}$ to all of $\mathbb R^2$.
Suppose, toward a contradiction, that there is a nonconstant closed forward chain.
Discard constant pieces and subdivide its straight segments at all segment endpoints,
isolated intersections, endpoints of overlaps, and crossings of grid lines.
Each resulting edge has a unique forward orientation: in an open cell,
the nonzero cell field fixes its direction, while an edge on the outer square boundary
has a unique incident cell. No edge lies along an internal switching segment,
because its normal velocity is nonzero. Thus, no edge can be traversed
in opposite directions. Identifying repeated copies of the same oriented edge
gives a finite embedded directed graph in which the chain is a nonempty closed
directed walk. This walk contains a simple directed cycle. A single straight edge
cannot form a loop, and a two-edge cycle would require opposite retracing.
Hence, the cycle is a simple closed polygonal curve $\Gamma$, bounding an interior region
$D$ of positive area. Let $D_\ell$ be the closures of the connected components of
$D\cap R_{ij}$ over all cells, so that $\overline D=\bigcup_\ell D_\ell$ is a finite decomposition.
We write $\partial D_\ell=\Gamma_\ell\cup Q_\ell\cup P_\ell$, where
$\Gamma_\ell=\partial D_\ell\cap\Gamma$, and $Q_\ell$ and $P_\ell$ are unions of
vertical and horizontal internal cut segments, respectively.
On each piece $D_\ell\subseteq\overline{R_{ij}}$, we use the corresponding field $\boldsymbol{F_{ij}}$
and invoke the flux form of Green's theorem (Lemma \ref{lem:green}), using a polygonal subdivision if necessary. $\boldsymbol{F_{ij}}$ has continuous first order partial derivatives on $\mathbb R^2$, which contains $D_\ell$, so we can apply Green's theorem. This gives
\begin{equation}
\label{eqt:green_local}
    \int_{\partial D_\ell} \boldsymbol{F_{ij}}\cdot \boldsymbol{n}\,ds
=
\iint_{D_\ell} \nabla\cdot \boldsymbol{F_{ij}}\,dA
=
-2\,\mathrm{Area}(D_\ell).
\end{equation}
We now sum this identity over all sets \(D_\ell\). The contributions along internal switching lines cancel. We observe that
\[ \int_{\partial D_\ell} \boldsymbol{F_{ij}}\cdot \boldsymbol{n}\,ds = \int_{\Gamma_\ell} \boldsymbol{F_{ij}}\cdot \boldsymbol{n}\,ds + \int_{Q_\ell} \boldsymbol{F_{ij}}\cdot \boldsymbol{n}\,ds+ \int_{P_\ell} \boldsymbol{F_{ij}}\cdot \boldsymbol{n}\,ds.\]
Along any vertical cut on $p=p_i$, the adjacent cells $R_{ij}$ and $R_{i+1,j}$ have the same horizontal component:
\[
L_{ij}(p_i,q)=x_j-p_i=L_{i+1,j}(p_i,q).
\]
Because their outward normals on the shared cut are opposite, their fluxes across it cancel. Likewise, along any horizontal cut on $q=q_j$, the adjacent cells $R_{ij}$ and $R_{i,j+1}$ have the same vertical component:
\[
M_{ij}(p,q_j)=y_i-q_j=M_{i,j+1}(p,q_j),
\]
so their fluxes across it cancel. Hence, the fluxes across all internal cuts cancel pairwise.

On the outer boundary portion \(\Gamma_\ell\), the vector field $\boldsymbol{F_{ij}}$ is tangent almost everywhere, so $\boldsymbol{F_{ij}}\cdot\boldsymbol{n}=0$ almost everywhere on $\Gamma_\ell$. The finitely many junctions, including cell vertices, do not affect the boundary integrals. Summing the left-hand side of
Equation~\eqref{eqt:green_local},
\[ \sum_\ell\int_{\partial D_\ell} \boldsymbol{F_{ij}}\cdot \boldsymbol{n}\,ds
= \sum_\ell\int_{\Gamma_\ell} \boldsymbol{F_{ij}}\cdot \boldsymbol{n}\,ds = 0.\]
The right-hand side is
\[
\sum_\ell-2\,\mathrm{Area}(D_\ell)=-2\,\mathrm{Area}(D)<0.
\]
This contradiction rules out every nonconstant closed forward chain.
\end{proof}
\paragraph{Step 4: crossings of one fixed switching segment are ordered.}
We fix an (open) switching segment \(S\). We treat a vertical switching segment $S:=S_{i,i+1;j}=\{(p,q):p=p_i,\ q_{j-1}<q<q_j\}.$ The proof for a horizontal segment is identical after exchanging \(p\) and \(q\). Embed the scalar interval $I:=(q_{j-1},q_j)$ onto \(S\) by $\sigma:I\to S,\ \sigma(s):=(p_i,s).$ Every crossing of \(S\) has the same orientation.

Let \(\phi_t\) denote the forward flow. We define the  \textit{first-return domain} \(D_S\subset I\) as follows. \(s\in D_S\) if there exists a finite time \(\tau(s)>0\) such that $\phi_{\tau(s)}(\sigma(s))\in S,$ and $\phi_t(\sigma(s))\notin S,\ t\in(0, \tau(s)),$
and the trajectory does not hit a cell vertex for $t\in[0, \tau(s)]$.
For \(s\in D_S\), we define the \textit{first-return map} \(\Pi_S:D_S \to I\) by
\[
\phi_{\tau(s)}(\sigma(s))=\sigma(\Pi_S(s))=(p_i,\Pi_S(s)).
\]
\begin{lem}[Regularity on a fixed itinerary]
    (i) The set \(D_S\) is open, i.e., \(D_S\) is a (countable) disjoint union of open intervals. (ii) On every connected component (maximal open interval) \(J\) of \(D_S\), the sequence of rectangles and switching segments visited before returning to \(S\) is constant. (iii) $\tau:J\to(0,\infty), \Pi_S:J\to I$
are \(C^1\).
\end{lem}
\begin{proof}
We fix \(s_0\in D_S\). By Lemma~\ref{lem:nonaccumulation}, only finitely many crossings occur before the finite return time \(\tau(s_0)\). They are all transverse, since the trajectory avoids cell vertices by definition of \(D_S\).

Standard continuous dependence for each smooth cell flow, together with the implicit-function theorem at every transverse crossing, implies that all sufficiently close initial points encounter the same sequence of switching segments in the same order. Their switching times and return time depend \(C^1\)-smoothly on the initial coordinate. Thus, \(D_S\) is open, and both \(\tau\) and \(\Pi_S\) are \(C^1\) locally.

The switching itinerary is locally constant and takes values in a discrete set. It is therefore constant on each connected component of \(D_S\).
\end{proof}
We next show that order is preserved on \(D_S\), including between different connected components.
\begin{lem}[Global order preservation]
\label{lem:order}
    For any \(s_1,s_2\in D_S\),
\[
s_1<s_2\quad\Longrightarrow\quad \Pi_S(s_1)<\Pi_S(s_2).
\]
\end{lem}
\begin{proof}
We fix $s_1<s_2$ in $D_S$ and write $r_i:=\Pi_S(s_i)$ for $i=1,2$. Let
$\Gamma_1,\Gamma_2$ be the corresponding first-return trajectory segments.
By Lemma~\ref{lem:no_orbit}, each segment is simple with distinct endpoints.
The segments with their endpoints removed are disjoint, and $r_1\ne r_2$:
uniqueness of backward continuation inside cells and through regular switching points, applied at an intersection reached at positive
times, would either identify the starting points or force an earlier
return to $S$.

Suppose $r_2<r_1$. The curve $C:=\Gamma_1\cup L$, where
$L:=\sigma([\min\{s_1,r_1\},\max\{s_1,r_1\}])$, is a Jordan curve.
Label its complementary components $U_+$ and $U_-$ by the outgoing
and incoming sides of the open closing segment, respectively;
we need not specify which component is bounded.

Near $\sigma(s_1)$, $C$ consists of the section pointing toward $r_1$
and an outgoing trajectory branch. The continuation of $S$ away
from $r_1$ therefore connects locally to the incoming side of $L$
and lies in $U_-$. At $\sigma(r_1)$, the trajectory branch is incoming,
so the continuation away from $s_1$ lies in $U_+$. These assignments
extend along the corresponding portions of $S\setminus L$, since
$C\cap S=L$. Between $s_1$ and $r_1$, departures enter $U_+$ and arrivals
come from $U_-$.

Thus, all six strict endpoint orders compatible with $s_1<s_2$ and $r_2<r_1$
give the following locations of short subarcs of $\Gamma_2$ just
after departure and just before return, excluding the endpoints:
\[
\begin{array}{c|cc}
\text{Endpoint order} & \text{Initial subarc} & \text{Final subarc} \\
\hline
s_1<s_2<r_2<r_1 & U_+ & U_- \\
s_1<r_2<s_2<r_1 & U_+ & U_- \\
s_1<r_2<r_1<s_2 & U_+ & U_- \\
r_2<s_1<s_2<r_1 & U_+ & U_- \\
r_2<s_1<r_1<s_2 & U_+ & U_- \\
r_2<r_1<s_1<s_2 & U_- & U_+
\end{array}.
\]
Figures~\ref{fig:transversal_order_a}--\ref{fig:transversal_order_f} illustrate the six endpoint
orders in the same sequence as the table.
The only remaining coincidences are $s_2=r_1$ or $r_2=s_1$, or both, which
force $s_1<r_1$. The same local endpoint geometry places departure from
$r_1$ in $U_+$ and arrival at $s_1$ in $U_-$, so these cases also have
initial and final subarcs in different components.

The connected trajectory segment $\Gamma_2\setminus\{\sigma(s_2),\sigma(r_2)\}$ must
therefore meet $C$. Meeting $L\subset S$ contradicts the first-return
condition; meeting $\Gamma_1$ contradicts disjointness, since its
endpoints also lie on $S$. Hence, $r_2<r_1$ is impossible, and $r_1\ne r_2$
gives $\Pi_S(s_1)=r_1<r_2=\Pi_S(s_2)$.
\end{proof}

\begin{figure}[!htb]
\centering
\begingroup
\def\transversalReturnArc{%
  (0,1)
  .. controls (1.25,1) and (2.15,1.2) .. (2.15,2.65)
  .. controls (2.15,4.1) and (1.4,4.85) .. (0.2,4.85)
  .. controls (-1,4.85) and (-1.35,3) .. (0,3)}
\begin{tikzpicture}[
  x=0.9cm,y=0.64cm,font=\footnotesize,
  section/.style={draw=black!50,dashed,line width=0.6pt},
  closing/.style={draw=black,line width=1.1pt},
  return arc/.style={
    draw=blue!65!black,line width=1.2pt,
    postaction={decorate},
    decoration={markings,
      mark=at position 0.08 with {\arrow{Stealth}},
      mark=at position 0.38 with {\arrow{Stealth}},
      mark=at position 0.70 with {\arrow{Stealth}},
      mark=at position 0.96 with {\arrow{Stealth}}}},
  second arc/.style={
    draw=red!70!black,densely dashed,line width=1pt,
    -{Stealth[length=1.7mm]}},
  endpoint/.style={circle,fill=black,inner sep=1.1pt},
  open endpoint/.style={circle,draw=black!50,fill=white,inner sep=1.1pt}
]
  \foreach \panel/\xoffset/\yoffset/\ssecond/\rsecond/\reflect/\ordering in {
    a/0/0/1.65/2.4/0/{s_1<s_2<r_2<r_1},
    b/5.7/0/2.4/1.65/0/{s_1<r_2<s_2<r_1},
    c/11.4/0/3.65/1.65/0/{s_1<r_2<r_1<s_2},
    d/0/-6.8/2.4/0.4/0/{r_2<s_1<s_2<r_1},
    e/5.7/-6.8/3.65/0.4/0/{r_2<s_1<r_1<s_2},
    f/11.4/-6.8/3.65/0.4/1/{r_2<r_1<s_1<s_2}%
  } {
    \begin{scope}[shift={(\xoffset,\yoffset)}]
      \ifnum\reflect=0\relax
        \path[fill=blue!7] \transversalReturnArc -- cycle;
        \draw[return arc] \transversalReturnArc;
        \def\sfirst{1}\def\rfirst{3}
        \def\insideheight{2.5}\def\outsideheight{1.75}\def\archeight{3.2}
      \else
        \begin{scope}[shift={(0,4)},yscale=-1]
          \path[fill=blue!7] \transversalReturnArc -- cycle;
          \draw[return arc] \transversalReturnArc;
        \end{scope}
        \def\sfirst{3}\def\rfirst{1}
        \def\insideheight{1.5}\def\outsideheight{2.3}\def\archeight{0.8}
      \fi
      \draw[section] (0,0) -- (0,4.2);
      \draw[closing] (0,1) -- (0,3);
      \node[open endpoint] at (0,0) {};
      \node[open endpoint] at (0,4.2) {};
      \node[right=3pt] at (0,4.2) {$S$};
      \node[right=3pt] at (0,2.025) {$L$};
      \node[text=blue!65!black] at (1.18,\insideheight) {$U_+$};
      \node at (-1.15,\outsideheight) {$U_-$};
      \node[right,text=blue!65!black] at (2.15,\archeight) {$\Gamma_1$};
      \node[endpoint,label={left:$s_1$}] at (0,\sfirst) {};
      \node[endpoint,label={right:$r_1$}] at (0,\rfirst) {};
      \draw[second arc] (0,\ssecond) -- (0.55,\ssecond);
      \draw[second arc] (-0.5,\rsecond) -- (0,\rsecond);
      \node[endpoint,label={left:$s_2$}] at (0,\ssecond) {};
      \node[endpoint,label={right:$r_2$}] at (0,\rsecond) {};
      \node[align=center] at (0.5,-1.28)
        {\textbf{\panellabel{fig:transversal_order_\panel}} $\ordering$};
    \end{scope}
  }
\end{tikzpicture}
\endgroup
\caption{Six schematic endpoint orders. Dashed red arrows show a
hypothetical $\Gamma_2$ departing and returning in different components
of $\mathbb R^2\setminus C$; shading marks $U_+$.}
\label{fig:transversal_order}
\end{figure}
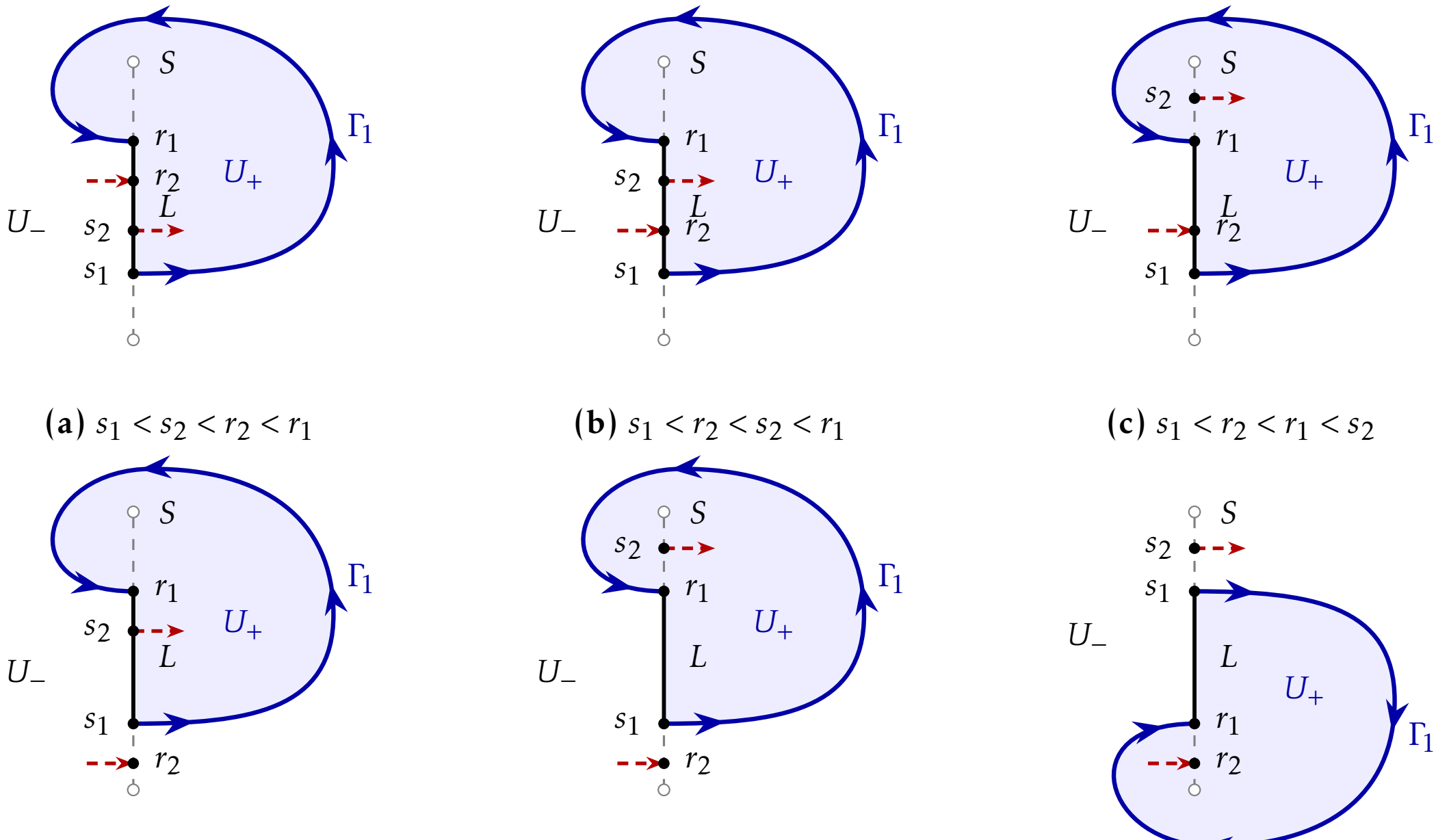

We obtain the ordering of repeated crossings.
\begin{lem}[Monotonicity of repeated crossings]
\label{lem:mn_cro}
    Suppose a trajectory crosses \(S\) infinitely many times. Let
$t_k$ be the time of the $k$-th crossing, and write
$\phi_{t_k}(z_0)=\sigma(s_k),\ s_k\in I.$ Then either $s_{k+1}>s_k$ for $k\geq 1$ or $s_{k+1}<s_k$ for $k\geq 1$. In particular, the sequence \((s_k)\) converges to a point of \(\overline I\) as $k\to\infty$.
\end{lem}

\begin{proof}
Because the trajectory crosses \(S\) infinitely often, it cannot hit a cell vertex between two successive crossings. Therefore,
$s_k\in D_S,\ s_{k+1}=\Pi_S(s_k)$ for every \(k\). If $s_2>s_1,$
then Lemma \ref{lem:order} gives $s_{k+1}>s_k$ for every \(k\) by induction. Similarly, if $s_2<s_1,$ then $s_{k+1}<s_k$ for every \(k\). The remaining possibility \(s_2=s_1\) implies that the trajectory returns to exactly the same point after a positive time. This implies a nonconstant periodic trajectory, which was ruled out by Lemma \ref{lem:no_orbit}. Thus, \((s_k)\) is strictly monotone. Since $s_k\in I=(q_{j-1},q_j),$
the sequence \((s_k)\) is bounded and hence converges to some point in $\overline I=[q_{j-1},q_j].$
\end{proof}
The same argument applies to a horizontal switching segment, using the \(p\)-coordinate as its parameter. Consequently, on every switching segment crossed infinitely often, the sequence of crossing points converges.

\paragraph{Step 5: infinitely many switches force convergence to some $(p_i,q_j)$.}
We now consider a trajectory \(\z_t=(p_t,q(t))\) with infinitely many switches. By Lemma~\ref{lem:nonaccumulation}, its switching times satisfy $t_k\to\infty$. Such a trajectory cannot reach a cell vertex, since it would then remain there forever. Let $\z_k:=(p(t_k),q(t_k))$ be the $k$-th crossing point, and let $S_k$ be the switching segment containing $\z_k$.

We first show that the recurrent transition graph is strongly connected.

We define the set of switching segments visited infinitely often
\[
\mathscr S_\infty:=\left\{S:\ S_k=S\text{ for infinitely many }k\right\}.
\]
For \(S,S'\in\mathscr S_\infty\), we place a directed edge $S\longrightarrow S'$ whenever $(S_k,S_{k+1})=(S,S')$ for infinitely many \(k\). We denote the resulting finite directed graph by $\mathscr G_\infty:=(\mathscr S_\infty,\mathscr E_\infty).$ Because there are only finitely many switching segments and finitely many ordered pairs of switching segments, there exists \(K_0\) such that, for every \(k\ge K_0\), $S_k\in\mathscr S_\infty,\ (S_k,S_{k+1})\in\mathscr E_\infty,$
since every segment or transition that occurs only finitely many times disappears from the event sequence after a sufficiently large index. Hence, $(S_{k})_{k\ge K_0}$ is an \textit{infinite directed walk} in \(\mathscr G_\infty\), and every vertex of \(\mathscr G_\infty\) occurs infinitely many times along this walk.
\begin{lem}
\label{lem:conn}
    \(\mathscr G_\infty\) is strongly connected.
\end{lem}
\begin{proof}
We consider its decomposition into strongly connected components and the associated condensation graph. The condensation graph is a finite directed acyclic graph. A directed walk in a finite acyclic graph cannot leave a component and later return to it. Consequently, the component containing \(S_k\) must eventually become constant as \(k\to\infty\). Since every vertex of \(\mathscr S_\infty\) occurs infinitely many times, every vertex must belong to this eventual component. Therefore, \(\mathscr G_\infty\) has only one strongly connected component.
\end{proof}
By Lemma \ref{lem:mn_cro}, the following limit exists for each \(S\in\mathscr S_\infty\), defining a map \(\z:\mathscr S_\infty\to[0,1]^2\):
\[
\z(S):=\lim_{\substack{k\to\infty\\S_k=S}}\z_k\in\overline S.
\]
We fix an edge $e=(S,S')\in\mathscr E_\infty.$
We choose indices \(k_\ell\to\infty\) such that $S_{k_\ell}=S,\ S_{k_\ell+1}=S'.$ For \(t\in(t_{k_\ell},t_{k_\ell+1})\), the trajectory lies in one open rectangle, denoted \(R_e\), whose boundary contains both \(S\) and \(S'\). The rectangle \(R_e\) is uniquely determined by the ordered pair \((S,S')\). Let
$\vc_e:=(a_e,b_e)\in\{0,1\}^2$ be the corner of the vector field in \(R_e\). On this portion of the trajectory, $\dot \z=\vc_e-\z.$
Let $\tau_\ell:=t_{k_\ell+1}-t_{k_\ell}>0.$ Using Equation~\eqref{eqt:seg}, we have $\z_{k_\ell+1}=\vc_e+\bigl(\z_{k_\ell}-\vc_e\bigr)e^{-\tau_\ell}.$
The sequence \((\tau_\ell)\) is bounded. To see this, we note that \(S'\) is an internal switching segment, so its closure does not contain the corner \(\vc_e\). Hence, $d_e:=\operatorname{dist}(\vc_e,\overline{S'})>0$, where \(\operatorname{dist}\) and \(\lVert\cdot\rVert\) denote Euclidean distance and norm, respectively. Since \(\z_{k_\ell+1}\in S'\),
\[
d_e\le \lVert \z_{k_\ell+1}-\vc_e\rVert=e^{-\tau_\ell}\lVert \z_{k_\ell}-\vc_e\rVert\le \sqrt{2}\,e^{-\tau_\ell}.
\]
Therefore, $\tau_\ell\le \log\frac{\sqrt{2}}{d_e}.$

After passing to a subsequence, we have $\tau_\ell\longrightarrow\tau_e\ge 0.$
It follows that $z(S')=\vc_e+\bigl(\z(S)-\vc_e\bigr)e^{-\tau_e}.$
Consequently,
\begin{equation}
    \label{eqt:limit_seg}
    \gamma_e(t):=\vc_e+\bigl(\z(S)-\vc_e\bigr)e^{-t},\qquad 0\le t\le\tau_e,
\end{equation}
is a forward line segment satisfying $\gamma_e(0)=\z(S),\ \gamma_e(\tau_e)=\z(S').$ Since \(\z(S)\) lies on the closure of an internal switching segment whereas \(\vc_e\) is a corner, \(\z(S)\neq \vc_e\). Hence, $\z(S)=\z(S')$ if and only if $\tau_e=0$.

Next, we show that all recurrent segment limits are equal.  We first show that the two endpoints of every recurrent edge coincide. We fix
$e_0:=(S_0,S_1)\in\mathscr E_\infty.$ Because \(\mathscr G_\infty\) is strongly connected, the edge \(e_0\) belongs to a directed cycle. There exist $S_0,S_1,\ldots,S_r=S_0$ with $r\ge 2$ such that $e_\ell:=(S_\ell,S_{\ell+1})\in\mathscr E_\infty, \ell=0,\ldots,r-1.$ For every \(e_\ell\), Equation~\eqref{eqt:limit_seg} gives a limiting forward line segment
$\gamma_{e_\ell}$ from \(\z(S_\ell)\) to \(\z(S_{\ell+1})\). Concatenating these segments gives a closed chain:
\[
\z(S_0)\overset{\gamma_{e_0}}{\longrightarrow}\z(S_1)\overset{\gamma_{e_1}}{\longrightarrow}\cdots\overset{\gamma_{e_{r-1}}}{\longrightarrow}\z(S_r)=\z(S_0).
\]
Each limiting segment $\gamma_{e_\ell}$ lies in $\overline{R_{e_\ell}}$, since its endpoints lie there and the closed cell is convex. Thus, the concatenation is a closed chain of forward cell-equation arcs covered by Lemma~\ref{lem:no_orbit}, even if an arc departs from a cell vertex where the selected flow would stop. By that lemma, the chain must be constant, so every arc in it is constant. In particular, $\z(S_0)=\z(S_1).$ Since \(e_0\) was arbitrary, $\z(S)=\z(S')$ for every $(S,S')\in\mathscr E_\infty.$ Because \(\mathscr G_\infty\) is strongly connected, any two vertices \(S,S'\in\mathscr S_\infty\) can be joined by a directed path. Applying this successively along that path shows that $\z(S)=\z(S')$ for all $S,S'\in\mathscr S_\infty.$ Hence, there exists exactly one point \(v\in[0,1]^2\) such that $\z(S)=v$ for every $S\in\mathscr S_\infty.$

We next show that the entire sequence of switching points converges to \(v\). Let \(\varepsilon>0\). For every \(S\in\mathscr S_\infty\), there exists an index \(N_S\) such that $\lVert \z_k-v\rVert<\varepsilon$ whenever $k\ge N_S$ and $S_k=S$.
Since \(\mathscr S_\infty\) is finite, let $N:=\max_{S\in\mathscr S_\infty}N_S.$ It follows that $S_k\in\mathscr S_\infty$ for all $k\ge N'=\max\{N,K_0\}$.
Therefore, $\lVert \z_k-v\rVert<\varepsilon$ for all $k\ge N'$. This proves $\z_k\longrightarrow v.$

We show that the trajectory crosses infinitely many vertical switching segments and infinitely many horizontal switching segments. Suppose, to the contrary, that there are only finitely many horizontal switches. After the last horizontal switch, \(q(t)\) remains in one fixed horizontal strip \(Q_j\), unless a cell vertex is reached. We choose \(T\) after the last horizontal switch, so that \(x(q_t)=x_j\) for all \(t\ge T\). The first coordinate then satisfies $\dot p=x_j-p$
throughout every later cell and also at every vertical switching instant. Thus, \(p_t=x_j+\bigl(p_T-x_j\bigr)e^{-(t-T)}\) is monotone and can cross each of the finitely many vertical switching lines at most once. Only finitely many vertical switches can occur, contradicting the assumption of infinitely many switches. Therefore, horizontal switches occur infinitely often. Similarly, vertical switches also occur infinitely often.

Hence, \(\mathscr S_\infty\) contains at least one vertical segment and at least one horizontal segment. Let $S^v\subset\{p=p_i\},\ S^h\subset\{q=q_j\}$
be recurrent vertical and horizontal segments. Since $v\in\overline{S^v}\cap\overline{S^h},$ $v=(p_i,q_j)$
for some $i,j$.

Finally, between \(\z_k\) and \(\z_{k+1}\), the trajectory is a line segment. Every point on that piece is a convex combination of its two endpoints, so
\[
\sup_{t_k\le t\le t_{k+1}}\|\z_t-v\|
\le
\max\{\|\z_k-v\|,\|\z_{k+1}-v\|\}.
\]
Since $t_k\to\infty$, this proves $\z_t\longrightarrow(p_i,q_j)$ as $t\to\infty$. Thus, every infinitely switching trajectory converges to $(p_i,q_j)$ for some \(i,j\).

\subsection{Proof of Theorem \ref{thm:2by2_lsc}}
\label{app:pf_2by2_lsc}

We fix a vertex $v=(p_i,q_j)$. Let \(x_j\) be the best response below \(q_j\), and let \(y_i\) be the best response to the left of \(p_i\). We recall that
$x_{j+1}=1-x_j,\ y_{i+1}=1-y_i.$ Within each of four neighboring rectangles of $v$, the best-response vector is constant:
\[
\begin{array}{c|cc}
 & p<p_i & p>p_i\\
\hline
q>q_j & (1-x_j,y_i) & (1-x_j,1-y_i)\\
q<q_j & (x_j,y_i) & (x_j,1-y_i)
\end{array}.
\]

\begin{lem}
\label{lem:vtx_cls}
    Every vertex $(p_i,q_j)\in \mathcal{V}$ is either (asymptotically) stable or saddle. In particular, $v$ is stable if \(x_j\neq y_i\), and saddle if \(x_j=y_i\).
\end{lem}
\begin{proof}
\textbf{Case 1: \(x_j\neq y_i\).} We consider first $(x_j,y_i)=(1,0)$. The four cell directions circulate around \(v\):
\[
\mathrm{SW}:\searrow,\qquad \mathrm{SE}:\nearrow,\qquad \mathrm{NE}:\nwarrow,\qquad \mathrm{NW}:\swarrow.
\]
The case \((x_j,y_i)=(0,1)\) has the opposite rotation.

The local return calculation in the proof of
Lemma~\ref{lem:nonaccumulation} gives
$0<r_{k+1}<r_k$ and $r_k\to0$ by
Equations~\eqref{eqt:local_spiral_return}--\eqref{eqt:local_spiral_radii}.
The intermediate offsets in Equation~\eqref{eqt:local_spiral_offsets}
(or its reflection for the opposite rotation) tend to zero as well,
so the entire four-segment loop converges to $(p_i,q_j)$.
These continuous ray-intersection maps also show that sufficiently
nearby nonstationary trajectories reach the lower section while
remaining close to the vertex, and all subsequent turns remain close because their
section offsets decrease. Together with $t_k\to\infty$ from
Lemma~\ref{lem:nonaccumulation}, this proves local asymptotic stability.

\textbf{Case 2: \(x_j=y_i\).} First suppose $x_j=y_i=1$.
In the southwest and northeast rectangles, both coordinates point toward \((p_i,q_j)\). In the other two rectangles, both point away.
It follows that the local stable branches in four neighboring rectangles of \((p_i,q_j)\) are
\begin{equation}
    \label{eqt:stable_branch}
    \left\{p_{i-1}<p<p_i,\ q_{j-1}<q<q_j:\ \frac{1-p_i}{1-p}=\frac{1-q_j}{1-q}\right\}, \quad \left\{p_{i+1}>p>p_i,\ q_{j+1}>q>q_j:\ \frac{p_i}{p}=\frac{q_j}{q}\right\}.
\end{equation}
For example, in the southwest rectangle,
\[
1-p_t=(1-p_0)e^{-t},\qquad 1-q(t)=(1-q_0)e^{-t},
\]
so the trajectory reaches \((p_i,q_j)\) in finite time $\log\frac{1-p_0}{1-p_i}$ and then stops. If the initial point does not lie on either branch in Equation~\eqref{eqt:stable_branch}, one switching segment is reached first. The trajectory then switches to one of the two outgoing rectangles and moves away from \((p_i,q_j)\).

For  $x_j=y_i=0$, the local stable branches are instead
\[
\left\{p_{i+1}>p>p_i,\ q_{j-1}<q<q_j:\ \frac{p_i}{p}=\frac{1-q_j}{1-q}\right\}, \quad \left\{p_{i-1}<p<p_i,\ q_{j+1}>q>q_j:\ \frac{1-p_i}{1-p}=\frac{q_j}{q}\right\}.
\]
In either case, the local stable set consists of the two corresponding branches together with the vertex $(p_i,q_j)$.
Thus, $x_j= y_i$ implies $(p_i,q_j)$ is a saddle.

For such a saddle $v=(p_i,q_j)$, we continue its two local stable
branches backward using the cell equations and the unique
continuation through regular switching points. This is a
geometric continuation, not an inverse of the equilibrium-stopped
evolution. If a continuation reaches the outer boundary, we stop
there. If it reaches another saddle $w$, we stop there and exclude
$w$ from the stable set of $v$: the stopping convention gives
$\phi_t(w)=w$ for all $t\ge0$. A stable cell vertex cannot be an
upstream endpoint: its local backward return map
$P^{-1}(r)=Cr/(C-r)>r$ expands instead of approaching zero.
A return to $v$ would produce a forbidden closed forward chain.

Let $\Gamma_v^1,\Gamma_v^2$ denote these maximal backward
continuations, with $v$ omitted, outer-boundary endpoints
included, and upstream saddle endpoints excluded.
Every nonconstant trajectory converging to $v$ eventually
lies on one of its two local incoming branches and reaches
$v$ in finite time. Uniqueness away from cell vertices
therefore gives
\[
\{\z\in[0,1]^2:\phi_t(\z)\to v\}
=\{v\}\cup\Gamma_v^1\cup\Gamma_v^2.
\]
On every compact time interval of its definition, each
backward continuation is Lipschitz, since the cell velocities
are bounded. A countable exhaustion by such intervals shows
that both branches, and hence the stable set, have
planar Lebesgue measure zero.
\end{proof}

We note also that an infinitely switching trajectory cannot converge to a saddle. Sufficiently near a saddle, every point not on its stable branches leaves a small neighborhood after at most one segment crossing, while a point on a stable branch reaches the saddle in finite time. Hence, the $(p_i,q_j)$ obtained in the infinite-switch argument must satisfy $x_j\neq y_i$.

The saddle stable set and its complement are
\[
\Sigma:=\bigcup_{b\in\mathcal V_{\mathrm{sad}}}W^s(b),
\qquad
\Omega:=\cX\setminus\Sigma.
\]
\paragraph{Basins of corner equilibria.}
For a corner equilibrium $c\in \mathcal C_\rho$, a nonempty attraction basin exists. The best-response vector in the cell
adjacent to such a corner equals $c$. Hence, a sufficiently small relative
rectangle $U_c\subset \cX$ about $c$ is positively invariant and, for every
$\z\in U_c$, $\phi_t(\z)=c+(\z-c)e^{-t}\longrightarrow c.$

If $c\notin\mathcal C_\rho$, at least one component of the selected
best-response vector at $c$ differs from the corresponding component of
$c$. The endpoint assumptions make that response constant in a relative
neighborhood of $c$, where the corresponding coordinate is driven toward
the opposite action. Thus, the trajectory starting at $c$ immediately
leaves it, and no trajectory can converge to it. Therefore, $c$ is not an
equilibrium and its stable set is empty.

\begin{lem}
\label{lem:boa}
The connected components\footnote{The connected components of a set are its maximal (with respect to set inclusion) connected subsets.} of \(\Omega\) are the attraction basins $B(a), a\in\mathcal E_{\mathrm{st}}.$
In particular, every connected component of \(\Omega\) contains exactly one stable equilibrium.
\end{lem}
Lemma \ref{lem:boa} entails that after the corner components are removed, every remaining component contains exactly one stable vertex.

To prove Lemma~\ref{lem:boa}, we first establish openness of the attraction basins.
\begin{lem}
\label{lem:rel_open}
    Let \(a \in \cE_{\mathrm{st}}\). Suppose there exists a relatively open neighborhood
$A_a\subset \cX$ such that $a\in A_a,$ and $\phi_t(A_a)\subseteq A_a$ for all $t\ge0$,
and $\phi_t(\z)\to a$ as $t\to\infty$ for every $\z\in A_a$. Then its global basin
$B(a)$ is relatively open in \(\cX\).
\end{lem}

\begin{proof}[Proof of Lemma \ref{lem:rel_open}]
Let $x\in B(a)$. If $x\in A_a$, then $A_a$ itself is a
relatively open neighborhood of $x$ contained in $B(a)$.
Otherwise, we choose a finite time $T>0$ such that
$\phi_T(x)\in A_a$, the endpoint is not on a switching line,
and the trajectory prefix
$\{\phi_t(x):0\le t\le T\}$ encounters no cell vertex.
Such a choice is possible: a trajectory converging to $a$
cannot encounter another cell vertex, because it would
stop there; if it reaches $a$ itself, it enters the open
neighborhood $A_a$ strictly before that time.

This compact prefix has only finitely many regular,
transverse switching crossings. The explicit cell flows
and their transverse crossing maps depend continuously
on the initial condition. Composing them along this finite
prefix therefore gives continuous dependence of its
endpoint at $x$. If $x$ is a regular switching point,
the two one-sided descriptions have the same limit:
the common nonzero normal velocity makes the time spent
in any additional initial cell tend to zero.
The same argument holds relative to $\cX$ at an outer-boundary
starting point.

Since $A_a$ is relatively open, there is a relatively open
neighborhood $U$ of $x$ such that $\phi_T(U)\subset A_a$.
For every $y\in U$, the semigroup property and the attraction
property of $A_a$ imply $\phi_t(y)=\phi_{t-T}(\phi_T(y))\longrightarrow a.$ Thus, $U\subset B(a)$, proving that $B(a)$ is relatively open.
\end{proof}

\begin{proof}[Proof of Lemma \ref{lem:boa}]
We first prove that the complement of the saddle stable sets is the union of the attraction basins
\[
\Omega
=
\bigcup_{a\in\mathcal E_{\mathrm{st}}}B(a),
\]
where the union is disjoint. Let \(z\in\Omega\). By Theorem~\ref{thm:ct_conv_2by2} and the definition of $\Omega$, $\lim_{t\to\infty}\phi_t(z)$ exists and is a stable equilibrium. Conversely, if \(z\in B(a)\) for \(a \in \mathcal E_{\mathrm{st}}\), then \(z\) cannot converge to a saddle. Therefore, \(z\in\Omega\). Finally, distinct basins are disjoint.

We show that \(\Omega\) is forward invariant. Let \(z\in\Omega\). We claim that $\phi_t(z)\in\Omega$ for all $t\ge0$. Suppose otherwise that, for some \(T\ge0\), $\phi_T(z)\in\Sigma.$ Then there exists a saddle \(s\in\mathcal V_{\mathrm{sad}}\) such that $\phi_T(z)\in W^s(s).$ It follows that
\[
\lim_{t\to\infty}\phi_{t+T}(z)
=
\lim_{t\to\infty}\phi_t(\phi_T(z))
=
s \in \mathcal V_{\mathrm{sad}}.
\]
This contradicts \(z\in\Omega\). Thus, \(\Omega\) is forward invariant.

We next show that every connected component of \(\Omega\) lies in one attraction basin. Let \(D \subset \Omega\) be a connected component of \(\Omega\). Since $\Omega
=
\bigcup_{a\in\mathcal E_{\mathrm{st}}}B(a)$ and the union is disjoint,
\[
D=\bigcup_{a\in\mathcal E_{\mathrm{st}}}
\bigl(D\cap B(a)\bigr),
\]
where the union is disjoint. By Lemma \ref{lem:rel_open}, for each \(a\in\mathcal E_{\mathrm{st}}\), the set \(B(a)\) is relatively open in \(\cX\), so $D\cap B(a)$ is relatively open in \(D\). These sets are pairwise disjoint. Since the set of stable equilibria is finite, if \(D\) intersected two distinct basins, say \(B(a_1)\) and \(B(a_2)\), then $D\cap B(a_1)$
and $D\cap\bigcup_{a\neq a_1}B(a)$ would be two disjoint, nonempty, relatively open subsets whose union is \(D\). This contradicts the connectedness of \(D\), so there exists a unique \(a_D\in\mathcal E_{\mathrm{st}}\) such that $D=D\cap B(a_D)$, i.e., $D\subseteq B(a_D).$ The limit map $L:\Omega\to\mathcal E_{\mathrm{st}}$, defined as $L(z):=\lim_{t\to\infty}\phi_t(z),$
is constant on each connected component of \(\Omega\).

We choose any \(z\in D\). We have $\phi_t(z)\longrightarrow a_D.$ The entire forward orbit remains in \(\Omega\). Moreover, for each \(T>0\), the set $\{\phi_t(z):0\le t\le T\}$ is connected, lies in \(\Omega\), and contains \(z\in D\). By maximality of the connected component \(D\), we have $\phi_t(z)\in D, \forall t\ge0$. Since connected components are closed relative to the original space,  \(D\) is closed in \(\Omega\). Since
$\phi_t(z)\in D, \phi_t(z)\to a_D,$
and \(a_D\in\Omega\), it follows that $a_D\in D.$ Every connected component contains its attracting equilibrium.

We now prove the reverse inclusion $B(a_D)\subseteq D$. Let $y\in B(a_D),$ and let \(D_y\) be the connected component of \(\Omega\) containing \(y\). Repeating the preceding argument for the trajectory starting from \(y\), we obtain $a_D\in D_y.$
But $a_D\in D.$ Two connected components of \(\Omega\) that have a common point must be equal, so $D_y=D.$
Therefore, \(y\in D\). Since \(y\in B(a_D)\) was arbitrary, $B(a_D)\subseteq D$,  which gives $D=B(a_D)$. Every connected component of \(\Omega\) is exactly the basin of one stable equilibrium.

\end{proof}

Let $\Omega_{\mathrm{int}}
:=
\Omega
\setminus
\bigcup_{c\in\mathcal C_\rho}B(c).$
Every component of \(\Omega_{\mathrm{int}}\) is a component of \(\Omega\) that is not a corner basin. By Lemma \ref{lem:boa}, it equals \(B(a)\) for one stable equilibrium $a\in\mathcal E_{\mathrm{st}}.$
Since it is not a corner basin, $a\in\mathcal V_{\mathrm{st}}.$
Consequently, each remaining region separated by the global saddle separatrices contains exactly one stable cell vertex. Moreover, that region is exactly the full attraction basin of that stable vertex.

\begin{rmk}[Basin boundaries]
\label{rmk:basin}
By Lemma \ref{lem:boa} and the openness of the attraction basins,
$\Sigma=\cX\setminus\bigcup_{a\in\mathcal E_{\mathrm{st}}}B(a)$
is closed, and $\partial_{\cX} B(a)\subset\Sigma$ for every
stable equilibrium $a$. Conversely, the saddle stable sets have planar measure zero,
and there are finitely many saddles, so $\Sigma$ has empty
relative interior in $\cX$. For any $z\in\Sigma$, we choose
$\z_k\in \cX\setminus\Sigma$ with $\z_k\to z$.
Because there are finitely many attraction basins, a subsequence
belongs to one fixed basin $B(a)$. Then
$z\in\overline{B(a)}\setminus B(a)=\partial_{\cX} B(a)$.
Consequently,
\[
\Sigma=\bigcup_{a\in\mathcal E_{\mathrm{st}}}\partial_{\cX} B(a).
\]
This argument does not require the individual stable sets
to be closed or their branches to terminate only at the
outer boundary.
\end{rmk}

We now construct explicit attracting neighborhoods of stable vertices for use in the discrete-time analysis.
Let \(v=(p_i,q_j)\) be a stable equilibrium. We choose \(r_0>0\)
sufficiently small that the local four-crossing calculation in
Equations~\eqref{eqt:local_spiral_offsets}--\eqref{eqt:local_spiral_return}
applies and the loop constructed below, together with its enclosed
region, lies in $(p_{i-1},p_{i+1})\times(q_{j-1},q_{j+1})$.
This neighborhood meets only the four incident cells and contains
no equilibrium other than \(v\).
Let \(\gamma_{r_0}\) be the trajectory segment making one complete turn from
$z(r_0):=(p_i,q_j-r_0)$ to $z(P(r_0))=(p_i,q_j-P(r_0)),$
where $P(r_0):=\frac{Cr_0}{C+r_0}<r_0.$
Join the two endpoints by the vertical segment
$I_{r_0}:=\{(p_i,q):q\in[q_j-r_0,q_j-P(r_0)]\}$.
The curve $\gamma_{r_0}\cup I_{r_0}$ is a simple closed curve
surrounding \(v\). Along \(\gamma_{r_0}\), the vector field is tangent
to the boundary except at crossing points. Along \(I_{r_0}\),
the vector field points into the bounded component.
The selected flow continues uniquely through regular switching points and
remains at \(v\). Lemma~\ref{lem:nonaccumulation} excludes accumulation
of switching times. Thus, the bounded component of
$\gamma_{r_0}\cup I_{r_0}$, denoted $A_v(r_0)$, is a positively
invariant open neighborhood of \(v\). The local contraction in
Equation~\eqref{eqt:local_spiral_radii} then gives convergence to \(v\)
as $t\to\infty$ for every trajectory starting in \(A_v(r_0)\).

Consequently, if a trajectory does not converge to a corner and does not lie on the stable set of a saddle, then it must converge to some stable \((p_i,q_j)\). Since \(A_v(r_0)\) is an open neighborhood of \(v\), there is a finite time \(T\) such that
$\z_t\in A_v(r_0), \forall t\ge T.$

\section{Proofs for Section \ref{sec:dt}}
\label{app:dt}

\subsection{Proof of Theorem \ref{thm:dt_conv_2by2}}

The proof uses the convergence and landscape results of continuous-time dynamics from Theorem ~\ref{thm:ct_conv_2by2} and~\ref{thm:2by2_lsc}. The complement of the stable sets of the saddle vertices is the disjoint union of the open basins of the stable equilibria. The saddle stable sets are one-dimensional separatrices that form the basin boundaries. We also use the fact, proved in the continuous-time argument, that there is no nonconstant closed chain of forward cell arcs. All estimates below are uniform over admissible pure-action tie choices.

We begin with the comparison estimate used throughout the proof.
For a step from $\z_t$ to $\z_{t+1}$ crossing a switching segment $S$,
we denote the \emph{virtual discrete crossing point} by the intersection
\[
\widehat\xi:=\z_t+\theta(\z_{t+1}-\z_t)\in S,
\qquad 0<\theta\le1.
\]
The associated \emph{actual crossing iterate} is the step endpoint
$w:=\z_{t+1}$, which need not lie on $S$.
An exact landing on $S$ has $\theta=1$ and $w=\widehat\xi$. We count it once, without counting subsequent steps along or departing
from the same segment as new crossings of that segment.
The notation $\xi$ without a hat denotes the crossing point of the
comparison continuous trajectory. Subscripts $\ell$ enumerate successive
crossings, rather than discrete periods.

Suppose that $M>N\ge1$ and $\vc_k=c\in\{0,1\}^2$, for $k=N,\ldots,M-1.$ Then $\z_{k+1}-c
=
(1-\gamma_{k+1})(\z_k-c),$
and therefore
\begin{equation*}
\z_M-c
=
\prod_{k=N}^{M-1}(1-\gamma_{k+1})(\z_N-c)
=
\frac{N+1}{M+1}(\z_N-c).
\end{equation*}
Thus, the discrete iterates and the continuous-time trajectory lie on the same ray within a cell. A geometric discrepancy can arise when a discrete step crosses a switching line before the best response is updated. Since
\begin{equation*}
\|\z_{t+1}-\z_t\|_\infty
=
\gamma_{t+1}\|\vc_t-\z_t\|_\infty
\le
\gamma_{t+1},
\end{equation*}
if a step at some period $t\ge N$ crosses a switching segment at the virtual crossing point $\widehat\xi$, then
\begin{equation*}
\|\z_{t+1}-\widehat\xi\|_\infty
\le \|\z_{t+1}-\z_t\|_\infty \le
\gamma_{t+1}
\le
\frac{1}{N+1}.
\end{equation*}
Thus, every overshoot is $O((N+1)^{-1})$. A bound on one overshoot is not enough because an error can move every later crossing point. The next lemma controls the accumulated error along a fixed finite itinerary. It assumes that every crossing lies in the relative interior of a switching segment and hence stays away from cell vertices.

\begin{lem}[Finite-itinerary comparison]
\label{lem:uni_comp}
Consider a continuous trajectory segment following a finite itinerary $R_0,S_1,R_1,\ldots,S_L,R_L,$ where each crossing belongs to the relative interior of the corresponding switching segment and no vertex is encountered. Let $K\subset R_0$ be a compact family of initial points corresponding to this itinerary. Then there exist an open neighborhood $K\subset U\subset R_0$ on which the trajectories follow the same itinerary, a sufficiently small $\delta_0>0$, finite constants $A,C$, and an integer $N_0\ge1$ such that the following holds.

If $N\ge N_0$ and the discrete iterates and the continuous-time trajectory start from the same $\widehat\xi_0=\z_N=\xi_0\in K$, then the discrete iterates follow the same itinerary. Let $\xi_\ell$ and $\widehat\xi_\ell$ be the continuous and virtual discrete crossing points on $S_\ell$, respectively. Then $\max_{1\le \ell\le L}
\|\widehat\xi_\ell-\xi_\ell\|_\infty
\le
\frac{C}{N+1}.$
More generally, if $N\ge N_0$, $\xi_0\in K$, $\widehat\xi_0=\z_N\in U$, and $\|\widehat\xi_0-\xi_0\|_\infty\le\delta$ for $0\le\delta\le\delta_0$,
then the discrete iterates follow the same itinerary and
\begin{equation*}
\max_{1\le \ell\le L}
\|\widehat\xi_\ell-\xi_\ell\|_\infty
\le
A\delta+\frac{C}{N+1}.
\end{equation*}
The neighborhood $U$ and constants $\delta_0,A,C,N_0$ are chosen uniformly over $K$, independently of the reference point, the allowed perturbation, and the starting period $N\ge N_0$.
\end{lem}

The proof of Lemma \ref{lem:uni_comp} is deferred to Appendix \ref{app:pf_lem}.
Lemma~\ref{lem:uni_comp} only compares finitely many crossings. Convergence to a stable vertex usually requires infinitely many turns, so an additional trapping argument is needed. The next lemma shows that late overshoots cannot push the discrete iterates out of a small attracting region.

\begin{lem}[Local capture near a stable equilibrium]
\label{lem:local_att}
Let $a\in\cE_{\text{st}}$. There exist relatively open neighborhoods $U_a^-\Subset U_a^+\subset B(a)$ and $N_a<\infty$ such that
\begin{equation}
N\ge N_a,\quad \z_N\in U_a^-
\quad\Longrightarrow\quad
\z_t\in U_a^+
\ \ \forall t\in\mathbb N,\ t\ge N,
\qquad
\lim_{t\to\infty}\z_t=a.
\end{equation}
\end{lem}

The proof of Lemma \ref{lem:local_att} is deferred to Appendix \ref{app:pf_lem}.
The local spiral calculation is the part of the continuous geometry that plays the role of the inward spiral in the argument of \citet{metrick1994fictitious}. Lemma~\ref{lem:local_att} gives capture only after the discrete iterates enter $U_a^-$. To use this fact for a limit point anywhere in $B(a)$, we must first follow the finite continuous-time trajectory segment from that limit point to $U_a^-$. This is the role of Lemma~\ref{lem:uni_comp} in the following lemma.

\begin{lem}[Global basin capture]
\label{lem:global_basin}
Let $a\in\cE_{\text{st}}$ and let $K\subset B(a)$ be compact. Then there exist a relatively open neighborhood $K\subset O_K\subset B(a)$ and $N_K<\infty$ such that for $N\ge N_K,\ \z_N\in O_K$, we have $\z_t\to a$ as the iteration index $t\to\infty$. Consequently, if $\omega_d(\z_0)\cap B(a)\neq\varnothing$, then $\z_t\to a$ as $t\to\infty$.
\end{lem}

The proof of Lemma \ref{lem:global_basin} is deferred to Appendix \ref{app:pf_lem}.
Lemma~\ref{lem:global_basin} shows that a nonconvergent sequence of discrete iterates cannot have a limit point in the interior of a basin of a stable equilibrium. The only points not covered are on saddle separatrices. Near a saddle, Lemma~\ref{lem:uni_comp} does not apply through the vertex because two switching lines are reached simultaneously. We therefore need a local description of what a small overshoot can do. The next lemma says that a late approach along an incoming branch has only two outcomes: either convergence to the saddle or departure close to one of its outgoing branches.

\begin{lem}[Passage near a saddle]
\label{lem:passage}
Let $s \in \V_{\mathrm{sad}}$. Choose a sufficiently small rectangle $Q_s$ centered at $s$, containing no switching lines other than
$p=p_i, q=q_j.$ The two local stable branches meet $\partial Q_s$ at two incoming gates $\iota_s^1,\iota_s^2.$ The two outgoing cell rays starting geometrically from $s$ meet $\partial Q_s$ at two outgoing gates $o_s^1,o_s^2.$ Suppose that the discrete process enters $Q_s$ at periods $N_k$ with entry points $\eta_k$, where $N_k\to\infty$ and $\eta_k\to\iota_s^\ell$ as $k\to\infty$. Then, after passing to a subsequence, one of the following occurs: either $\z_t\to s$ as the iteration index $t\to\infty$, or the process exits $Q_s$ at points $\zeta_k$ satisfying $\zeta_k\to o_s^r$ as $k\to\infty$ for one fixed $r\in\{1,2\}$.
\end{lem}

The proof of Lemma \ref{lem:passage} is deferred to Appendix \ref{app:pf_lem}.
Lemma~\ref{lem:passage} controls one saddle visit, but it allows the outgoing branch of one saddle to lie on the stable set of another. To rule out an endless sequence of such visits, we record the possible saddle-to-saddle transitions in a finite directed graph. A directed cycle would produce a closed chain of forward cell arcs, which is excluded by the continuous-time geometry.

\begin{lem}[Acyclicity of the saddle-transition graph]
\label{lem:acyclic}
Let $\mathcal V_{\mathrm{sad}}$ denote the finite set of saddle vertices. For every outgoing gate $o_s^r$, Theorem \ref{thm:2by2_lsc} implies that, since $o_s^r$ is not an equilibrium, either $o_s^r\in B(a)$ for some stable equilibrium $a$, or $o_s^r\in W^s(s')$ for another saddle $s'$. In the second case, introduce a directed edge $s\longrightarrow s'.$
Then the resulting finite directed graph $G_{\mathrm{sad}}$ contains no directed cycle.
\end{lem}

The proof of Lemma \ref{lem:acyclic} is deferred to Appendix \ref{app:pf_lem}.
The preceding lemmas now cover all parts of the continuous-time landscape. A limit point in an attraction basin gives convergence by Lemma~\ref{lem:global_basin}. A limit point on a saddle separatrix leads, by Lemma~\ref{lem:passage}, either to convergence at a saddle or to an outgoing branch. Lemma~\ref{lem:acyclic} ensures that, unless the process converges to a saddle, repeating this argument reaches an attraction basin after finitely many saddle transitions. The next lemma collects these conclusions and reduces the rest of the proof to a finite-set argument.

We set $\mathcal E:=\mathcal C\cup\mathcal V$, the finite set of all corners and cell vertices. This set includes corners that are not $\rho$-NE.

\begin{lem}[Location of the discrete limit set]
\label{lem:lim}
Let $\omega_d(\z_0)$ be the limit set of the sequence of discrete iterates starting from $z_0$, defined by
\[
\omega_d(\z_0):=
\left\{w:\exists\,(t_k)_{k\ge1}\subset\mathbb N,\quad
t_k\to\infty\ \text{and}\ \z_{t_k}\to w
\ \text{as }k\to\infty\right\}.
\]
Then $\omega_d(\z_0)\subset\mathcal E.$
\end{lem}

The proof of Lemma \ref{lem:lim} is deferred to Appendix \ref{app:pf_lem}.

It remains to show that the finite set $\omega_d(\z_0)\subset\mathcal E$ contains a single point. This last step uses only the vanishing increments of the discrete process. From the step-size estimate, $\|\z_{t+1}-\z_t\|_\infty
\le
\gamma_{t+1}
\longrightarrow0.$
Since $\cX$ is compact, $\omega_d(\z_0)$ is nonempty and compact. We claim that $\omega_d(\z_0)$ is connected. Suppose otherwise that
$\omega_d(\z_0)=A\cup B,$ where $A$ and $B$ are nonempty disjoint compact sets. Then $d:=\operatorname{dist}(A,B)>0.$ We choose $0<\varepsilon<\frac d3.$ Since $\omega_d(\z_0)$ is the set of all subsequential limits, $\operatorname{dist}\bigl(\z_t,\omega_d(\z_0)\bigr)\to0.$
Hence, for all sufficiently large $t$, $\z_t\in
N_\varepsilon(A)\cup N_\varepsilon(B).$
Both neighborhoods are visited infinitely often, because both $A$ and $B$ contain limit points. Their mutual distance is at least $d-2\varepsilon>0.$ On the other hand, for all sufficiently large $t$, $\|\z_{t+1}-\z_t\|_\infty
<
d-2\varepsilon.$
Thus, the sequence cannot pass from $N_\varepsilon(A)$ to $N_\varepsilon(B)$, or conversely, at sufficiently large times. This is a contradiction. Therefore, $\omega_d(\z_0)$ is connected. By Lemma~\ref{lem:lim}, $\omega_d(\z_0)\subset\mathcal E,$ and $\mathcal E$ is finite. The only connected subsets of a finite set are singletons. Hence, $\omega_d(\z_0)=\{z_\infty\}$ for some $\z_\infty\in\mathcal E.$ A bounded sequence with a unique limit point converges to that point. Therefore, $\z_t\longrightarrow z_\infty\in\mathcal C\cup\mathcal V.$

Since $\mathcal V=\cP\times\cQ\subset\mathcal E_\rho$, it remains to exclude $\z_\infty\in\mathcal C\setminus\mathcal C_\rho$. At such a corner, some player has a strict best response opposite to its coordinate. By sign constancy on the endpoint intervals, this response is constant near the corner. The coordinatewise identity above would then make that coordinate converge to the opposite action, contradicting $\z_t\to \z_\infty$. Thus, $\z_\infty\in\mathcal E_\rho$, which proves Theorem~\ref{thm:dt_conv_2by2}.

\subsection{Proofs of technical lemmas}
\label{app:pf_lem}

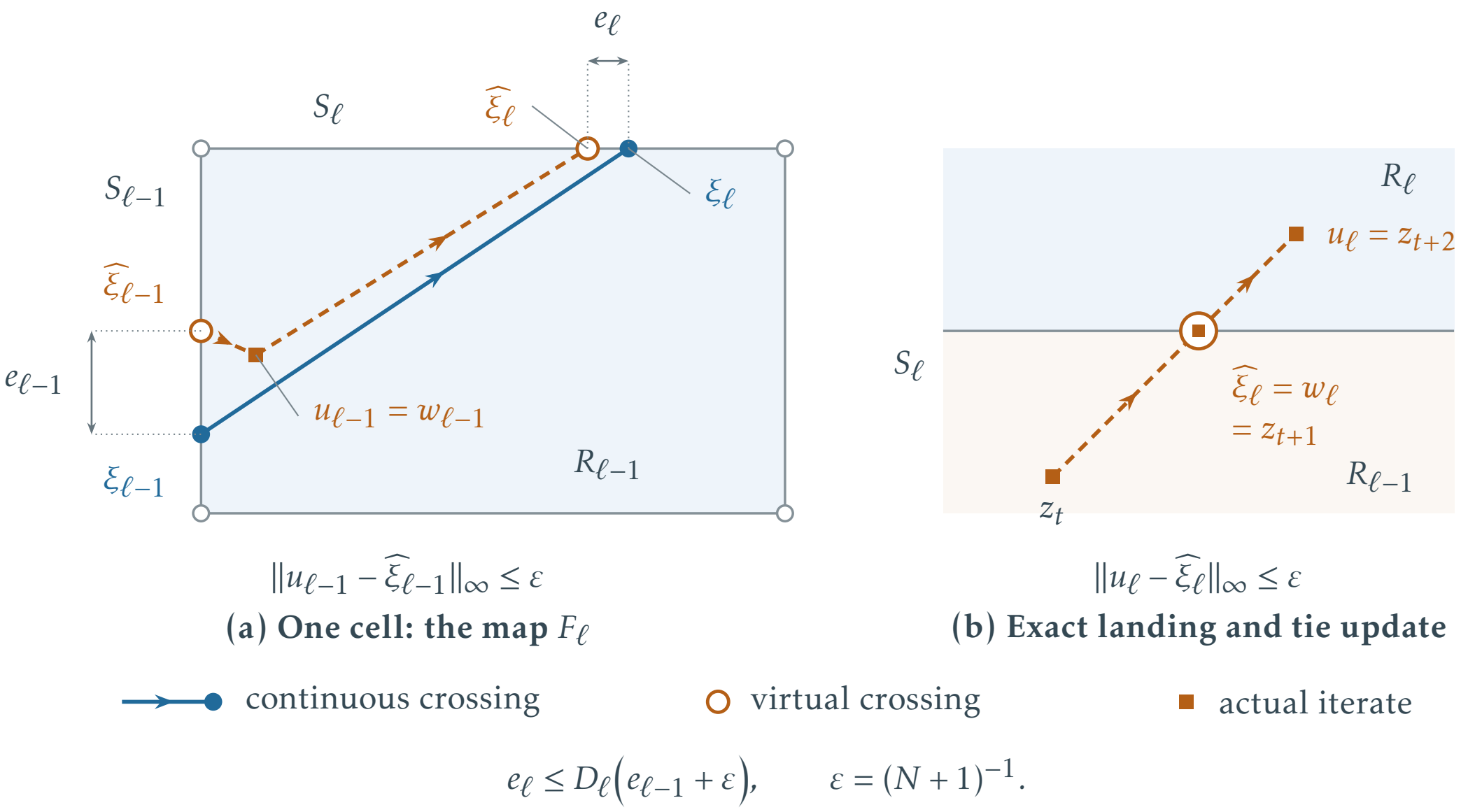
\begin{figure}[!htb]
\centering
\begingroup
\definecolor{itRef}{HTML}{216A9A}
\definecolor{itDisc}{HTML}{B45F16}
\definecolor{itInk}{HTML}{344853}
\definecolor{itWash}{HTML}{EEF4FA}
\begin{tikzpicture}[
  x=1.15cm,y=1.15cm,font=\footnotesize,text=itInk,
  boundary/.style={draw=itInk!60,line width=.65pt},
  reference/.style={draw=itRef,line width=1.05pt,
    postaction={decorate},decoration={markings,
      mark=at position .57 with {\arrow{Stealth[length=1.8mm]}}}},
  discrete/.style={draw=itDisc,line width=1.05pt,densely dashed,
    postaction={decorate},decoration={markings,
      mark=at position .58 with {\arrow{Stealth[length=1.8mm]}}}},
  ref point/.style={circle,fill=itRef,inner sep=1.7pt},
  virtual point/.style={circle,draw=itDisc,fill=white,
    line width=.9pt,inner sep=2pt},
  iterate/.style={rectangle,fill=itDisc,inner sep=1.9pt},
  vertex/.style={circle,draw=itInk!60,fill=white,
    line width=.65pt,inner sep=1.5pt},
  measure/.style={draw=itInk!75,line width=.5pt,
    {Stealth[length=1.1mm]}-{Stealth[length=1.1mm]}},
  leader/.style={draw=itInk!65,line width=.4pt}
]
  \pgfmathsetmacro{\itRefExit}{8*(3-.65)/(6-.65)}
  \pgfmathsetmacro{\itEntryY}{1.5-.45*3.5/8}
  \pgfmathsetmacro{\itDiscExit}{.45+(3-\itEntryY)*(8-.45)/(6-\itEntryY)}

  \node[font=\footnotesize\bfseries] at (1.7,-.95)
    {\panellabel{fig:finite_itinerary_cell} One cell: the map $F_\ell$};
  \fill[itWash] (0,0) rectangle (4.8,3);
  \draw[boundary] (0,0) rectangle (4.8,3);
  \foreach \x/\y in {0/0,0/3,4.8/0,4.8/3}
    \node[vertex] at (\x,\y) {};
  \node[anchor=east] at (-.14,2.65) {$S_{\ell-1}$};
  \node[anchor=south] at (1.05,3.08) {$S_\ell$};
  \node at (3.35,.4) {$R_{\ell-1}$};

  \coordinate (refin) at (0,.65);
  \coordinate (virtualin) at (0,1.5);
  \coordinate (actualin) at (.45,\itEntryY);
  \coordinate (refout) at (\itRefExit,3);
  \coordinate (virtualout) at (\itDiscExit,3);
  \draw[reference] (refin) -- (refout);
  \draw[discrete] (virtualin) -- (actualin);
  \draw[discrete] (actualin) -- (virtualout);
  \node[ref point] at (refin) {};
  \node[ref point] at (refout) {};
  \node[virtual point] at (virtualin) {};
  \node[virtual point] at (virtualout) {};
  \node[iterate] at (actualin) {};

  \node[anchor=north east,text=itRef] at (-.15,.51) {$\xi_{\ell-1}$};
  \node[anchor=south east,text=itDisc] at (-.15,1.61) {$\widehat\xi_{\ell-1}$};
  \draw[leader] (actualin) -- (.8,.8);
  \node[anchor=west,text=itDisc] at (.8,.8) {$u_{\ell-1}=w_{\ell-1}$};
  \draw[leader] (virtualout) -- (2.72,3.35);
  \node[anchor=east,text=itDisc] at (2.72,3.35) {$\widehat\xi_\ell$};
  \draw[leader] (refout) -- (4.02,2.63);
  \node[anchor=west,text=itRef] at (4.02,2.63) {$\xi_\ell$};

  \draw[measure] (-.9,.65) -- (-.9,1.5)
    node[midway,left=3pt] {$e_{\ell-1}$};
  \draw[leader,densely dotted] (refin) -- (-.9,.65);
  \draw[leader,densely dotted] (virtualin) -- (-.9,1.5);
  \draw[leader,densely dotted] (refout) -- (\itRefExit,3.72);
  \draw[leader,densely dotted] (virtualout) -- (\itDiscExit,3.72);
  \draw[measure] (\itDiscExit,3.72) -- (\itRefExit,3.72)
    node[midway,above=3pt] {$e_\ell$};
  \node at (1.7,-.5)
    {$\|u_{\ell-1}-\widehat\xi_{\ell-1}\|_\infty\le\varepsilon$};

  \node[font=\footnotesize\bfseries] at (8.2,-.95)
    {\panellabel{fig:finite_itinerary_tie} Exact landing and tie update};
  \begin{scope}[shift={(8.2,1.5)}]
    \fill[itWash] (-2.1,0) rectangle (2.1,1.5);
    \fill[itDisc!5] (-2.1,-1.5) rectangle (2.1,0);
    \draw[boundary] (-2.1,0) -- (2.1,0);
    \node[anchor=north east] at (-2.12,-.04) {$S_\ell$};
    \node at (1.65,1.25) {$R_\ell$};
    \node at (1.5,-1.2) {$R_{\ell-1}$};
    \draw[discrete] (-1.2,-1.2) -- (0,0);
    \draw[discrete] (0,0) -- (.8,.8);
    \node[iterate] at (-1.2,-1.2) {};
    \node[below=3pt] at (-1.2,-1.2) {$z_t$};
    \node[virtual point,inner sep=3.4pt] at (0,0) {};
    \node[iterate,inner sep=1.6pt] at (0,0) {};
    \node[anchor=north west,align=left,text=itDisc] at (.15,-.16)
      {$\widehat\xi_\ell=w_\ell$\\$=z_{t+1}$};
    \node[iterate] at (.8,.8) {};
    \node[anchor=west,text=itDisc] at (.93,.75) {$u_\ell=z_{t+2}$};
  \end{scope}
  \node at (8.2,-.5)
    {$\|u_\ell-\widehat\xi_\ell\|_\infty\le\varepsilon$};

  \draw[reference] (-.65,-1.55) -- (.1,-1.55);
  \node[ref point] at (.1,-1.55) {};
  \node[anchor=west] at (.25,-1.55) {continuous crossing};
  \node[virtual point] at (4.25,-1.55) {};
  \node[anchor=west] at (4.4,-1.55) {virtual crossing};
  \node[iterate] at (8.1,-1.55) {};
  \node[anchor=west] at (8.25,-1.55) {actual iterate};
  \node at (4.65,-2.2)
    {$e_\ell\le D_\ell\bigl(e_{\ell-1}+\varepsilon\bigr),
      \qquad \varepsilon=(N+1)^{-1}.$};
\end{tikzpicture}
\endgroup
\caption{Schematic finite-itinerary comparison for $\ell\ge2$:
(a) crossing-error propagation through the cell map $F_\ell$;
(b) one extra update after an exact landing.}
\label{fig:finite_itinerary}
\end{figure}

\begin{proof}[Proof of Lemma \ref{lem:uni_comp}]
For $\ell=1,\ldots,L$, let $F_\ell$ map an entry point to the
intersection of its ray toward the focal corner of $R_{\ell-1}$
with the line supporting $S_\ell$. We define $F_1$ near $K$ and the
later maps on planar neighborhoods of the reference crossings on
$S_{\ell-1}$. Transversality and compactness give uniform local
Lipschitz constants $D_\ell\ge1$ in the infinity norm.
Since the reference crossings stay away from segment endpoints,
there are an open neighborhood $K\subset U\subset R_0$ carrying
the same continuous itinerary and a common margin $\eta>0$ such
that entry errors below $\eta$ preserve the intended exits and
stay within the map neighborhoods. We choose $\eta$ small enough that
a crossing step and one subsequent step, each with infinity norm
at most $\eta$, cannot reach a second switching line near these exits.

We put $\varepsilon:=(N+1)^{-1}$ and let $u_\ell$ be the first
iterate in the open cell $R_\ell$ after the virtual crossing
$\widehat\xi_\ell$. For a strict overshoot, $u_\ell=w_\ell$.
After an exact landing, $w_\ell=\widehat\xi_\ell$ and one further
update enters $R_\ell$:
the normal component of the drift ($x(q)-p$ on a vertical segment,
$y(p)-q$ on a horizontal one) is unique and points across the segment, while
the tie choice affects only the tangential component.
Thus, $u_\ell-\widehat\xi_\ell$ is either the remainder of the
crossing step or one full update. Both have infinity norm at most
$\varepsilon$, so, as long as the geometric margin is respected, $\|u_\ell-\widehat\xi_\ell\|_\infty
\le \varepsilon.$
See Figure~\ref{fig:finite_itinerary_cell} for a strict overshoot and Figure~\ref{fig:finite_itinerary_tie} for an exact landing.

We set $u_0:=\widehat\xi_0=\z_N$ and
$e_\ell:=\|\widehat\xi_\ell-\xi_\ell\|_\infty$, so $e_0\le\delta$.
The telescoping identity shows that the discrete iterates starting from
$u_{\ell-1}$ lie on the exact cell ray and reach or cross its transverse
exit in finitely many steps. Thus,
\[
\widehat\xi_\ell=F_\ell(u_{\ell-1}),
\qquad
\xi_\ell=F_\ell(\xi_{\ell-1}),
\qquad
e_\ell\le D_\ell(e_{\ell-1}+\varepsilon).
\]
Iterating this inequality over the $L$ crossings gives uniform
constants $A\ge1$ and $C\ge0$ with
$e_\ell\le A\delta+C\varepsilon$ for every $\ell$.
We choose $\delta_0>0$ and an integer $N_0\ge1$ so that $A\delta_0+\frac{C+1}{N_0+1}<\eta.$
For $N\ge N_0$, $\widehat\xi_0\in U$, and $\delta\le\delta_0$,
induction now keeps every entry error below $\eta$, including
overshoots and tie updates. Hence, all crossings occur in the
prescribed order and satisfy the claimed bound. Taking $\delta=0$
gives the same-start estimate. All choices are uniform over $K$.
\end{proof}

\begin{proof}[Proof of Lemma \ref{lem:local_att}]
First let $v=(\alpha,\beta)$ be a stable vertex.
We choose a small open rectangle $\mathcal Q$ about $v$ containing
no other switching lines. If necessary, we reflect $p\mapsto1-p$ so that
the local responses are $x=\mathbf 1_{\{q>\beta\}},
\
y=\mathbf 1_{\{p<\alpha\}}.$
We extend this rule to the entire square, allowing the corresponding
response to range over $[0,1]$ at a tie. We define
\[
H(p,q):=
(1-\beta)(\alpha-p)^+
+\beta(p-\alpha)^+
+\alpha(\beta-q)^+
+(1-\alpha)(q-\beta)^+,
\]
where $r^+:=\max\{r,0\}$. Its positive coefficients make $H$
equivalent to the distance from $v$. Along any absolutely continuous
auxiliary solution $\dot \z=(x,y)-z$, off the switching lines, we have
\[
\frac{d}{du}H(p(u),q(u))
=(\beta-y)(x-p)+(x-\alpha)(y-q)
=-H(p(u),q(u)).
\]
The identity holds almost everywhere at ties as well: an absolutely
continuous function has derivative zero almost everywhere on each
level set, so the nonzero normal velocities exclude positive time
on either switching line away from $v$; on $\{z=v\}$, both $H$ and
its derivative vanish almost everywhere. Thus, $H(p(u),q(u))=e^{-u}H(p_0,q_0).$
Small sublevel sets contained in $\mathcal Q$ are therefore
invariant for the original continuous dynamics and lie in $B(v)$.

Let $\widetilde \z_t$ follow the auxiliary discrete recursion with
the same step sizes and arbitrary admissible pure-action tie choices.
We set $\sigma_{N,m}:=\sum_{t=N}^{m-1}\gamma_{t+1}$, and interpolate
linearly by $\overline \z_N(\sigma_{N,m})=\widetilde \z_m$.
On $[0,2]$, these interpolations stay in the square and are $1$-Lipschitz
in the infinity norm. As $N\to\infty$, any sequence therefore has
a uniformly convergent subsequence whose limit solves the auxiliary
inclusion. Indeed, the left-endpoint states differ from the
interpolations by at most $\gamma_{N+1}\to0$, and the response
primitives also have uniformly convergent subsequences. Their limits
have derivatives in $[0,1]^2$ almost everywhere, with the prescribed unique responses
wherever the corresponding limiting coordinate is off its switching
line. Passing to the integral equations proves the claim.
Compactness and the decay identity now give $\eta_N\to0$ such that
\[
\sup_{0\le u\le2}
\left|H(\overline \z_N(u))
-e^{-u}H(\widetilde \z_N)\right|
\le\eta_N,
\]
uniformly over initial states and tie choices: otherwise a
subsequential limit would violate the decay identity.

We choose $h>0$ so small that
$\{\z\in[0,1]^2:H(z)\le2h\}\subset\mathcal Q$, and set
\[
U_v^-:=\{\z\in[0,1]^2:H(\z)<h\},
\qquad
U_v^+:=\{\z\in[0,1]^2:H(\z)<2h\}.
\]
Then $U_v^-\Subset U_v^+\subset B(v)$. For $M(N):=\min\{m>N:\sigma_{N,m}\ge\log2\},$ the divergent step-size sum ensures finiteness, and
$\log2\le\sigma_{N,M(N)}\le\log2+\gamma_{N+1}<2$ for $N\ge1$.
Hence,
\[
\max_{N\le m\le M(N)}H(\widetilde \z_m)
\le H(\widetilde \z_N)+\eta_N,
\qquad
H(\widetilde z_{M(N)})
\le\frac12H(\widetilde \z_N)+\eta_N.
\]
We choose an integer $N_v\ge1$ so that $\eta_N<h/4$ for all $N\ge N_v$.
Each block starting with $H\le h$ then stays in $U_v^+$ and
ends with $H<3h/4$. We couple the original and auxiliary recursions
using the same actions in $\mathcal Q$; they coincide through any
first exit iterate, which these bounds exclude. Starting at
$N\ge N_v$ with $\z_N\in U_v^-$, we iterate the blocks and thereby
keep the original iterates in $U_v^+$.
With $N_0=N$, $N_{k+1}=M(N_k)$, and $b_k=H(z_{N_k})$, we have
$0\le b_k\le h$ and $b_{k+1}\le\frac12b_k+\eta_{N_k}$.
Since $N_k\to\infty$, taking limsup gives $b_k\to0$, and $\max_{N_k\le t\le N_{k+1}}H(\z_t)
\le b_k+\eta_{N_k}\longrightarrow0.$ Thus, $\z_t\to v$. All estimates are uniform over admissible tie
choices, including exact line and vertex hits.
The construction allows arbitrarily small $h$ and applies to the
other stable orientation by undoing the reflection.

For a stable corner $a\in\mathcal C_\rho$, we choose small relative rectangles
$U_a^-\Subset U_a^+$ about $a$ inside its terminal best-response region.
The focal point is $a$, so both rectangles lie in $B(a)$, and
convexity makes $U_a^+$ invariant under the discrete update.
For $t\ge N\ge1$, $\z_t-a=\frac{N+1}{t+1}(\z_N-a)\longrightarrow0.$
\end{proof}

\begin{proof}[Proof of Lemma \ref{lem:global_basin}]
We choose the local capture neighborhood $U_a^-$ from Lemma \ref{lem:local_att}. We fix $z\in K$. If $z\in U_a^-$, we take $V_z:=U_a^-$ and $N_z:=N_a$, and the conclusion follows directly from Lemma \ref{lem:local_att}.
Otherwise, we use the finite-prefix argument below. Since $z\in B(a)$,
$\phi_t(z)\to a$.
Hence, there exists a finite $T_z$ such that $\phi_{T_z}(z)\in U_a^-.$
If necessary, perturb $T_z$ slightly so that the endpoint does not lie on a switching line. Before reaching $U_a^-$, the continuous-time trajectory cannot encounter a saddle  vertex: under the continuous convention it would then remain there and could not converge to $a$. Thus, the finite trajectory segment from $z$ to $U_a^-$ consists of finitely many regular switches. Lemma~\ref{lem:uni_comp} and continuous dependence on initial conditions therefore imply that there exist a neighborhood
$V_z\subset B(a)$
and an integer $N_z$ such that, whenever $N\ge N_z,
\
\z_N\in V_z,$
the discrete iterates follow the corresponding finite continuous-time trajectory segment up to an $O((N+1)^{-1})$ error and enter $U_a^-$.

If $z$ itself is a regular switching point, there are at most two one-sided itineraries. The possible first-step tie-breaking error is again $O((N+1)^{-1})$, so the same conclusion follows after shrinking $V_z$. Once $U_a^-$ is reached, Lemma~\ref{lem:local_att} implies convergence to $a$. The family $\{V_z:z\in K\}$ is an open cover of $K$. By compactness, it admits a finite subcover. Taking the maximum of the corresponding starting times proves the claim.

For the consequence, we choose $z^\ast\in\omega_d(\z_0)\cap B(a).$
Applying the preceding statement to $K=\{z^\ast\}$ gives a capture neighborhood $O_{z^\ast}$. Since some subsequence satisfies $z_{t_k}\to z^\ast, t_k\to\infty,$
one sufficiently late $z_{t_k}$ lies in $O_{z^\ast}$. Hence, the entire subsequent tail converges to $a$.
\end{proof}

\begin{proof}[Proof of Lemma \ref{lem:passage}]
We consider, for example, the saddle type
$x_j=y_i=1$.
Then the southwest and northeast cells are incoming, while the southeast and northwest cells are outgoing. The case $x_j=y_i=0$ is obtained by exchanging the two row actions, which replaces $p$ by $1-p$. Along an incoming cell, the two incident switching lines are reached simultaneously if and only if the initial point lies on the continuous stable branch. Hence, by continuity of the two ray-intersection maps, initial points tending to an incoming gate generate first switching points tending to $s$. We shrink $Q_s$ so that $\min\{p,1-p,q,1-q\}\ge\kappa>0$ throughout $Q_s$.
Thus, both incoming velocity components have magnitude at least
$\kappa$. If a step lands exactly on an incident line or at $s$,
we consider the first subsequent iterate strictly off both incident
lines, if it exists. A tie step leaves every incident line containing
its starting point. If two consecutive iterates lie on incident
lines, they lie on different lines and both are within
$\gamma_{t+1}$ of $s$. Consequently, an infinite run of such landings
converges to $s$, while a finite run of successive line landings
ends within $O(\gamma_t)$ of $s$, with no accumulated error.
An isolated line landing contributes only one additional
$O(\gamma_t)$ step. Any change in the nominal single- or
double-crossing classification therefore occurs within
$O(\gamma_t)$ of $s$.

For iterates strictly off both incident lines, there are two possibilities.

First, suppose exactly one incident switching line is crossed first. Then the discrete process enters an outgoing cell at a point tending to $s$. In an outgoing cell, both coordinates move away from the two incident switching lines. The process therefore leaves $Q_s$ without returning to either switching line, and continuity of the outgoing ray-intersection map yields
$\zeta_k\to o_s^r$
for one of the two outgoing gates.

The remaining possibility is that one discrete step crosses both incident switching lines. In that case, both coordinates of the endpoint differ from those of $s$ by at most the step size, and hence $\|\z_{t+1}-s\|_\infty
\le
\gamma_{t+1}.$
The process then lies in the opposite incoming cell.

After a double crossing or a tie passage returning to an incoming
cell, the iterate is within $O(\gamma_t)$ of $s$. Since each incoming
velocity component has magnitude at least $\kappa$ and
$\gamma_{t+1}=1/(t+2)$, the number of incoming-cell steps before
the next line hit or crossing is uniformly bounded for sufficiently
late $t$; all intervening iterates therefore remain within
$O(\gamma_t)$ of $s$. If an outgoing cell is eventually reached,
the preceding exit argument applies. Otherwise, double crossings
and exact-line landings continue, and the preceding bounds give
$\z_t\to s$.
\end{proof}

\begin{proof}[Proof of Lemma \ref{lem:acyclic}]
Suppose, toward a contradiction, that $s_0\to s_1\to\cdots\to s_r=s_0$
is a directed cycle. For each edge $s_\ell\to s_{\ell+1}$, we concatenate the outgoing cell ray from $s_\ell$ to the corresponding outgoing gate $o_{s_\ell}^{r_\ell}$, and the continuous stable trajectory from $o_{s_\ell}^{r_\ell}$ to $s_{\ell+1}$. By Theorem~\ref{thm:2by2_lsc}, the second trajectory reaches $s_{\ell+1}$ in finite time through finitely many cell crossings. Concatenating all the edges therefore produces a nonconstant closed chain of forward cell-equation arcs. This contradicts Lemma~\ref{lem:no_orbit}, which excludes such geometric chains, including departures from stopped saddle vertices. Hence, $G_{\mathrm{sad}}$ is acyclic.
\end{proof}

\begin{proof}[Proof of Lemma \ref{lem:lim}]
Suppose, toward a contradiction, that $w\in\omega_d(\z_0)\setminus\mathcal E.$
By Theorem \ref{thm:ct_conv_2by2}, there are two possibilities.

Case 1: $w$ belongs to an attraction basin. Then
$w\in B(a)$
for some stable equilibrium $a$. Lemma~\ref{lem:global_basin} immediately gives $\z_t\to a,$ which contradicts $w\neq a$. Hence, this case is impossible.

Case 2: $w$ lies on a saddle separatrix. Then $w\in W^s(s_0)\setminus\{s_0\}$ for some saddle $s_0$. We choose sufficiently small saddle boxes $Q_s$ as in Lemma~\ref{lem:passage}, with pairwise disjoint closures,
and shrink $Q_{s_0}$ further so that $w\notin\overline{Q_{s_0}}$. We choose
$t_k\to\infty,
\
\z_{t_k}\to w.$
The continuous trajectory from $w$ follows a finite regular itinerary until it enters the local box $Q_{s_0}$ through one of its incoming gates. If $w$ is a regular switching point, we use the two
one-sided finite itineraries and the first-step tie estimate
from the proof of Lemma~\ref{lem:uni_comp}. The corresponding
crossing maps extend continuously to $w$, and an exact initial
tie contributes only an $O(\gamma_{t_k+1})$ error. Lemma~\ref{lem:uni_comp} therefore implies that the corresponding late discrete tails enter $Q_{s_0}$ at points
$\eta_k\to\iota_{s_0}^{\ell_0}$. We apply Lemma~\ref{lem:passage}. If the process converges to $s_0$, then
$\omega_d(\z_0)=\{s_0\}$,
contradicting $w\neq s_0$. Otherwise, after passing to a subsequence, the exit points satisfy
$
\zeta_k\to o_{s_0}^{r_0}$.
If
$
o_{s_0}^{r_0}\in B(a)
$
for some stable equilibrium $a$, then Lemma~\ref{lem:global_basin} forces $\z_t\to a$, again a contradiction. Therefore, $o_{s_0}^{r_0}\in W^s(s_1)$ for some saddle $s_1$. Equivalently, $s_0\to s_1$ is an edge of $G_{\mathrm{sad}}$.

Applying Lemma~\ref{lem:uni_comp} to the finite stable-separatrix itinerary from $o_{s_0}^{r_0}$ to $Q_{s_1}$, followed by Lemma~\ref{lem:passage} at $s_1$, yields the same dichotomy. Hence, unless the discrete process has already converged to a saddle or entered an attraction basin, we obtain a directed path $s_0\to s_1\to s_2\to\cdots$ in $G_{\mathrm{sad}}$. But $G_{\mathrm{sad}}$ is finite and acyclic by Lemma~\ref{lem:acyclic}. Thus, the construction must terminate after finitely many saddles at an outgoing gate belonging to an attraction basin. Lemma~\ref{lem:global_basin} then forces convergence to the corresponding stable equilibrium. Every alternative contradicts the assumption that
$w\in\omega_d(\z_0)\setminus\mathcal E$. Therefore, $\omega_d(\z_0)\subset\mathcal E.$
\end{proof}

\section{Convergence in \texorpdfstring{$2 \times n$}{2 x n} Games}
\label{app:conv_2byn}

\subsection{Revisiting convergence of classical FP in \texorpdfstring{$2 \times n$}{2 x n} games}

\subsubsection{An example}
\label{app:exm_fp_2byn}
We first illustrate why convergence of $(p_t,\mathbf a^\top\q_t)$ does not by itself imply convergence of the mixed-strategy profile $(p_t,\q_t)$ under unrestricted mixed best-response selections.
We take column payoffs $G_1(p)=1-p$, $G_2(p)=p$, $G_3(p)=\frac{1}{2}$ and row advantages $\mathbf a=(1,1,-1)^\top$. Every pure strategy has a unique best response, but at $p=\frac{1}{2}$ all columns tie, and every $\q\in\Delta_3$ with $q_3=\frac{1}{2}$ gives a Nash equilibrium. In rescaled time, we set $h_t:=(\sin t-\cos t)/8$ and $\q_t:=(1/4+h_t,1/4-h_t,\frac{1}{2})^\top$. We choose mixed best responses $b_R(t)=\frac{1}{2}\in\operatorname{BR}^1(\q_t)$ and $b_C(t)=(1/4+\sin t/4,1/4-\sin t/4,\frac{1}{2})^\top\in\operatorname{BR}^2(p_t)$. Since $\mathbf a^\top\q_t=0$, the row selection is valid and keeps $p_t=\frac{1}{2}$ through $\dot p=b_R-p=0$; the column selection is valid because all columns tie and satisfies $\dot\q=b_C-\q$. Thus, the full trajectory is periodic within the equilibrium set, although its projection is fixed at $(\frac{1}{2},0)$.

\subsubsection{Proof of Proposition~\ref{prop:fp_2byn}}
\label{app:pf_fp_2byn}
We use the selection convention in Section~\ref{sec:rev_fp}. When the projected state satisfies
$(p,\mathbf a^\top\q)=(x_i,0)$ with $a_i a_{i+1}<0$, we choose
the constant mixed row response $x_i\in\operatorname{BR}^1(\q)$ and mixed column response $\q^{i,*}\in\operatorname{BR}^2(x_i)$.
Under this convention, a projected trajectory that reaches a mixed-equilibrium vertex
remains there thereafter. The full column strategy satisfies
$\dot\q=\q^{i,*}-\q$ and converges to $\q^{i,*}$.

\noindent\textbf{Step 1: Projection onto \((K+1)\times2\) grid.} We denote the row-player payoff advantage as
\[
y:=d^1(\q)=\mathbf{a}^\top \q
=
\sum_{j=1}^n a_jq_j
=
(A\q)_1-(A\q)_2.
\]
Inactive coordinates satisfy $\dot q_j=-q_j$, so their initial mass decays exponentially in rescaled time.
The row player's mixed best-response correspondence in scalar coordinates is
\[
\operatorname{BR}^1(\q)
=
\begin{cases}
\{1\}, & y>0,\\
\left[0,1\right], & y=0,\\
\{0\}, & y<0.
\end{cases}
\]
When \(p\in I_j\), player 2 uniquely plays column \(j\), and hence $\dot \q=\e_j-\q.$ Consequently, $\dot y
=
\mathbf{a}^\top(\e_j-\q)
=
a_j-y.$
Thus, the projected dynamics \(z=(p,y)\) are
\begin{equation}
\label{eqt:prj_dy}
    \dot p
=
\begin{cases}
1-p, & y>0,\\
-p, & y<0,
\end{cases}
\qquad
\dot y=a_j-y,\quad p\in I_j.
\end{equation}
We define the upper and lower cells $R_j^+:=I_j\times(0,\overline a),
\
R_j^-:=I_j\times(\underline a,0),$
where $\underline a:=\min_{1\leq\ell\leq n} a_\ell,
\
\overline a:=\max_{1\leq\ell\leq n} a_\ell.$ Thus, $y(t)\in[\underline a,\overline a]$ for every initial $\q_0\in\Delta_n$. The vector fields in these cells are
\[
F_j^+(p,y):=(1-p,a_j-y),\qquad
F_j^-(p,y):=(-p,a_j-y).
\]
Inside a fixed cell the solution is $\z_t
=
c_j^\pm
+
e^{-(t-t_0)}
\bigl(\z_{t_0}-c_j^\pm\bigr),$
where $c_j^+=(1,a_j), c_j^-=(0,a_j).$ Within each cell, the trajectory traces a straight line segment directed toward its focal point. Figure~\ref{fig:2byn_lsc} shows these directions and how the saddle separatrices change slope when they cross a vertical threshold.

Furthermore, there is no sliding away from the vertices. On \(p=x_i\), the normal velocity is \(1-x_i>0\) above the axis and \(-x_i<0\) below it, independently of which adjacent column is selected. On \(y=0\) inside \(I_j\), the normal velocity is \(a_j\neq0\), independently of the row-player tie-breaking rule. Thus, every trajectory crossing at a regular switching point is transverse.
At a non-NE vertex $v_i=(x_i,0)$, $\dot y$ has the common strict sign
of $a_i,a_{i+1}$ in a sufficiently small neighborhood. Thus, a trajectory
reaching $v_i$ immediately and uniquely continues along $F_{i+1}^+$
if both advantages are positive, or $F_i^-$ if both are negative.

\paragraph{Step 2: which switching vertices represent mixed NE.}

At \(p=x_i\), player 2's mixed best-response set is $\operatorname{BR}^2(x_i)
=
\operatorname{co}\{\e_i,\e_{i+1}\}.$ A mixed strategy supported on these two columns has the form $\q
=
\lambda \e_i+(1-\lambda)\e_{i+1}.$ Player 1 is indifferent if and only if $a^\top \q
=
\lambda a_i+(1-\lambda)a_{i+1}
=
0.$
Such a \(\lambda\in(0,1)\) exists if and only if \(0\) lies strictly between \(a_i\) and \(a_{i+1}\), i.e., $a_i a_{i+1}<0.$ The unique solution is $\lambda
=
\frac{-a_{i+1}}{a_i-a_{i+1}},$
and therefore
\begin{equation}
\label{eqt:qi}
    \q^{i,*}
=
\frac{-a_{i+1}}{a_i-a_{i+1}}\e_i
+
\frac{a_i}{a_i-a_{i+1}}\e_{i+1}.
\end{equation}
Since \(x_i\in(0,1)\), both row actions are used, and both are best responses because $a^\top \q^{i,*}=0.$
Since \(\q^{i,*}\) is supported on the two column best responses at \(x_i\), we obtain a mixed NE $(x_i,\q^{i,*}).$ Conversely, if
$a_i a_{i+1}>0,$ every mixed strategy supported on columns \(i\) and \(i+1\) gives a payoff difference with the same nonzero sign. Hence, no mixed best response of the column player at \(x_i\) can make player 1 indifferent. Therefore, \((x_i,0)\) has no NE preimage. Thus, $(x_i,0)$ is the projection of a mixed NE if and only if $a_i a_{i+1}<0.$

The following steps follow the proof structure of Theorem~\ref{thm:ct_conv_2by2}.
\paragraph{Step 3: exclusion of nonconstant closed chains of forward cell arcs.}
The cellwise proof of Lemma~\ref{lem:no_orbit} applies to $F_j^\pm$:
each field has divergence $-2$, and the matching normal components
checked in Step 1 make internal fluxes cancel. Any nonconstant closed
chain of forward cell arcs contains a simple positive-area subcycle,
whose boundary flux is zero by tangency, contradicting Green's theorem.
Junctions at vertices, including non-NE passages and geometric departures
from stopped saddles, do not affect the flux integrals.

A projected trajectory reaching a mixed-equilibrium vertex remains
there under the selection convention, while the explicit column
evolution above gives convergence of the full strategy profile.
Otherwise, every vertex visit is a non-NE passage.
The proof of Lemma~\ref{lem:nonaccumulation} applies here: the projected
velocity is bounded, and regular switches are transverse by Step 1.
Near a projected mixed equilibrium, we restrict to a neighborhood in the two adjacent
column strips. An affine change of the vertical coordinate sending
$\min\{a_i,a_{i+1}\}$ to $0$ and $\max\{a_i,a_{i+1}\}$ to $1$
puts the local equations in exactly the form used in that lemma, with
the same time variable and stopping convention. At a non-NE vertex,
$y$ is strictly monotone nearby, so it crosses $0$ at most once. On each
side of $y=0$, $p$ is monotone, allowing only finitely many switches
while the trajectory stays in this neighborhood. Thus, there are finitely many
cell arcs between any two finite-time visits. A repeated visit to
the same vertex would therefore give a nonconstant closed
chain, excluded by Step 3 in the proof of Theorem \ref{thm:ct_conv_2by2}. Since there are only finitely many vertices,
we may discard a finite initial time interval containing all vertex
visits. The remaining return and switching arguments concern this
vertex-avoiding tail.

\paragraph{Step 4: crossings of a fixed switching segment are ordered.}

We fix an open switching segment \(S\). It may be either $\{x_i\}\times(0,\overline a)$, $\{x_i\}\times(\underline a,0)$ or $I_j\times\{0\}$. Every crossing of \(S\) has the same orientation: (i) an upper vertical segment is always crossed from left to right; (ii) a lower vertical segment is always crossed from right to left; (iii) a horizontal segment \(I_j\times\{0\}\) is crossed upward if \(a_j>0\) and downward if \(a_j<0\).

We parameterize \(S\) by a scalar coordinate \(s\), and define the first-return map \(\Pi_S\) on points whose trajectories return to \(S\) without hitting a cell vertex. On the vertex-avoiding tail, the proofs of Lemmas~\ref{lem:order} and~\ref{lem:mn_cro} apply: regular switches have unique continuations in both time directions, and Step 3 excludes closed chains. Thus, whenever the tail crosses \(S\) infinitely often, its crossing coordinates are strictly monotone and bounded and hence converge.

\paragraph{Step 5: trajectories with finitely many switches.} Suppose first that the projected trajectory crosses switching lines only finitely many times. After its final regular switch, it lies forever in the closure of a single cell, unless it has arrived at a cell vertex. We consider an upper cell \(R_j^+\). There, $p_t\longrightarrow1, \ y(t)\longrightarrow a_j$. If \(j<K+1\), then \(p_t\) reaches the right boundary \(p=x_j<1\) in finite time, giving another switch. Hence, a final upper cell must be \(R_{K+1}^+\). If $a_{K+1}<0,$
then \(y(t)\), starting positive and converging to a negative number, crosses \(y=0\) in finite time. Hence, necessarily $a_{K+1}>0.$ Thus,
$(p_t,y(t))
\longrightarrow
(1,a_{K+1}).$

Moreover, player 2 eventually plays only column \(K+1\), so $\q_t\longrightarrow \e_{K+1}$. Therefore, the original FP trajectory converges to the pure NE $(1,\e_{K+1})$.

Reflecting $(p,y)\mapsto(1-p,-y)$ and reversing the column order reduces the lower-cell case to the upper-cell case. A final lower cell must be \(R_1^-\), and it can contain the entire tail only if $a_1<0.$ Then
$(p_t,\q_t)
\longrightarrow
(0,\e_1),$
which is the other possible pure NE. Finally, suppose the trajectory reaches a cell vertex \((x_i,0)\). If it remains there, stationarity requires a mixed best response of the column player with zero \(a\)-average. By Step 2, this is possible only when $a_i a_{i+1}<0.$ If $a_i a_{i+1}>0,$ the vertical component has a strict common sign for every mixed strategy supported on the two adjacent column best-response actions, so the trajectory must leave the vertex. Consequently, every finite-switch limit is either a pure-NE projection or a projected mixed equilibrium.

\paragraph{Step 6: trajectories with infinitely many switches.} Now suppose the projected trajectory switches infinitely often.
By the nonaccumulation argument in Step 3, the switching times satisfy
$t_\ell\longrightarrow\infty.$ Let \(S_\ell\) denote the switching segment crossed at time \(t_\ell\), and let $z_\ell:=(p(t_\ell),y(t_\ell)).$ We define the finite set of switching segments visited infinitely often $\mathcal S_\infty
:=
\{S:S_\ell=S\text{ for infinitely many }\ell\}.$

We place an edge \(S\to S'\) when the transition $S_\ell=S,
\ S_{\ell+1}=S'$ occurs infinitely often. After discarding finitely many transitions, the itinerary uses only recurrent edges and visits every recurrent segment infinitely often. The proof of Lemma~\ref{lem:conn} therefore gives a strongly connected graph.

By Step 4, each recurrent segment has a crossing limit. We apply the
limiting-arc and cycle argument from the proof of Theorem~\ref{thm:ct_conv_2by2}
(Equation~\eqref{eqt:limit_seg} and the following paragraphs).
Its hypotheses hold here: the focal points $(0,a_j)$ and $(1,a_j)$
lie off the closures of internal switching segments because
$x_i\in(0,1)$ and $a_j\ne0$. Together with the bounded projected domain
and the exponential cell solutions, this gives the same traversal-time
bounds and limiting cell arcs. Step 3 excludes nonconstant closed
geometric chains, including departures from stopped vertices.
The cited argument therefore gives a common limit $v$ for all recurrent
segments, and hence $z_\ell\longrightarrow v$.
Between consecutive switching points the trajectory is a straight line segment, so every intermediate point is a convex combination of its endpoints. Therefore, $(p_t,y(t))
\longrightarrow
v.$

\paragraph{Step 7: the common limit is the projection of a mixed NE.}

There must be infinitely many vertical switches and infinitely many horizontal switches. If horizontal switches were eventually absent, then the row best response would be constant, and \(p_t\) would be monotone toward \(0\) or \(1\). It could cross each of the finitely many vertical lines at most once. Similarly, if vertical switches were eventually absent, then the column best response would be a fixed \(j\), and $y(t)
=
a_j
+
\bigl(y(T)-a_j\bigr)e^{-(t-T)}$
could cross the horizontal axis at most once. Thus, \(\mathcal S_\infty\) contains both a vertical segment and a horizontal segment. Since all recurrent segments have the same limiting point, $v=(x_i,0)$ for some \(i\). It remains to exclude $a_i a_{i+1}>0.$ Suppose, for example, $a_i>0,\ a_{i+1}>0.$ We choose \(\eta>0\) satisfying $0<\eta
<
\frac12\min\{a_i,a_{i+1}\}.$ In a sufficiently small neighborhood of \((x_i,0)\), the only possible column best responses are \(i\) and \(i+1\). At \(p=x_i\), every mixed column best response $b\in\operatorname{co}\{e_i,e_{i+1}\}$ satisfies
$a^\top b
\ge
\min\{a_i,a_{i+1}\}.$ Therefore, whenever \(|y|<\eta\),
\[
\dot y
=
a^\top b-y
>
\min\{a_i,a_{i+1}\}-\eta
>
0.
\]
Hence, \(y\) is strictly increasing near the limit and can cross \(y=0\) at most once. This contradicts the existence of infinitely many horizontal switches. The case $a_i<0,
a_{i+1}<0$
is identical with all inequalities reversed. Consequently, $a_i a_{i+1}<0.$ Thus, every infinitely switching trajectory converges in the projected plane to the projection of a mixed NE.

Furthermore, an infinite-switch limit must have $a_i>0>a_{i+1},$
the stable spiral orientation. If $a_i<0<a_{i+1},$ the vertex is a saddle, which is reached by only two stable rays.

\paragraph{Step 8: convergence of the column mixed strategy.}

Suppose $p_t\longrightarrow x_i, \ y(t)=a^\top \q_t\longrightarrow0.$ For all sufficiently large \(t\), \(p_t\) lies between the neighboring thresholds \(x_{i-1}\) and \(x_{i+1}\). Therefore, the only column actions that can be selected are \(i\) and \(i+1\). For every $\ell\notin\{i,i+1\},$
$\dot q_\ell=-q_\ell$ for all sufficiently large \(t\), and hence
$q_\ell(t)\longrightarrow0$. Therefore, $q_i(t)+q_{i+1}(t)\longrightarrow1.$ Moreover, $a_iq_i(t)+a_{i+1}q_{i+1}(t)\longrightarrow0.$ These two limiting relations imply
\[
q_i(t)
\longrightarrow
\frac{-a_{i+1}}{a_i-a_{i+1}},\quad
q_{i+1}(t)
\longrightarrow
\frac{a_i}{a_i-a_{i+1}}.
\]
Thus, $\q_t\longrightarrow \q^{i,*},$ and $(p_t,\q_t)
\longrightarrow
(x_i,\q^{i,*}),$
which is the mixed NE identified in Step 2.  This completes the proof.

\subsubsection{Landscape}
\label{app:cls_fp_lsc_2byn}
Let $\phi_t$ denote the selected projected flow under the convention
in Section~\ref{sec:rev_fp}.
Under Assumption~\ref{asp:fp_2byn} and the stopping convention, we provide a geometric analysis to characterize the global landscape of classical FP in $2\times n$ games (Figure~\ref{fig:2byn_lsc}). In particular, we characterize the open attraction basins of stable equilibria and the piecewise-linear saddle separatrices between them.

Let $X:=[0,1]\times[\underline a,\overline a]$ be the projected rectangle, and write $v_i:=(x_i,0)$ for $i=1,\ldots,K$. We first characterize the saddle vertices and the non-NE vertices.

\paragraph{Saddle vertex.}
When $a_i<0<a_{i+1}$, the two local stable branches in the adjacent column strips are
\begin{equation}
\label{eqt:sad_cr_nw}
    \Gamma_i^{NW}
:=
\left\{
(p,y)\in X:x_{i-1}<p<x_i,\ y>0,\quad
y=\frac{-a_i(x_i-p)}{1-x_i}
\right\},
\end{equation}
\begin{equation}
\label{eqt:sad_cr_se}
\Gamma_i^{SE}
:=
\left\{
(p,y)\in X:x_i<p<x_{i+1},\ y<0,\quad
-y=\frac{a_{i+1}(p-x_i)}{x_i}
\right\}.
\end{equation}

These segments form the local portions of the saddle's stable branches.
Extending them backward through successive cells traces the global
piecewise-linear separatrices, with each segment determined by the focal
point of its cell.
A point on either branch reaches \(v_i\) in finite time. Every sufficiently close point outside $\{v_i\}\cup\Gamma_i^{NW}\cup\Gamma_i^{SE}$ leaves a neighborhood of \(v_i\).

\paragraph{Non-NE vertices.}
If \(a_i,a_{i+1}>0\), then in a sufficiently small neighborhood of \(v_i\), $\dot y\geq \frac12\min\{a_i,a_{i+1}\}>0.$
Thus, every nearby trajectory passes upward through a neighborhood of the vertex. Similarly, if \(a_i,a_{i+1}<0\), every nearby trajectory passes downward. Such a point can be a kink or double-switch point on a trajectory or even on a global saddle separatrix, but it does not generate its own basin, stable manifold, or equilibrium.

We define the set of stable mixed-NE projections $\cE_{\mathrm{mix}}^{\mathrm{st}}
:=
\{v_i:a_i>0>a_{i+1}\},$
the set of saddle mixed-NE projections $\cE_{\mathrm{mix}}^{\mathrm{sad}}
:=
\{v_i:a_i<0<a_{i+1}\},$
and the stable pure-NE projections
\[
\cE_{\mathrm{pure}}
:=
\bigl\{(0,a_1):a_1<0\bigr\}
\cup
\bigl\{(1,a_{K+1}):a_{K+1}>0\bigr\}.
\]
We set $\cE_{\mathrm{st}}
:=
\cE_{\mathrm{pure}}
\cup
\cE_{\mathrm{mix}}^{\mathrm{st}}.$ For each saddle \(v\in \cE_{\mathrm{mix}}^{\mathrm{sad}}\), we define its stable set $W^s(v)
:=
\left\{
z\in X:
\lim_{t\to\infty}\phi_t(z)=v
\right\},$
and let $\Sigma
:=
\bigcup_{v\in \cE_{\mathrm{mix}}^{\mathrm{sad}}}W^s(v),
\
\Omega:=X\setminus\Sigma.$
For \(a\in \cE_{\mathrm{st}}\), we define its (attraction) basin $B(a)
:=
\left\{
z\in X:
\lim_{t\to\infty}\phi_t(z)=a
\right\}.$

The following proposition extends the classification of limits and the description of attraction basins in Theorem~\ref{thm:2by2_lsc} to the projected classical FP dynamics in $2\times n$ games.
\begin{prop}[Landscape of classical FP in 2 $\times n$ games]
\label{prop:lsc_2byn}
Under Assumption \ref{asp:fp_2byn}, every nonconstant selected projected trajectory $\z_t$ converges as $t\to\infty$ to a point in
$\cE_{\mathrm{st}}\cup \cE_{\mathrm{mix}}^{\mathrm{sad}}.$ In particular, it converges to either
\begin{itemize}
\item a saddle vertex \(v\in \cE_{\mathrm{mix}}^{\mathrm{sad}}\) after finitely many cell switches only if it belongs to the two-branch stable set \(W^s(v)\). %
\item a stable mixed-NE projection, approached through infinitely many switches.
\item a pure-NE projection with finitely many switches.
\end{itemize}

In addition, the attraction basins are pairwise disjoint, and $\Omega
=
\bigcup_{a\in\cE_{\mathrm{st}}}B(a), \operatorname{Comp}(\Omega)
=
\{B(a):a\in\cE_{\mathrm{st}}\}, \Sigma
=
\bigcup_{a\in\cE_{\mathrm{st}}}\partial_X B(a).$

\end{prop}
The basin identities in Proposition~\ref{prop:lsc_2byn} follow from the local-attraction and continuous-dependence argument of Lemma~\ref{lem:rel_open}, followed by the connected-component argument of Lemma~\ref{lem:boa}. The saddle branches have zero area, so the boundary argument of Remark~\ref{rmk:basin} identifies their union $\Sigma$ with the union of the boundaries of the attraction basins.

\paragraph{Lifting the landscape to the original \(2\times n\) strategy space.}
Let $\mathcal X:=[0,1]\times\Delta_n, F(p,\q):=(p,\mathbf a^\top \q).$
Under the equilibrium-stopping convention, the projected trajectory \(F(p_t,\q_t)\) is determined by the projected initial condition and follows Equation~\eqref{eqt:prj_dy}. For a stable induced equilibrium \(v\), let \(\z_v^*\) be the corresponding original NE. If \(v=(0,a_1)\), $\z_v^*=(0,\e_1)$. If \(v=(1,a_{K+1})\), $\z_v^*=(1,\e_{K+1}).$ If \(v=(x_i,0)\), $\z_v^*=(x_i,\q^{i,*})$ with \(\q^{i,*}\) given by Equation~\eqref{eqt:qi}.

We define the lifted saddle set $\widetilde\Sigma:=F^{-1}(\Sigma),$
and the lifted attraction basins $\widetilde B(\z_v^*)
:=
F^{-1}(B(v)).$ %
Then $\mathcal X
=
\widetilde\Sigma
\cup
\bigcup_{v\in\cE_{\mathrm{st}}}
\widetilde B(\z_v^*),$
where the sets in this decomposition are pairwise disjoint.
Moreover, each \(\widetilde B(z_v^*)\) is relatively open and connected, and \(\widetilde\Sigma\) is a codimension-one piecewise-linear set with zero \(n\)-dimensional measure in \([0,1]\times\Delta_n\). For Lebesgue-almost every original initial condition, the selected trajectory converges to a stable pure or stable mixed NE as $t\to\infty$.

For a mixed sink, once \(p_t\) is sufficiently close to \(x_i\), only columns \(i\) and \(i+1\) are selected. Therefore, all other coordinates of \(q(t)\) decay to zero, while
$q_i(t)+q_{i+1}(t)\to1$ and $a_iq_i(t)+a_{i+1}q_{i+1}(t)\to0.$
This  implies $\q_t\to \q^{i,*}.$

For a column switching threshold $x_i$ outside the simple-tie
setting of Assumption \ref{asp:fp_2byn}, let
$J_i:=J(x_i)$ be the set of tied pure
column-action indices. The Nash equilibria with row
coordinate $x_i$ are exactly $\{x_i\}\times Q_i^*$, where $Q_i^*:=\{\q\in\Delta(J_i):a^\top \q=0\}.$
At $p=x_i$,
the full best-response inclusion has the projected component
\[
\dot y\in\operatorname{co}\{a_j:j\in J_i\}-y,
\qquad y=a^\top \q.
\]
Additional tied columns can therefore change the projected
dynamics and create new projected equilibria. The convergence
and landscape results above retain Assumption \ref{asp:fp_2byn}. This
equilibrium characterization does not extend those results
to ties involving more than two columns.

\subsection{Convergence of NL-FP}

\subsubsection{Proof of Theorem \ref{thm:conv_2byn}}
\label{app:pf_ct_2byn}
We use the selection convention in Section~\ref{sec:grid_2byn}:
at a projected mixed-equilibrium vertex $(r_i,c_\ell)$,
the projection remains fixed and the column strategy satisfies
$\dot\q_t=\q^{i,\ell}-\q_t$.
For the projected dynamics, write
\[
\underline\alpha:=\min_j\alpha_j,
\qquad
\overline\alpha:=\max_j\alpha_j,
\qquad
\beta_k:=\alpha_{j_k},\quad k\in[K+1].
\]
In the risk-neutral specialization $\alpha=a$, this gives
$\beta_k=a_{j_k}=A_{1j_k}-A_{2j_k}$.
We set $c_0:=\underline\alpha,\ c_{L+1}:=\overline\alpha,\
H_\ell:=(c_\ell,c_{\ell+1}), \
\ell=0,\ldots,L.$ We write $\chi(s)=\{\xi_\ell\},
\ s\in H_\ell, \ \xi_\ell\in\{0,1\}.$
By Assumption~\ref{asp:fin_sw}, every row crossing is a switch. Hence,
$\xi_{\ell-1}\ne\xi_\ell$, or equivalently
$\xi_\ell=1-\xi_{\ell-1}$, for every $\ell=1,\ldots,L$.
Each interior crossing is simple, with $\operatorname{BR}^2(r_i)
=
\operatorname{co}\{\e_{j_i},\e_{j_{i+1}}\}.$ We note that the index \(j\) denotes a physical column action, whereas \(k\) denotes a \(p\)-strip. The same column action may be optimal on several disjoint strips, so the sequence \((\beta_k)\) may repeat.
\\

\noindent\textbf{Step 1: projection onto \((K+1)\times (L+1)\) grid.}
The projection $(p,\q)\mapsto(p,s)$, where $s=\alpha^\top\q\in[\underline\alpha,\overline\alpha]$, maps the strategy space to a rectangle and sends a pure column \(j\) to height \(\alpha_j\). When \(p\in I_k\), the column player chooses \(j_k\), and therefore $\dot\q=\e_{j_k}-\q.$ It follows that $\dot s
=
\alpha^\top\dot\q
=
\alpha^\top(\e_{j_k}-\q)
=
\beta_k-s.$
When \(s\in H_\ell\), the row player chooses \(\xi_\ell\), so $\dot p=\xi_\ell-p.$ In the (projected) cell $R_{k\ell}:=I_k\times H_\ell,$
the dynamics are
\[
\dot p=\xi_\ell-p,
\qquad
\dot s=\beta_k-s.
\]
We define the focal point of \(R_{k\ell}\) by $f_{k\ell}:=(\xi_\ell,\beta_k)$. The explicit solution inside the cell is
\[
p_t
=
\xi_\ell+
\bigl(p_{t_0}-\xi_\ell\bigr)e^{-(t-t_0)},\quad
s_t
=
\beta_k+
\bigl(s_{t_0}-\beta_k\bigr)e^{-(t-t_0)}.
\]
Consequently,
\[
(p_t,s_t)
=
(\xi_\ell,\beta_k)
+
e^{-(t-t_0)}
\bigl((p_{t_0},s_{t_0})-f_{k\ell}\bigr).
\]
Within each cell, the trajectory traces a straight line segment toward its cell focal point, as in the proof of Theorem \ref{thm:ct_conv_2by2}. There is no sliding along any open switching segment: on $s=c_\ell$ inside $I_k$, the normal velocity is $\beta_k-c_\ell=\alpha_{j_k}-c_\ell\neq0$, while on $p=r_i$ inside $H_\ell$, it is $\xi_\ell-r_i\neq0$. The only places at which both coordinates can simultaneously be indifferent are the cell vertices $v_{i\ell}:=(r_i,c_\ell).$

\paragraph{Step 2: no nonconstant closed chain of forward cell arcs.}
We use an argument analogous to the cellwise flux proof of
Lemma~\ref{lem:no_orbit} to exclude any nonconstant closed
chain of forward cell arcs. Each projected cell field
$F_{k\ell}(p,s)=(\xi_\ell-p,\beta_k-s)$  has divergence $-2$. Across $p=r_i$,
adjacent fields share the nonzero normal component
$\xi_\ell-r_i$; across $s=c_\ell$, they share
$\beta_k-c_\ell\ne0$. Thus, internal normal fluxes cancel,
and no nonconstant cell arc lies along an internal switching
segment. As in Lemma~\ref{lem:no_orbit}, subdivision at
intersections reduces any nonconstant finite closed chain to
a simple polygon bounding a region $D$ of positive area.
Each nonconstant edge has a unique forward orientation, so
opposite retracing is impossible. Summing the cellwise flux
identities gives $0=-2\operatorname{Area}(D)<0$, since the
field is tangent to each polygonal boundary arc. The finitely
many junctions at cell vertices do not affect these integrals.
This includes non-equilibrium double crossings and geometric
departures from vertices where the selected dynamics would stop.

\paragraph{Step 3: local classification at \((r_i,c_\ell)\).}
Let $\xi_\ell^-:=\xi_{\ell-1}$ be the row response below \(c_\ell\), and let $\xi_\ell^+:=\xi_\ell=1-\xi_\ell^-$
be the response  above it. Near \(v_{i\ell}=(r_i,c_\ell)\), the four fields are
\[
\begin{array}{c|c|c}
& p<r_i & p>r_i\\
\hline
s>c_\ell
&
(\xi_\ell-p,\beta_i-s)
&
(\xi_\ell-p,\beta_{i+1}-s)
\\[1mm]
s<c_\ell
&
(\xi_{\ell-1}-p,\beta_i-s)
&
(\xi_{\ell-1}-p,\beta_{i+1}-s)
\end{array}.
\]
There are two main cases.

The first case is same-side focal heights. If $\beta_i>c_\ell,
\
\beta_{i+1}>c_\ell,$
then \(\dot s>0\) throughout a sufficiently small neighborhood of the vertex. Every nearby trajectory passes upward through \(s=c_\ell\). If $\beta_i<c_\ell,
\
\beta_{i+1}<c_\ell,$
then every nearby trajectory passes downward. These are non-equilibrium crossings.

The other case is opposite-side focal heights. When $(\beta_i-c_\ell)(\beta_{i+1}-c_\ell)<0$, the vertex is the projection of a mixed $\rho$-NE. Its type is determined jointly by the direction in which \(\chi\) crosses at \(c_\ell\), and which side of \(c_\ell\) contains the left focal height \(\beta_i\):
\[
\begin{array}{c|c|c}
\text{Row cross at }c_\ell
&
\beta_i>c_\ell>\beta_{i+1}
&
\beta_i<c_\ell<\beta_{i+1}
\\
\hline
0\text{ below},\ 1\text{ above}
&
\text{stable }
&
\text{saddle}
\\
1\text{ below},\ 0\text{ above}
&
\text{saddle}
&
\text{stable}
\end{array}.
\]

\paragraph{Step 4: return map around stable vertex, and saddle branches.} For a stable vertex, suppose $\xi_\ell^-=0,\
\xi_\ell^+=1$. Let $\beta_i=c_\ell+A,
\
\beta_{i+1}=c_\ell-B,
\
A,B>0.$
Starting from the lower vertical section $(r_i,c_\ell-\rho),$ we can show that $\frac1{\rho_{m+1}}
=
\frac1{\rho_m}
+
\frac1{C_{i\ell}},$
where $C_{i\ell}
:=
\frac{AB(1-r_i)}{A+B}>0.$ Therefore, \(\rho_m\to0\) as $m\to\infty$, and nearby trajectories spiral inward. The other stable orientation, $\xi_\ell^-=1$ and $\xi_\ell^+=0$, reduces to this calculation under the reflection $p\mapsto1-p$, which exchanges the adjacent column intervals and replaces $r_i$ by $1-r_i$.

For a saddle vertex, suppose $\xi_\ell^-=0,
\
\xi_\ell^+=1,$
and $\beta_i=c_\ell-A,
\
\beta_{i+1}=c_\ell+B.$ The two local stable branches are
\[
s-c_\ell
=
\frac{A(r_i-p)}{1-r_i},
\
p<r_i,\ s>c_\ell;\qquad
c_\ell-s
=
\frac{B(p-r_i)}{r_i},
\
p>r_i,\ s<c_\ell.
\]
Points on these branches reach \(v_{i\ell}\) in finite time. Every sufficiently close point not on the branches exits a small neighborhood. Consequently, an infinitely switching trajectory can converge only to a stable vertex, not to a saddle vertex.

\paragraph{Step 5: finitely many switches.}
Suppose the projected trajectory has only finitely many switches.
The local arguments in Steps~3--4 give nonaccumulation as in
Lemma~\ref{lem:nonaccumulation}. A repeated visit to a non-NE vertex
would therefore form a finite closed chain, excluded by Step~2.
Since there are finitely many vertices, we discard a finite prefix
containing all non-NE vertex visits. If the remaining trajectory
reaches a mixed-equilibrium vertex, its projection stays there by
the selection convention, and Step~8 gives convergence of the
full strategy profile. Otherwise, after its last regular switch,
it remains forever in the closure of one cell \(R_{k\ell}\).
Inside that cell it converges toward $(\xi_\ell,\beta_k)$. If \((\xi_\ell,\beta_k)\) is not in the closure of \(R_{k\ell}\), the line segment toward \(f_{k\ell}\) must cross another cell boundary in finite time, contradicting the assumption. Thus, $(\xi_\ell,\beta_k)\in\overline{R_{k\ell}}.$
Because \(\xi_\ell\in\{0,1\}\), this is possible only in a boundary column strip. If \(\xi_\ell=0\), then \(k=1\). If \(\xi_\ell=1\), then \(k=K+1\). The possible non-vertex limits are $(0,\e_{j_1})$ if $0\in\chi(\alpha_{j_1})$, or $(1,\e_{j_{K+1}})$ if $1\in\chi(\alpha_{j_{K+1}})$. %
By the endpoint conditions in Assumption \ref{asp:mean_rho}, the
column response is uniquely $j_1$ throughout $[0,r_1)$
and uniquely $j_{K+1}$ throughout $(r_K,1]$.
Thus, the corresponding column dynamics remain valid
on the outer edges $p=0$ and $p=1$.
If $p_t\to0$, then for some $T$ and all $t\ge T$,
\[
\dot \q_t=\e_{j_1}-\q_t,
\qquad
\q_t=\e_{j_1}+e^{-(t-T)}\bigl(\q_T-\e_{j_1}\bigr)
\longrightarrow \e_{j_1}.
\]
If $p_t\to1$, endpoint uniqueness gives $\dot\q_t=\e_{j_{K+1}}-\q_t$ for all sufficiently large $t$, and therefore $\q_t\to\e_{j_{K+1}}$ as $t\to\infty$.

\paragraph{Step 6: infinitely many switches.}
After discarding a finite prefix, the trajectory avoids vertices:
Steps~3--4 give nonaccumulation as in
Lemma~\ref{lem:nonaccumulation}, and Step~2 excludes repeated
non-NE vertex visits, while the projection remains fixed after
reaching an equilibrium vertex under the selection convention.
Regular switches have unique forward and backward continuations, while Step~2 excludes closed forward chains. The first-return-map and order-preservation arguments of Lemmas~\ref{lem:order} and~\ref{lem:mn_cro} therefore show that crossing coordinates on each infinitely visited open switching segment are strictly monotone and bounded. The subgraph of segments and consecutive transitions
occurring infinitely often is strongly connected by
Lemma~\ref{lem:conn}. Since $r_i\in(0,1)$ and $\beta_k\ne c_\ell$,
focal points avoid internal switching sections, so the exponential
cell solutions give bounded passage times. The limiting-arc
argument of Equation~\eqref{eqt:limit_seg}, together with Step~2,
therefore forces all recurrent crossing limits to coincide.
Both vertical and horizontal segments must recur, since otherwise
one best response becomes constant and only finitely many
switches remain. Thus, the common limit is $v=(r_i,c_\ell)$,
and the intervening straight trajectory segments also converge
to $v$.

\paragraph{Step 7: The adjacent focal heights must cross \(c_\ell\).}
Suppose, for example, $\beta_i>c_\ell,
\
\beta_{i+1}>c_\ell.$ We choose $0<\varepsilon
<
\frac12
\min\{
\beta_i-c_\ell,
\beta_{i+1}-c_\ell
\}.$
In a sufficiently small neighborhood of \(v\), whenever $|s-c_\ell|<\varepsilon,$
one has $\dot s>\varepsilon.$ Thus, \(s_t\) can cross \(c_\ell\) at most once in that neighborhood, contradicting infinitely many horizontal switches converging to \(v\). If both heights lie below $c_\ell$, then $\dot s$ is bounded above by a negative constant near $v$, so $s_t$ can cross $c_\ell$ at most once there. It follows that $(\beta_i-c_\ell)(\beta_{i+1}-c_\ell)<0.$
Therefore, \(v\) is a mixed-equilibrium crossing. Finally, the saddle geometry cannot support infinitely many switches converging to the vertex. Hence, an infinite-switch limit must be a stable mixed crossing, satisfying $(\xi_\ell-\xi_{\ell-1})(\beta_{i+1}-\beta_i)<0.$

\paragraph{Step 8: convergence of the column mixed strategy.}
If the projection reaches a mixed-equilibrium vertex
$(r_i,c_\ell)$ at a finite time $\tau$,
the selection convention gives
\[
\q_t=\q^{i,\ell}+e^{-(t-\tau)}(\q_\tau-\q^{i,\ell})
\longrightarrow\q^{i,\ell}
\qquad\text{as }t\to\infty.
\]
It remains to consider an asymptotic approach to the vertex.
Suppose $(p_t,s_t)\to(r_i,c_\ell)$. By the simple-crossing
assumption, only column actions $j_i,j_{i+1}$ can be selected
for sufficiently large $t$. Hence, $\dot q_h=-q_h$ for
$h\notin\{j_i,j_{i+1}\}$, so these coordinates decay
exponentially. Since $\sum_hq_h=1$ and $s=\alpha^\top\q$,
\[
q_{j_i}+q_{j_{i+1}}\to1,\qquad
\beta_iq_{j_i}+\beta_{i+1}q_{j_{i+1}}\to c_\ell.
\]
Because $\beta_i\ne\beta_{i+1}$, these relations uniquely
determine the limiting mixed strategy $\q^{i,\ell}$ in
Theorem~\ref{thm:conv_2byn}. Thus, $(p_t,\q_t)\to(r_i,\q^{i,\ell})$, completing the proof.

\subsubsection{Landscape of continuous-time NL-FP}
\label{app:lsc_2byn}
Let $\phi_t$ denote the selected projected flow under the convention
in Section~\ref{sec:grid_2byn}.
We now use the local classification above to describe the attraction basins and their saddle-separatrix boundaries.
In this subsection, let
$X:=[0,1]\times[\underline\alpha,\overline\alpha]$
denote the projected rectangle.
For each row crossing, we define its crossing orientation by $\delta_\ell:=\xi_\ell-\xi_{\ell-1}\in\{-1,1\}.$ We define the set of mixed-equilibrium crossings $\V_{\mathrm{eq}}
:=
\left\{
v_{i\ell}:=(r_i,c_\ell):
(\beta_i-c_\ell)(\beta_{i+1}-c_\ell)<0
\right\}.$
We further partition them into $\V_{\mathrm{st}}
:=
\left\{
v_{i\ell}\in \V_{\mathrm{eq}}:
\delta_\ell(\beta_{i+1}-\beta_i)<0
\right\},$
and $\V_{\mathrm{sad}}
:=
\left\{
v_{i\ell}\in \V_{\mathrm{eq}}:
\delta_\ell(\beta_{i+1}-\beta_i)>0
\right\}.$
\(\V_{\mathrm{st}}\) and \(\V_{\mathrm{sad}}\) are the stable and saddle cases in the local classification of Theorem \ref{thm:conv_2byn}. The induced pure equilibria are
\[
\cE_{\mathrm{pure}}
:=
\left\{(0,\beta_1):\chi(\beta_1)=\{0\}\right\}
\cup
\left\{(1,\beta_{K+1}):\chi(\beta_{K+1})=\{1\}\right\}.
\]
We set $\cE_{\mathrm{st}}:=\cE_{\mathrm{pure}}\cup \V_{\mathrm{st}}.$ For \(a\in \cE_{\mathrm{st}}\) and \(v\in \V_{\mathrm{sad}}\), we define
\[
B(a):=\{\z\in X:\lim_{t\to\infty}\phi_t(\z)=a\},
\qquad
W^s(v):=\{\z\in X:\lim_{t\to\infty}\phi_t(\z)=v\},
\]
and $\Sigma:=\bigcup_{v\in \V_{\mathrm{sad}}}W^s(v),
\
\Omega:=X\setminus\Sigma.$
Then we have the following landscape theorem.

\begin{thm}[Landscape of continuous-time $\rho$-FP]
\label{thm:lsc_2byn}
Assume that the column player satisfies
Assumption~\ref{asp:mean_rho} and the row player
satisfies Assumption~\ref{asp:fin_sw}. Every selected projected
trajectory $\z_t$ converges as $t\to\infty$ to a point in
$\cE_{\mathrm{st}}\cup\V_{\mathrm{sad}}$.
For a nonconstant trajectory, the possible outcomes are:
\begin{itemize}
\item a saddle crossing $v\in\V_{\mathrm{sad}}$
after finitely many cell switches if and only if
$\z_0\in W^s(v)$;
\item a stable mixed crossing with infinitely many switches;
\item an induced pure equilibrium with
finitely many switches.
\end{itemize}
The set $\Sigma$ is a closed, finite union of piecewise-linear
saddle separatrices of measure zero. The attraction basins are
relatively open and path-connected, and
\[
\Omega
=
\bigcup_{a\in\cE_{\mathrm{st}}}B(a),
\qquad
\operatorname{Comp}(\Omega)
=
\{B(a):a\in\cE_{\mathrm{st}}\},
\qquad
\Sigma
=
\bigcup_{a\in\cE_{\mathrm{st}}}\partial_X B(a).
\]
\end{thm}

Under $\Pi(p,\q):=(p,\alpha^\top\q),$ the original attraction basins are the inverse images of the projected basins. The lifted set \(\Pi^{-1}(\Sigma)\) has zero \(n\)-dimensional measure in $[0,1]\times\Delta_n.$ Near a stable mixed crossing $(r_i,c_\ell)$, we choose a
positively invariant projected neighborhood where only
$j_i,j_{i+1}$ can be selected. Then $m_t:=\sum_{h\notin\{j_i,j_{i+1}\}}q_h(t)=e^{-t}m_0.$
Since $\beta_i\ne\beta_{i+1}$, normalization and
$s=\alpha^\top\q$ give, for a fixed constant $C>0$,
\[
\|\q_t-\q^{i,\ell}\|_1
\le C\bigl(|s_t-c_\ell|+e^{-t}m_0\bigr),
\qquad t\ge0.
\]
Thus, projected local asymptotic stability implies full-state
local asymptotic stability. At induced pure equilibria,
the fixed local responses give exponential stability directly.
Consequently, almost every original initial condition converges to a locally asymptotically stable pure or mixed \(\rho\)-NE as $t\to\infty$.

\begin{proof}[Proof of Theorem \ref{thm:lsc_2byn}]
\noindent\textbf{Step 1: local classification.}

\noindent\textbf{Stable mixed crossings.} This is covered by Steps~3--4 of the proof of Theorem~\ref{thm:conv_2byn}.

\paragraph{Saddle crossings.}
Let \(v=v_{i\ell}\in \V_{\mathrm{sad}}\), and put $r:=r_i,\
c:=c_\ell,\
A:=|\beta_i-c|,\
B:=|\beta_{i+1}-c|.$
The two local stable branches have the unified expressions
\[
\Gamma^-_{i\ell}
=
\left\{
(p,s):
p<r,\quad
s-c
=
\delta_\ell\frac{A(r-p)}{1-r}
\right\},\quad
\Gamma^+_{i\ell}
=
\left\{
(p,s):
p>r,\quad
s-c
=
-\delta_\ell\frac{B(p-r)}r
\right\},
\]
restricted to a sufficiently small neighborhood of \(v\).

\paragraph{Induced pure equilibria.}

We consider \(a=(0,\beta_1)\in \cE_{\mathrm{pure}}\). Since \(\beta_1\) is not a row crossing level and its row response is \(0\), a sufficiently small relative neighborhood \(U_a\subset X\) lies in the terminal best-response region. There, $\phi_t(\z)
=a+e^{-t}(\z-a).$
Choosing \(U_a\) to be a small relative rectangle makes it positively invariant. Hence, \(a\) is locally asymptotically stable. At $a=(1,\beta_{K+1})$, the local field $(1-p,\beta_{K+1}-s)$ gives $\phi_t(\z)=a+e^{-t}(\z-a)$ on a small invariant relative rectangle, proving local asymptotic stability. Theorem \ref{thm:conv_2byn} gives convergence to the displayed equilibrium set. The local results above distinguish the three possible convergence behaviors.

\paragraph{Step 2: saddle stable sets are finite piecewise-linear separatrices.} We now fix \(v\in \V_{\mathrm{sad}}\). We continue each of its two local stable branches backward. At a regular switching point, continuation is unique by transversality. Continuation is also unique through a non-equilibrium double-switch vertex. Near such a vertex, \(s\) has a strict direction, so the trajectory must pass through it. The resulting branch is piecewise linear and cannot intersect itself.

We claim that a backward branch cannot have infinitely many cell crossings. Suppose otherwise. We reverse time along that branch. The reversed cell dynamics are $\dot \z=\z-\textbf{f}_{k\ell}.$ Their divergence is \(+2\), their internal normal fluxes still match, and regular switches are still transverse. Thus, the ordered-crossing and recurrent-segment argument used in Theorem \ref{thm:conv_2byn} applies to this bounded reversed trajectory as well.  However, no vertex can be such a backward infinite-switch limit. A through-crossing has a strict transverse direction. A saddle permits only finitely many local crossings along a trajectory remaining near it. At a stable mixed vertex, the reversed return map is $u
\longmapsto
\frac{Cu}{C-u}
>
u$ for sufficiently small \(u>0\). Backward turns expand rather than approach the vertex. This is a contradiction. Thus, each backward branch has finitely many cell crossings. After its last crossing, the reversed cell solution $\z_{\tau}
=
\textbf{f}+e^\tau(\z_0-\textbf{f})$
cannot remain in the bounded set \(X\) indefinitely unless it is constant. Therefore the maximal branch terminates at the outer boundary or at a saddle encountered in backward continuation. If the backward endpoint is a saddle $w\ne v$, exclude $w$
from $W^s(v)$, since $\phi_t(w)=w$; it belongs to $\Sigma$
through $W^s(w)$. With this endpoint convention, every point
of $W^s(v)\setminus\{v\}$ lies on one of the two maximal
backward continuations of the local incoming branches,
and conversely every point on those continuations converges
to $v$. Thus, they exhaust $W^s(v)\setminus\{v\}$.

There are finitely many saddles, so the finite-branch
conclusion above makes $\Sigma$ a finite union of line
segments and their endpoints, with planar measure zero.
Its closedness follows below from openness of the attraction
basins. Individual sets $W^s(v)$ need not be closed.

\paragraph{Step 3: Openness, connectedness, and boundaries of the attraction basins.} %

Each stable equilibrium $a$ has a relatively open invariant
attraction neighborhood $A_a$ by Step~1. For a point in
$B(a)\setminus A_a$, we choose a finite trajectory prefix
ending inside $A_a$ and avoiding mixed-equilibrium crossings.
Regular switches have continuous passage maps by
transversality. At a non-equilibrium crossing, both adjacent
focal heights lie strictly on the same side of the row
threshold, so $\dot s$ has a common strict sign in a small
neighborhood. Its passage map is continuous as well:
only finitely many crossing orders are possible, and their
ray-intersection maps agree when an intermediate segment
shrinks to zero. %
Hence, nearby initial conditions also enter $A_a$, and the
finite-prefix argument of Lemma~\ref{lem:rel_open} proves that $B(a)$
is relatively open, without using global continuity of
$\phi_t$.

By Theorem \ref{thm:conv_2byn}, $X\setminus\Sigma$ is the disjoint union
of these attraction basins. Each basin is path-connected:
for any $z\in B(a)$, the path
$u\mapsto\phi_{u/(1-u)}(z)$ on $0\le u<1$, extended by
the value $a$ at $u=1$, joins $z$ to $a$ inside $B(a)$.
Thus, the connected components of $X\setminus\Sigma$ are
exactly the attraction basins, as in Lemma \ref{lem:boa}.
Their openness makes $\Sigma$ closed, and its measure-zero
property from Step~2 gives
$\Sigma=\bigcup_{a\in \cE_{\mathrm{st}}}\partial_X B(a)$
by the argument of Remark \ref{rmk:basin}.

\end{proof}

\subsubsection{Proof of Theorem \ref{thm:conv_dt_2byn}}
\label{app:pf_dt_2byn}

The proof follows the geometric route in the $2\times 2$ case in Appendix \ref{app:dt}. The only additional passage argument concerns non-equilibrium cell crossings. Lemma~\ref{lem:uni_comp} excludes vertices, whereas in the projected \(2\times n\) system a non-equilibrium crossing may lie on a trajectory or even on a saddle separatrix. At every mixed-equilibrium crossing, an affine rescaling gives exactly the local \(2\times2\) recursion, so the local-capture and saddle-passage lemmas can be reused.

As in Section~\ref{sec:dt}, let
$\gamma_{t+1}=1/(t+2)$.
For the projected state $\z_t:=(p_t,s_t)$, where
$s_t=\alpha^\top\q_t$, the recursion is
\[
\z_{t+1}
=
\z_t+\gamma_{t+1}(\f_t-\z_t),
\qquad
\textbf{f}_t:=(x_t,\alpha_{J_t}).
\]

Here $x_t\in\operatorname{BR}^1_\rho(\q_t)=\chi(s_t)\cap\{0,1\}$ and $J_t\in\operatorname{BR}^2_\rho(p_t)=J(p_t)$
may depend on the iteration and the preceding history.
We use the proofs of Lemmas~\ref{lem:uni_comp},
\ref{lem:local_att}, and~\ref{lem:passage} uniformly over
these choices. At a regular vertical switching point
$p_t=r_i$, with $s_t\in(c_\ell,c_{\ell+1})$, $p_{t+1}-r_i=\gamma_{t+1}(\xi_\ell-r_i)\neq0,$
independently of the column tie choice. At a regular
horizontal switching point $s_t=c_\ell$, with $p_t\in I_k$, $s_{t+1}-c_\ell=\gamma_{t+1}(\beta_k-c_\ell)\neq0,$
independently of the row tie choice. A tie at a regular switching point therefore
changes only a tangential displacement of order
$\gamma_{t+1}$, so the finite-itinerary estimates are uniform.
At a non-equilibrium vertex, the common strict sign of
$\beta_i-c_\ell$ and $\beta_{i+1}-c_\ell$ gives the same
uniformity in Lemma~\ref{lem:F14} below.
At a mixed-equilibrium vertex, the affine rescaling below
gives binary local responses. The local-capture proof is
already uniform over tie choices, while the saddle-passage
proof uses only purity and the vanishing bounds for exact
landings and double crossings. No consistency across
different visits to a tie is used. Hence, the subsequent
finite-prefix and saddle-transition arguments apply to
every admissible sequence of pure best responses.

On a cell \(R_{k\ell}\), its focal point is $\f_{k\ell}=(\xi_\ell,\beta_k)$. If the selected focal point remains equal to \(\textbf{f}\) for $t=N,\ldots,T-1,$
then
\[
\z_T-\f
=
\prod_{t=N}^{T-1}(1-\gamma_{t+1})(\z_N-\f)
=
\frac{N+1}{T+1}(\z_N-\f).
\]
The discrete iterates follow exactly the same ray as the continuous trajectory until the response changes. We note that this is the identity used in Appendix~\ref{app:dt},
with a general focal point in place of a corner.

Since \(\z_t,\f_t\in X\), $\|\z_{t+1}-\z_t\|_\infty
\le
M\gamma_{t+1},$ where $M:=\max\{1,\overline\alpha-\underline\alpha\}$.
For \(N\ge1\), we define $\varepsilon_N:=\frac{M}{N+1}$.  If a step at period \(t\ge N\) crosses a switching line at the virtual crossing point \(\widehat\zeta\), then $\|\z_{t+1}-\widehat\zeta\|_\infty
\le
\varepsilon_N.$ Consequently, every late overshoot vanishes.

On any fixed finite itinerary consisting of regular switches, Lemma~\ref{lem:uni_comp} applies with modified constants. The same proof applies with general focal points in place of the corners of the unit square. For a sufficiently small initial discrepancy \(\delta\), with the perturbed point in the neighborhood carrying the chosen itinerary, the crossing errors are bounded by $A\delta+C\varepsilon_N.$
The same error recursion $e_{m+1}
\le
L_m(e_m+\varepsilon_N)$ holds
and only finitely many such inequalities are composed.

We provide a small extension of Lemma~\ref{lem:uni_comp}.

\begin{lem}
\label{lem:F14}
   A sequence of discrete iterates starting at a sufficiently late period follows any fixed finite continuous-time trajectory segment that avoids mixed-equilibrium crossings, with endpoint error $C\bigl(\delta+\varepsilon_N\bigr)$
even when the continuous-time trajectory segment passes through non-equilibrium cell vertices. At such a vertex, the discrete and continuous crossing orders need not agree.
\end{lem}
\begin{proof}[Proof of Lemma \ref{lem:F14}]
    Finite trajectory segments avoiding cell vertices are already covered by Lemma~\ref{lem:uni_comp}. We consider a non-equilibrium vertex $v=(r_i,c_\ell)$ with $(\beta_i-c_\ell)(\beta_{i+1}-c_\ell)>0$.
Suppose first that both focal heights exceed \(c_\ell\). We choose a sufficiently small neighborhood \(Q\) of \(v\) and \(\kappa>0\) such that $\beta_i-s\ge\kappa,
\
\beta_{i+1}-s\ge\kappa$ on $Q$.
Then every continuous passage satisfies $\dot s\ge\kappa,$
and every discrete step starting in \(Q\) satisfies $s_{t+1}-s_t\ge\kappa\gamma_{t+1}.$ Thus, \(s\) has a strict common direction through the neighborhood. We choose a small incoming section below \(c_\ell\) and an outgoing section above it, transverse to the reference passage. Nearby trajectories reach the outgoing section without leaving a slightly larger neighborhood. During this passage, the horizontal threshold is crossed at most once. Before that crossing, the row response is constant, so \(p\) is monotone and crosses \(p=r_i\) at most once. Afterward, the row response is again constant, so there is at most one further vertical crossing. Hence, there are only finitely many possible local itineraries. Their ray-intersection maps have bounded derivatives: the horizontal normal velocities are bounded below by \(\kappa\), while the vertical normal velocities at \(p=r_i\) have magnitude at least $\min\{r_i,1-r_i\}>0.$

When two crossings merge at \(v\), the intervening ray segment has zero length. The passage maps for the possible itineraries agree at this limiting passage: \(s\) crosses \(c_\ell\) in a fixed direction, and the row response on the outgoing side determines the subsequent direction of \(p\). Thus, the local passage map is continuous and piecewise Lipschitz. For a discrete passage, each ordinary overshoot contributes at most \(\varepsilon_N\). If one step crosses both incident lines, its endpoint satisfies $\|\z_{t+1}-v\|_\infty\le\varepsilon_N.$
Inserting or deleting the corresponding short intermediate pieces therefore changes the passage endpoint by \(O(\varepsilon_N)\). Applying the finite-map estimate to the finitely many possible crossing orders gives
\[
\|\widehat P(w)-P(w_0)\|_\infty
\le
C\bigl(\|w-w_0\|_\infty+\varepsilon_N\bigr),
\]
where \(w_0\) is the reference entry point. The same reasoning covers a prefix starting at \(v\).

The case in which both focal heights lie below \(c_\ell\) is identical, with \(s\) decreasing. Composing these local passage estimates with the finite-itinerary estimates for portions avoiding cell vertices finishes the proof.
\end{proof}
This extension compares passage endpoints, not identical itineraries through double crossings. Next, we reuse local capture and saddle passage. We fix a mixed-equilibrium crossing $v=(r_i,c_\ell)$ with $(\beta_i-c_\ell)(\beta_{i+1}-c_\ell)<0.$
We define $b_v:=\min\{\beta_i,\beta_{i+1}\},
\
d_v:=|\beta_{i+1}-\beta_i|>0,$
and introduce the affine coordinate $y:=\frac{s-b_v}{d_v},
\
\theta_v:=\frac{c_\ell-b_v}{d_v}\in(0,1).$
We choose the local neighborhood small enough that it contains no other switching lines and lies in $0<p<1, 0<y<1.$ In these coordinates, the projected recursion becomes
\[
\begin{aligned}
p_{t+1}
=
p_t+\gamma_{t+1}(x_t-p_t),\quad
y_{t+1}
=
y_t+\gamma_{t+1}(b_t-y_t),
\end{aligned}
\]
where $x_t,b_t\in\{0,1\}.$ The row response switches at \(y=\theta_v\), and the column response switches at \(p=r_i\). This is exactly the local \(2\times2\) recursion of Appendix \ref{app:dt}, with either its stable or its saddle orientation.

\paragraph{Local capture at stable equilibria.}

For a stable mixed crossing $v=(r_i,c_\ell)$, let $T_v(p,s):=\left(p,\frac{s-b_v}{d_v}\right),
\
\widetilde v:=T_v(v)=(r_i,\theta_v).$
The preceding affine change of coordinates conjugates the
local continuous fields and the discrete recursion to the
stable $2\times2$ system, with the same step sizes and with
admissible tie choices mapped to admissible tie choices. We apply the local-capture construction in the proof of Lemma \ref{lem:local_att}
at $\widetilde v$, choosing its neighborhoods
$\widetilde U_v^-\Subset\widetilde U_v^+$ inside the
transformed four-cell neighborhood. We set
$U_v^\pm:=T_v^{-1}(\widetilde U_v^\pm)$ in the projected
rectangle. The proof of Lemma \ref{lem:local_att} gives trapping and
convergence in these neighborhoods, uniformly over all
admissible tie choices. In particular, it remains applicable
when the projected tie choices depend on the mixed-strategy
profile $(p,\q)$. No comparison of crossing itineraries near $v$
is required. For an induced pure equilibrium $a$, we choose nested small convex relative neighborhoods inside its terminal
best-response region. The focal point is $a$, and $\z_t-a=\frac{N+1}{t+1}(\z_N-a),\ t\ge N\ge1$.
Consequently, for every $a\in \cE_{\mathrm{st}}$, there exist
$U_a^-\Subset U_a^+\subset B(a)$ and $N_a<\infty$ such that
\[
N\ge N_a,\quad \z_N\in U_a^-
\quad\Longrightarrow\quad
\z_t\in U_a^+\ \text{for all }t\in\mathbb N,\ t\ge N,
\qquad \lim_{t\to\infty}\z_t=a.
\]

\paragraph{Passage near a saddle.}

Lemma \ref{lem:passage} likewise applies to every saddle crossing. We choose a small rectangle \(Q_v\), with incoming gates $\iota_v^1,\iota_v^2$ and geometric outgoing gates $o_v^1,o_v^2.$
If entry periods satisfy $N_m\to\infty$ and entry points satisfy $\eta_m\to\iota_v^\ell$ as $m\to\infty$, then either the discrete process converges to \(v\) as its iteration index $t\to\infty$, or, after taking a subsequence, its exit points satisfy $\zeta_m\to o_v^r$ as $m\to\infty$ for one \(r\in\{1,2\}\). The first switching point of an incoming ray tends to \(v\) as $m\to\infty$. If only one incident line is crossed, the process enters an outgoing cell near \(v\), and then follows an outgoing ray to the boundary of \(Q_v\). If a step crosses both lines, its endpoint is within \(\varepsilon_N\) of \(v\), in the opposite incoming cell.

By the affine rescaling, Lemma~\ref{lem:passage} also covers
exact landings on either incident line or at $v$.
Isolated tie landings add $O(\gamma_t)$ errors. Consecutive
line landings and double crossings remain within
$O(\gamma_t)$ of $v$. The incoming velocity bounds control
intervening iterates, so the process either converges to $v$
or enters an outgoing cell near $v$.
The locally Lipschitz ray-intersection maps therefore give
\begin{equation}
    \operatorname{dist}\bigl(
\zeta,
\{o_v^1,o_v^2\}
\bigr)
\le
C_v
\bigl(
\|\eta-\iota_v^\ell\|_\infty
+
\varepsilon_N
\bigr),
\end{equation}
whenever an exit occurs. A long sequence of double crossings does not accumulate a macroscopic error. Each double crossing returns the process to the current vanishing overshoot scale.

We now establish global capture for the projected discrete-time iterates, extending Lemma~\ref{lem:global_basin} from the unit square to $X$. Let \(a\in \cE_{\mathrm{st}}\) and let \(K\subset B(a)\) be compact. Then there exist a relatively open neighborhood $K\subset O_K\subset B(a)$ and \(N_K<\infty\) such that
\begin{equation}
\label{eqt:F6}
    N\ge N_K,\quad \z_N\in O_K
\quad\Longrightarrow\quad
\lim_{t\to\infty}\z_t=a.
\end{equation}
We fix \(w\in K\). If $w\in U_a^-$, the local-capture conclusion above already applies on a neighborhood of $w$. Otherwise, we use the finite-prefix argument below. By continuous-time convergence, $\phi_t(w)\to a.$
We choose a finite continuous-time trajectory segment ending strictly inside \(U_a^-\). It cannot encounter a mixed-equilibrium crossing before reaching that neighborhood: under the continuous equilibrium-stopping convention, such a trajectory would remain at that crossing. The prefix may encounter non-equilibrium crossings, but we have covered them. Thus, for sufficiently late starting periods and initial points sufficiently close to \(w\), the discrete iterates enter \(U_a^-\). The local capture conclusion established above then gives convergence to \(a\).

A finite subcover of \(K\) makes the starting-time requirement uniform. This proves Equation~\eqref{eqt:F6}, by the same compactness argument as Lemma \ref{lem:global_basin}. Consequently,
\begin{equation}
\label{eqt:F7}
    \omega_d(\z_0)\cap B(a)\ne\varnothing
\quad\Longrightarrow\quad
\lim_{t\to\infty}\z_t=a,
\end{equation}
where
\[
\omega_d(\z_0):=
\left\{w:\exists\,(t_m)_{m\ge1}\subset\mathbb N,\quad
t_m\to\infty\ \text{and}\ \z_{t_m}\to w
\ \text{as }m\to\infty\right\}.
\]
Thus, a nonconvergent sequence of discrete iterates cannot have a limit point inside any continuous-time attraction basin.

We show that saddle transitions cannot continue indefinitely by constructing the saddle-transition graph exactly as in Lemma \ref{lem:acyclic}. For each saddle \(v\) and each outgoing gate \(o_v^r\), Theorems \ref{thm:conv_2byn}-\ref{thm:lsc_2byn} imply that either $o_v^r\in B(a)$ for some $a\in \cE_{\mathrm{st}}$ or $o_v^r\in W^s(w)$  for some saddle $w$. In the latter case, we introduce the edge $v\longrightarrow w.$ This finite graph is acyclic. Indeed, a directed cycle would concatenate the geometric outgoing rays and the finite incoming separatrix trajectories into a nonconstant closed chain of forward cell arcs. Such a chain is excluded by  Green's-theorem argument, which is the proof of Lemma \ref{lem:acyclic}.

We define the finite projected equilibrium set $\widehat \cE_\rho:=\cE_{\mathrm{st}}\cup \V_{\mathrm{sad}}.$ We can show that $\omega_d(\z_0)\subseteq\widehat \cE_\rho.$ Suppose otherwise. We choose $w\in\omega_d(\z_0)\setminus\widehat \cE_\rho.$
By Theorem \ref{thm:lsc_2byn}, either \(w\) lies in an attraction basin or $w\in W^s(v_0)\setminus\{v_0\}$ for a saddle \(v_0\).
On the other hand, the case in which $w$ lies in an attraction basin is impossible by Equation~\eqref{eqt:F7}: it would imply convergence to a stable equilibrium, contradicting the choice of \(w\).

In the second case, we choose $N_m\to\infty,
\
\z_{N_m}\to w,$
and take a small saddle box \(Q_{v_0}\) not containing \(w\). The continuous trajectory from \(w\) reaches an incoming gate of \(Q_{v_0}\) through finitely many cell passages.  The corresponding late discrete iterates enter \(Q_{v_0}\) at points tending to that gate.

We apply the saddle-passage conclusion. If the discrete iterates converge to \(v_0\), then \(w\) cannot be a distinct limit point. Otherwise, a subsequence of exits converges to some outgoing gate \(o_{v_0}^{r_0}\). If that gate belongs to an attraction basin, Equation~\eqref{eqt:F7} forces convergence, again a contradiction. Therefore, $o_{v_0}^{r_0}\in W^s(v_1)$ for another saddle \(v_1\), so $v_0\longrightarrow v_1.$
We repeat the finite-prefix comparison and saddle-passage argument from that outgoing gate. Unless convergence has already occurred, this construction produces a directed path $v_0\longrightarrow v_1\longrightarrow v_2\longrightarrow\cdots$
in the saddle-transition graph. Because the graph is finite and acyclic, such a path cannot continue indefinitely. At a terminal saddle, every outgoing gate belongs to an attraction basin, which forces capture.

Every alternative contradicts the existence of $w\in\omega_d(\z_0)\setminus\widehat \cE_\rho.$
Hence, $\omega_d(\z_0)\subseteq\widehat \cE_\rho$ holds. This is the argument of Lemma \ref{lem:lim}.

The remaining argument from Appendix \ref{app:dt} applies unchanged. The projected rectangle \(X\) is compact, and $\|\z_{t+1}-\z_t\|_\infty\longrightarrow0.$ Therefore \(\omega_d(\z_0)\) is nonempty, compact, and connected. To recall the connectedness argument, if it were the union of two nonempty disjoint compact sets at positive distance, then every sufficiently late iterate would lie near one of those sets. Since both neighborhoods are visited infinitely often, the sequence would have to move between them. Vanishing increments make such a transition impossible without producing an additional limit point outside their union. By $\omega_d(\z_0)\subseteq\widehat \cE_\rho$, this connected set is contained in the finite set \(\widehat \cE_\rho\). Hence, $\omega_d(\z_0)=\{\z_\infty\}$ for some \(\z_\infty\in\widehat \cE_\rho\), and $\z_t\longrightarrow \z_\infty.$

The last step is to lift convergence to the original strategy profile.  Suppose first that $\z_t=(p_t,s_t)\longrightarrow(r_i,c_\ell).$
For all sufficiently large \(t\), the only column actions that can be selected are \(j_i\) and \(j_{i+1}\). Thus, for some \(N\ge1\) and every $h\notin\{j_i,j_{i+1}\},$ $q_{t,h}
=
\frac{N+1}{t+1}q_{N,h}
\longrightarrow0.$ Consequently, $q_{t,j_i}+q_{t,j_{i+1}}\longrightarrow1.$ Since \(\alpha^\top \q_t\to c_\ell\), we also obtain $\beta_iq_{t,j_i}
+
\beta_{i+1}q_{t,j_{i+1}}
\longrightarrow c_\ell.$
The straddling condition implies \(\beta_i\ne\beta_{i+1}\), so these two equations uniquely determine the limit:
\[
\q_t
\longrightarrow
\frac{\beta_{i+1}-c_\ell}{\beta_{i+1}-\beta_i}\,\e_{j_i}
+
\frac{c_\ell-\beta_i}{\beta_{i+1}-\beta_i}\,\e_{j_{i+1}}
=
\q^{i,\ell}.
\]
Therefore, $(p_t,\q_t)\longrightarrow(r_i,\q^{i,\ell})$. If the projected limit is $(0,\beta_1)$ or $(1,\beta_{K+1})$, the column action is eventually fixed at $j=j_1$ or $j=j_{K+1}$, respectively. For a sufficiently large iteration $N$ and every $t\ge N$, the recursion gives
\[
\q_t-\e_j=\frac{N+1}{t+1}(\q_N-\e_j)\longrightarrow0
\qquad\text{as }t\to\infty.
\]
Every resulting profile is a \(\rho\)-NE in the set specified by Theorem \ref{thm:conv_2byn}. This completes the proof.

\section{Proof of nonconvergence in a \texorpdfstring{$2\times 3$}{2 x 3} game}
\label{app:2by3}
We prove Proposition~\ref{prop:cvar_nonconvergence} by constructing an exact nonconstant periodic trajectory in a game with integer payoffs. We consider the bimatrix game
\[
\begin{array}{c|ccc}
 & 1 & 2 & 3\\ \hline
1 &(9,0)&(0,1)&(-3,3)\\
0 &(5,0)&(-1,-2)&(2,-7).
\end{array}
\]
The row player uses lower-tail CVaR with tail mass $\alpha=\frac15$, whereas the column player is risk-neutral.

Let \(p\in[0,1]\) denote the empirical frequency with which the row
player uses action \(1\), and let \(\q\in\Delta_3\) denote the empirical distribution of the column player's actions. With pure selections, Equation~\eqref{eqt:cfp_2byn} gives the continuous-time dynamics
\begin{equation}
\dot p=x-p,
\qquad
\dot \q=\e_J-\q,
\qquad
x\in\operatorname{BR}^1_\rho(\q),
\quad
J\in\operatorname{BR}^2_\rho(p).
\label{eq:clean_dynamics}
\end{equation}

\paragraph{The nonlinear row player.}
Let the row player's payoff vectors for actions \(1\) and \(0\) be $\vu=(9,0,-3), \
\vv=(5,-1,2),$
respectively. We define the CVaR advantage $d(\q)
:=
C_{1/5}(\vu;\q)-C_{1/5}(\vv;\q).$
The payoff orderings are $u_3<u_2<u_1, v_2<v_3<v_1.$ On the region \(q_3<1/5\), the two CVaR values are
\begin{align}
C_{1/5}(\vu;\q)
&=
\begin{cases}
9-45q_2-60q_3,
& q_2+q_3\le \frac15,\\[1mm]
-15q_3,
& q_2+q_3\ge \frac15,
\end{cases}
\label{eq:clean_cvar_u}\\
C_{1/5}(\vv;\q)
&=
\begin{cases}
5-30q_2-15q_3,
& q_2+q_3\le \frac15,\\[1mm]
2-15q_2,
& q_2+q_3\ge \frac15,\quad q_2\le\frac15,\\[1mm]
-1,
& q_2\ge\frac15.
\end{cases}
\label{eq:clean_cvar_v}
\end{align}
Subtracting \eqref{eq:clean_cvar_v} from
\eqref{eq:clean_cvar_u} gives
\begin{equation}
d(\q)=
\begin{cases}
4-15q_2-45q_3,
& q_2+q_3\le \frac15,\\[1mm]
15(q_2-q_3)-2,
& q_2+q_3\ge\frac15,\quad q_2\le\frac15,\\[1mm]
1-15q_3,
& q_2\ge\frac15.
\end{cases}
\label{eq:clean_dq}
\end{equation}
The expressions in \eqref{eq:clean_dq} agree on both regime
boundaries. On \(q_2+q_3=1/5\), the first two expressions are
both \(1-30q_3\), and on \(q_2=1/5\), the last two expressions are both
\(1-15q_3\). Hence, \(d\) is continuous on this region.

We consider the segment from \(\e_1\) into the interior of the simplex,
\begin{equation}
\q(\lambda)
:=
\left(
1-\frac{\lambda}{3},
\frac{23\lambda}{90},
\frac{7\lambda}{90}
\right),
\qquad
0\le\lambda\le1.
\label{eq:clean_ray}
\end{equation}
Along this segment,
\[
q_2(\lambda)+q_3(\lambda)=\frac{\lambda}{3},
\qquad
q_2(\lambda)=\frac{23\lambda}{90},
\qquad
q_3(\lambda)=\frac{7\lambda}{90}<\frac15.
\]
The two CVaR-regime boundaries are therefore $q_2(\lambda)+q_3(\lambda)=\frac15$ iff $\lambda=\frac35$, and $q_2(\lambda)=\frac15$ iff $\lambda=\frac{18}{23}$.
Substituting \eqref{eq:clean_ray} into \eqref{eq:clean_dq} gives
\begin{equation}
d(\q(\lambda))
=
\begin{cases}
\dfrac23(6-11\lambda),
&0\le\lambda\le\frac35,\\[2mm]
\dfrac23(4\lambda-3),
&\frac35\le\lambda\le\frac{18}{23},\\[2mm]
\dfrac16(6-7\lambda),
&\frac{18}{23}\le\lambda\le1.
\end{cases}
\label{eq:clean_dlambda}
\end{equation}
Consequently, \(d(\q(\lambda))\) has exactly three zeros, $\lambda_1=\frac6{11},\lambda_2=\frac34,\lambda_3=\frac67.$
Their ordering relative to the CVaR kink points is $\frac6{11}
<
\frac35
<
\frac34
<
\frac{18}{23}
<
\frac67.$
The sign pattern, as \(\lambda\) increases, is $+\;,-\;,+\;,-.$
Equivalently, $d(\q(\lambda))>0$ for $\lambda\in
\left[0,\frac6{11}\right)
\cup
\left(\frac34,\frac67\right)$, and $d(\q(\lambda))<0$ for $\lambda\in
\left(\frac6{11},\frac34\right)
\cup
\left(\frac67,1\right].$

\paragraph{The risk-neutral column player.}

The expected payoffs of the column player's three actions are $G_1(p)=0,
\
G_2(p)=3p-2,
\
G_3(p)=10p-7.$
The relevant pairwise differences are
$G_2(p)-G_1(p)=3p-2,
\
G_3(p)-G_2(p)=7p-5,
\
G_3(p)-G_1(p)=10p-7.$
It follows that the column player's pure-action best-response correspondence is
\begin{equation}
\operatorname{BR}^2_\rho(p)
=
\begin{cases}
\{1\},
&0\le p<\frac23,\\[1mm]
\{1,2\},
&p=\frac23,\\[1mm]
\{2\},
&\frac23<p<\frac57,\\[1mm]
\{2,3\},
&p=\frac57,\\[1mm]
\{3\},
&\frac57<p\le1.
\end{cases}
\label{eq:clean_column_br}
\end{equation}

\paragraph{Exact closed orbit.}

We set $p_0=\frac23, \z_0=\q(1)$, where \(\z_k\) denotes the column-distribution waypoint after the
\(k\)-th segment. We consider the action sequence %
\begin{equation}
\bigl((x_k,J_k)\bigr)_{k=0}^7
=
\bigl(
(0,1),(1,1),(0,1),(1,1),
(1,2),(1,3),(0,3),(0,2)
\bigr),
\label{eq:clean_action_itinerary}
\end{equation}
and the contraction factors $(r_k)_{k=0}^7
=
\left(
\frac67,\frac78,\frac8{11},\frac{11}{18},
\frac67,\frac{21}{22},\frac{55}{56},\frac{14}{15}
\right).$ We define the segment durations $\tau_k:=-\log r_k>0$ and
switching times $t_0:=0,\
t_{k+1}:=t_k+\tau_k,\ k=0,\ldots,7.$
We write $(p_k,\z_k):=(p(t_k),\q(t_k))$ for the waypoints,
$k=0,\ldots,8$. Thus, segment $k$ connects waypoint $k$ to
waypoint $k+1$, with actions $(x_k,J_k)$ fixed on
$(t_k,t_{k+1})$. Solving \eqref{eq:clean_dynamics} on this
segment gives
\begin{equation}
p_{k+1}
=
x_k+r_k(p_k-x_k),
\qquad
\z_{k+1}
=
\e_{J_k}+r_k(\z_k-\e_{J_k}),
\qquad k=0,\ldots,7.
\label{eq:clean_segment}
\end{equation}
Applying \eqref{eq:clean_segment} successively gives
\begin{equation}
\frac23
\longrightarrow
\frac47
\longrightarrow
\frac58
\longrightarrow
\frac5{11}
\longrightarrow
\frac23
\longrightarrow
\frac57
\longrightarrow
\frac8{11}
\longrightarrow
\frac57
\longrightarrow
\frac23,
\label{eq:clean_p_sequence}
\end{equation}
for the \(p\)-coordinate, and
\begin{align*}
\z_0&=
\left(
\frac23,\frac{23}{90},\frac7{90}
\right),\
\z_1=
\left(
\frac57,\frac{23}{105},\frac1{15}
\right),\
\z_2=
\left(
\frac34,\frac{23}{120},\frac7{120}
\right), \
\z_3=
\left(
\frac9{11},\frac{23}{165},\frac7{165}
\right), \\
\z_4&=
\left(
\frac89,\frac{23}{270},\frac7{270}
\right),\
\z_5=
\left(
\frac{16}{21},\frac{68}{315},\frac1{45}
\right),\
\z_6=
\left(
\frac8{11},\frac{34}{165},\frac1{15}
\right),\
\z_7=
\left(
\frac57,\frac{17}{84},\frac1{12}
\right).
\end{align*}
Finally,
\begin{equation}
\z_8
=
\left(
\frac23,\frac{23}{90},\frac7{90}
\right)
=\z_0.
\label{eq:clean_z8}
\end{equation}
Every fixed-action trajectory segment is the line segment joining its
two endpoints. Since $\max_{0\le k\le8}(\z_k)_3=\frac1{12}<\frac15,$
the entire orbit lies in the region \(q_3<1/5\). Thus,
\eqref{eq:clean_dq} is valid on every segment of the orbit.

For fixed \(x_k\), the function $p_t=x_k+(p_k-x_k)e^{-(t-t_k)}$
is monotone on the \(k\)-th segment. Therefore, on the interiors of
the first four segments, \(p<2/3\); on the fifth segment,
\(2/3<p<5/7\); on the sixth and seventh segments, \(p>5/7\); and on
the eighth segment, \(2/3<p<5/7\). Hence, the prescribed column-player
actions $1,1,1,1,2,3,3,2$ agree with \eqref{eq:clean_column_br} throughout the interiors of all
eight segments.

It remains to verify the prescribed row-player actions. During the
first four segments, \(J_k=1\), and hence \(\z\) remains on the segment
\eqref{eq:clean_ray}. More precisely,
\[
\z_0=\q(1),
\qquad
\z_1=\q\left(\frac67\right),
\qquad
\z_2=\q\left(\frac34\right),
\qquad
\z_3=\q\left(\frac6{11}\right),
\qquad
\z_4=\q\left(\frac13\right).
\]
The segment parameter decreases as $1
\longrightarrow
\frac67
\longrightarrow
\frac34
\longrightarrow
\frac6{11}
\longrightarrow
\frac13.$
The signs of \(d\) on the four open segments are $-,\ +,\ -,\ +.$ Thus, the row-player actions on the first four segments are $0,\ 1,\ 0,\ 1.$

For the segment from \(\z_4\) to \(\z_5\), we define $\z_{45}(\theta)
:=
(1-\theta)\z_4+\theta \z_5,
\
0\le\theta\le1.$ This segment crosses the two CVaR kink lines at $(\z_{45}(\theta))_2+(\z_{45}(\theta))_3=\frac15$ when $\theta=\frac7{10},$ and $(\z_{45}(\theta))_2=\frac15$ when $\theta=\frac{217}{247}.$
At the endpoints and the two kink points, $d(\z_{45}(0))=\frac{14}{9}, d\left(\z_{45}\left(\frac7{10}\right)\right)=\frac3{10},$
and $d\left(\z_{45}\left(\frac{217}{247}\right)\right)
=\frac{163}{247}, d(\z_{45}(1))=\frac23.$ All four values are strictly positive. Since \(d(\z_{45}(\theta))\)
is affine between successive kink points, it follows that $d(\z_{45}(\theta))>0,
\ 0\le\theta\le1.$
Hence, no additional row-player switch occurs on the segment
\(\z_4\to \z_5\).

On the remaining three segments, $q_2\ge\frac{17}{84}>\frac15,$
so the third expression in \eqref{eq:clean_dq} applies: $d(\q)=1-15q_3.$
On \(\z_5\to \z_6\), the coordinate \(q_3\) increases from \(1/45\) to
\(1/15\), so $d(\q)>0$ on the open segment and \(d(\z_6)=0\). On \(\z_6\to \z_7\), \(q_3\) increases from \(1/15\) to \(1/12\), so
$d(\q)<0$ on the open segment. Finally, on \(\z_7\to \z_8\), \(q_3\) remains between \(7/90\) and \(1/12\), both of which are strictly larger than
\(1/15\), so \(d(\q)<0\) throughout that segment. Thus, the row-player
actions on the final four segments are $1,\ 1,\ 0,\ 0.$
This verifies the full action sequence
\eqref{eq:clean_action_itinerary}.

By \eqref{eq:clean_p_sequence} and \eqref{eq:clean_z8}, $p_8=p_0,
\ \z_8=\z_0.$
Moreover, we have $\prod_{k=0}^7r_k=\frac14.$ Since \(\tau_k=-\log r_k\), the total return time is $T
=
\sum_{k=0}^7\tau_k
=
-\log\left(\prod_{k=0}^7r_k\right)
=
\log4.$
Repeating the eight-segment concatenation periodically therefore
produces an absolutely continuous solution satisfying
\[
(p_{t+\log4},\q_{t+\log4})
=
(p_t,\q_t)
\qquad
\text{for every }t\ge0.
\]
The solution is nonconstant, so continuous-time
NL-FP does not converge in general $2\times n$ games.

\end{document}